\documentclass[aps,showpacs,twocolumn,superscriptaddress]{revtex4-2}
\usepackage{graphicx}% Include figure files
\usepackage{dcolumn}% Align table columns on the decimal point
\usepackage{bm}% bold math
\usepackage{multirow}
\usepackage[normalem]{ulem} % \sout{old text} for strikeout
\usepackage[dvipsnames]{xcolor} % For blue in-text comments
\usepackage{hyperref}
\hypersetup{
  colorlinks=true,       
  linkcolor=blue,         
  citecolor=cyan,         % color of links to bibliography
}
\usepackage{mathrsfs}
\usepackage[utf8]{inputenc}
\usepackage{mathtools}
\usepackage{orcidlink}
\usepackage{amsmath}
\usepackage{amssymb}
\usepackage{enumitem} 

\begin{document}

\title{Magnetically charged spinning particles around a regular black hole: constraints on the monopole charge from quasi-periodic oscillations}
\author{Shokhzod Jumaniyozov \orcidlink{0009-0009-6254-5608}}
\email{sh.jumaniyozov@newuu.uz}
\affiliation{New Uzbekistan University, Movarounnahr Street 1, Tashkent 100000, Uzbekistan}
\affiliation{Tashkent State Technical University, Tashkent 100095, Uzbekistan}

\author{Javlon Rayimbaev\orcidlink{0000-0001-9293-1838}}
\email{javlonrayimbaev6@gmail.com}
\affiliation{Kimyo International University in Tashkent, Shota Rustaveli Street 156, Tashkent 100121, Uzbekistan}

\author{Chengxun~Yuan~\orcidlink{0000-0002-2308-6703}}
\email{corresponding author: yuancx@hit.edu.cn}
\affiliation{School of Physics, Harbin Institute of Technology, Harbin 150001, People’s Republic of China}

\author{Ahmadjon~Abdujabbarov\orcidlink{0000-0002-6686-3787}} \email{ahmadjon@astrin.uz}
\affiliation{School of Physics, Harbin Institute of Technology, Harbin 150001, China}
\affiliation{Institute of Fundamental and Applied Research, National Research University TIIAME, Kori Niyoziy 39, Tashkent 100000, Uzbekistan}
\affiliation{University of Tashkent for Applied Sciences, Str. Gavhar 1, Tashkent 100149, Uzbekistan}

\author{Faisal Javed\orcidlink{0000-0001-6970-1305}}
\email{faisaljaved.math@gmail.com}
\affiliation{School of Mathematical and Statistics, Zaozhuang University, Zaozhuang 277160, China}
\affiliation{Research Center of Astrophysics and Cosmology, Khazar University, Baku, AZ1096, 41 Mehseti Street, Azerbaijan}
\affiliation{Department of Mathematics, School of Science, University of Management and Technology,  Lahore, 54000, Pakistan}

\author{Satimbay~Palvanov\orcidlink{0000-0002-2694-6545}}
\email{satimbay@yandex.ru}
\affiliation{National University of Uzbekistan, Tashkent 100174, Uzbekistan}

\date{\today}

\begin{abstract}
We investigate the dynamics of magnetically charged spinning test particles in the spacetime of the Bardeen regular black hole, sourced by nonlinear electrodynamics and featuring a magnetic monopole charge parameter $g$. Using the Mathisson-Papapetrou-Dixon equations with the Tulczyjew spin condition, extended by the Coulomb-like force between the particle's magnetic charge and the monopole, we derive the effective potential for equatorial motion and analyze circular orbits, the innermost stable circular orbit (ISCO), superluminality constraints and high-energy collisions. Solving the ISCO conditions perturbatively around Schwarzschild we obtain exact coefficients, $r_{\rm ISCO}=6M-\tfrac{2\sqrt6}{3}s-\tfrac{19}{12}g^2/M+\dots$, and show that the magnetic coupling $\lambda g$ shifts $r_{\rm ISCO}$ only at higher order while it changes the ISCO energy and angular momentum at first order. We compute radiative efficiencies (up to $\simeq 8\%$ in the explored range, against $5.72\%$ for Schwarzschild) and the orbital and radial epicyclic frequencies, including the ISCO and the 3:2 resonance, whose frequency at fixed mass is shifted by $5$--$11\%$ for $g/M=0.5$--$0.7$. All effective-potential results are verified against a direct integration of the full MPD system, and a superluminality bound for circular orbits is derived. Finally, we show that the centre-of-mass energy of colliding particles is finite for every sub-extremal Bardeen black hole, as for any black hole with a non-degenerate horizon, with a horizon value $2(1+\mathcal L_{\rm cr}^2/r_h^2)$ that actually grows with $g$ (from $10$ for Schwarzschild to about $25$ at the extremal charge), and that it diverges as $(r-r_h)^{-1}$ only on the degenerate horizon $g=4M/3\sqrt3$ for a critical charged particle. These results identify distinctive, quantitatively predicted signatures of the Bardeen geometry in accretion-disk and quasi-periodic-oscillation observables.
\end{abstract}

%\pacs{04.50.-h, 04.40.Dg, 97.60.Gb}

\maketitle

\section{Introduction}
\label{sec:intro}

Singularities in spacetime represent one of the most profound shortcomings of classical general relativity, signaling the theory's breakdown under extreme conditions~\cite{Senovilla:2014gza,Hawking:1970zqf}. To address this issue, researchers have proposed various regular black hole models, often arising from modified gravity theories or nonlinear couplings between gravity and matter fields. A pioneering example is the Bardeen black hole~\cite{1968qtr..conf...87B}, later reinterpreted as an exact solution of Einstein's equations coupled to nonlinear electrodynamics~\cite{Cai:2025ejd,Ayon-Beato:1999kuh,NOORIGASHTI2026117244}. This spherically symmetric and static spacetime replaces the central singularity with a regular de Sitter-like core, supported by a magnetic monopole charge parameter $g$.  In the limit $g \to 0$, the metric recovers the Schwarzschild solution, while finite $g$ ensures that all curvature invariants remain bounded at $r=0$. The Bardeen model has since inspired a broad class of regular black holes and serves as an important phenomenological tool for exploring potential quantum gravity effects that might regularize singularities~\cite{Dymnikova:2004zc,Cordeiro:2025ydg,2021A&AT...32...83N}.

The motion of test particles around black holes directly probes gravity in the strong-field regime. While geodesic equations suffice for structureless particles, astrophysically realistic objects such as neutron stars or magnetized compact bodies possess intrinsic spin and possibly charge, leading to significant deviations from geodesic trajectories. The dynamics of spinning test particles is governed by the Mathisson-Papapetrou-Dixon (MPD) equations~\cite{Mathisson:1937zz,Papapetrou:1951pa,Dixon-1970}, which incorporate spin-curvature coupling in the pole-dipole approximation. This coupling gives rise to gravitomagnetic effects analogous to frame-dragging in rotating spacetimes, resulting in modified circular orbits, shifted ISCOs, and constraints against superluminal motion~\cite{PhysRevD.6.1476,Semerak_1999}. In regular black holes, the absence of a singularity further modifies these phenomena, often introducing natural bounds on spin-induced accelerations.

Magnetic charges, though not yet observed, are theoretically motivated by grand unified theories and could play a role in extreme astrophysical environments. The interaction between a magnetically charged test particle and the monopole field of the Bardeen black hole introduces a generalized Lorentz force that can be either attractive or repulsive, depending on the relative sign of the charges. Combined with spin-curvature coupling, this force produces rich orbital phenomenology that differs markedly from electrically charged or uncharged cases~\cite{Blaschke:2022pgf,Nascimento:2025get,Rayimbaev:2022mrk}.

Another key aspect of black hole physics is the possibility of ultra-high-energy particle collisions near the horizon. The Bañados-Silk-West (BSW) mechanism~\cite{Banados:2009pr} shows that, near the degenerate horizon of an extremal rotating black hole, center-of-mass energies can diverge for particles on near-critical trajectories; analogous statements hold for extremal charged black holes~\cite{Zaslavskii:2010jd}, whereas for non-extremal black holes the collision energy at the horizon is bounded~\cite{Berti:2009bk}. Extensions to spinning and charged particles have revealed further enhancements or restrictions. For regular black holes it has been argued that the modified near-horizon structure turns potentially infinite collision energies into finite ones~\cite{Ma:2025tzm}. Here we examine which property of the horizon is actually responsible for the finiteness in the Bardeen case.

Motivated by these considerations, this work systematically investigates the combined influence of particle spin and magnetic charge on dynamics around the Bardeen regular black hole~\cite{doi:10.1142/S0218271822500559,Zeng:2025wlb}. Using the MPD equations with the Tulczyjew spin-supplementary condition and incorporating magnetic interactions through the generalized Lorentz force, we derive the effective potential for equatorial motion. We analyze circular orbits and the ISCO, explore constraints on superluminal trajectories, and compute critical angular momenta and center-of-mass collision energies near the horizon. Our results reveal how prograde spin shifts the ISCO and the critical orbits inward, while the magnetic coupling mainly modifies the energy and angular momentum of the orbits; we give analytic expansions, radiative efficiencies and orbital frequencies, verify the effective-potential analysis against the full MPD equations, and show that the finite collision energy of the Bardeen black hole is a consequence of its non-degenerate horizon, with a divergence appearing only at the extremal charge $g=4M/3\sqrt3$.

This manuscript is organized as follows: Section~\ref{sec:spacetime} reviews the magnetically charged Bardeen spacetime and fixes the coupling and the sign convention of the magnetic interaction. In Section~\ref{sec:eom}, we derive and discuss the equations of motion for spinning test particles with magnetic charge using the MPD equations. Section~\ref{sec:effpot} presents the effective potential governing the radial motion and analyzes the properties of circular orbits, including the ISCO, its analytic expansion, radiative efficiencies and characteristic frequencies. In Section~\ref{sec:superluminal}, we analyze the superluminal constraints. Section~\ref{sec:collisions} explores the critical angular momentum and the center-of-mass energy of collisions between such particles near the horizon, including the degenerate-horizon limit. Section~\ref{sec:discussion} discusses physical scales and limits of validity, and Section~\ref{sec:conclusion} summarizes the key findings. Appendix~\ref{app:mpd} contains the verification of our results against the full MPD equations.

Throughout this work, we adopt the metric signature $(-,+,+,+)$ and geometrized units with $G = c = 1$. Greek indices run from 0 to 3, and Latin indices run from 1 to 3.

\section{Magnetically charged Bardeen Black Hole Spacetime}\label{sec:spacetime}

In this work, we consider the spacetime geometry of a nonlinear electrodynamics regular black hole solution originally proposed by Bardeen~\cite{1968qtr..conf...87B}. This model describes a spherically symmetric, static black hole whose singularity at the origin is regularized by introducing a magnetic monopole. The line element in spherical coordinates $(t,r,\theta,\phi)$ is given by
\begin{equation}
ds^2 = -f(r)\,dt^2 + f(r)^{-1}\,dr^2 + r^2 \bigl(d\theta^2 + \sin^2\theta\, d\phi^2\bigr),
\label{eq:metric}
\end{equation}
where the lapse function is as follows
\begin{equation}
f(r) = 1 - \frac{2Mr^2}{(r^2 + g^2)^{3/2}}.
\label{eq:lapse}
\end{equation}
Here, $M$ denotes the black hole mass, and $g$ is the magnetic monopole charge parameter. In the limit $g\to 0$, the metric reduces to the Schwarzschild solution, while for finite $g$ the curvature invariants remain finite at $r=0$, rendering the center regular~\cite{Wang:2025oek,Mahanta:2025uph}. An event horizon exists only for a sufficiently small monopole charge: the condition $f(r_h)=0$ admits real, positive roots only up to the extremal value obtained from the simultaneous solution of $f(r_h)=0$ and $f'(r_h)=0$, which yields $g/M \simeq 0.7698$ (at $r_h/M \simeq 1.0887$); beyond this value the black hole horizon disappears, and the geometry describes a horizonless regular compact object. Throughout this work we therefore restrict our analysis to the sub-extremal regime, and, unless stated otherwise, fix $g/M = 0.5$, which lies safely within it.

The electromagnetic field supporting this geometry arises from nonlinear electrodynamics coupled to gravity. Following the formal duality construction of Ref.~\cite{Rayimbaev:2022hrn}, the four-potential is written as

\begin{equation}
    A_{\mu} = \left( A^{*}_{t}, 0, 0, A_{\phi} \right), \quad A^{*}_{t} = \frac{i g}{r}, \quad A_{\phi} = g \cos \theta.
    \label{elmag}
\end{equation}

Here the asterisk does not denote complex conjugation: $A^*_t$ is a purely formal, auxiliary quantity that encodes the magnetic-charge coupling in the same algebraic form as an electric potential, so that the equations of motion for a magnetically charged particle can be obtained from the electrically charged ones by the substitution $q_e A_t \to q_m A^*_t$. The physical field is purely magnetic, with the only genuine (real) component of the four-potential being $A_\phi = g\cos\theta$; the associated field strength tensor $F_{\mu\nu} = \partial_\mu A_\nu - \partial_\nu A_\mu$ then corresponds to a magnetic monopole field with magnetic charge $g$, and the explicit factor of $i$ multiplying $q_m$ in Eq.~\eqref{s209} below is what restores a real, physical conserved energy.

\paragraph*{Coupling, equatorial consistency and sign convention.}
The test charge couples to the \emph{dual} field strength $\tilde F_{\mu\nu}=\tfrac12\varepsilon_{\mu\nu\alpha\beta}F^{\alpha\beta}$ of the monopole, whose only non-vanishing components are $\tilde F_{tr}=-\tilde F_{rt}=g/r^{2}$, i.e.\ $\tilde F=d\tilde A$ with $\tilde A_t=g/r=-iA^{*}_t$; this is precisely the role of the formal potential $A^{*}_t$ in Eq.~\eqref{elmag}. The magnetic force on the particle is therefore purely radial and Coulomb-like, $\propto q_m g/r^{2}$. In particular its $\theta$-component vanishes identically, while the term $q_mA_\phi=q_m g\cos\theta$ in the angular momentum, Eq.~\eqref{s210}, vanishes on $\theta=\pi/2$. Since the spin--curvature force has no $\theta$-component either for a spin tensor with $S^{\theta\alpha}=0$ in a static, spherically symmetric background, the equatorial plane is an invariant subspace of the MPD system and the reduction to equatorial motion used below is exactly consistent (we also verified this by integrating the complete four-dimensional MPD system, Appendix~\ref{app:mpd}). The situation would be different for an \emph{electrically} charged particle in the monopole field, which feels $F_{\theta\phi}=-g\sin\theta$ and, as in Poincar\'e's classic problem~\cite{Poincare:1896}, moves on cones rather than in a plane; such orbits are not considered here. We treat the dual field as the Maxwell-like field of the monopole acting on the test charge, neglecting corrections from the nonlinear constitutive relation of the source.

\emph{Sign convention.} With $\tilde A_t=+g/r$ the conserved energy, Eq.~\eqref{s209}, contains $+q_mg/r$ and, for a static particle far from the black hole, $\mathcal E\simeq\sqrt{f}-\lambda g/r$. A positive product $\lambda g$ therefore \emph{lowers} the energy (binding, ``attractive''), while $\lambda g<0$ raises it (``repulsive''). We use this convention in all statements and figures. In terms of the physical rule that like magnetic charges repel, it corresponds to measuring the particle's charge-to-mass ratio as $-\lambda$; the opposite convention is obtained by $\lambda\to-\lambda$ throughout.

This spacetime provides a suitable arena for investigating the influence of magnetic monopole charges on the geodesic motion of test particles, particularly spinning and magnetically charged ones, as well as high-energy particle collisions near the horizon.

\section{Equations of motion for a magnetically charged spinning test particle}
\label{sec:eom}

The dynamics of a magnetically charged spinning test particle in the curved spacetime of the Bardeen black hole differ significantly from those of non-spinning or uncharged particles because of the interplay among the particle's intrinsic spin, its magnetic charge, and the spacetime geometry regularized by the magnetic monopole field. To describe this motion, we use the MPD equations in the pole-dipole approximation, extended to include the electromagnetic interactions of a charged particle. These equations capture both the spin-curvature coupling and the generalized Lorentz force arising from the magnetic interaction. They are particularly well-suited to studying astrophysical scenarios in which spin and charge effects can substantially alter trajectories and orbital stability.

The MPD equations for a magnetically charged spinning particle in the presence of a magnetic monopole field are given by~\cite{Jumaniyozov:2025nnj,Oteev:2025csb}
\begin{eqnarray}
\frac{D p^\alpha}{d \tau} &=& -\frac{1}{2} R^\alpha_{\;\beta \delta \sigma} u^\beta S^{\delta \sigma} + q_m F^\alpha_{\;\beta} u^\beta,
\label{eq:mpd_mom}
\\
\frac{D S^{\alpha \beta}}{d \tau} &=& p^\alpha u^\beta - p^\beta u^\alpha,
\label{eq:mpd_spin}
\end{eqnarray}
where $D/d\tau \equiv u^\mu \nabla_\mu$ denotes the covariant derivative along the particle's worldline parameterized by proper time $\tau$, $p^\alpha$ is the four-momentum, $u^\alpha$ is the four-velocity, $S^{\alpha \beta}$ is the antisymmetric spin tensor ($S^{\alpha \beta} = -S^{\beta \alpha}$), $R^\alpha_{\;\beta \delta \sigma}$ is the Riemann curvature tensor, $q_m$ is the particle's magnetic charge (its dimensionless specific value, $\lambda \equiv q_m/m$, is introduced in Eq.~\eqref{eq:lambda_def} below), and $F_{\alpha \beta} = \partial_\alpha A_\beta - \partial_\beta A_\alpha$ is the electromagnetic field strength tensor derived from the vector potential in Eq.~\eqref{elmag}.

The first term on the right-hand side of Eq.~\eqref{eq:mpd_mom}, $-\frac{1}{2} R^\alpha_{\;\beta \delta \sigma} u^\beta S^{\delta \sigma}$, represents the spin--curvature coupling (gravitomagnetic force), while the second term, $\lambda F^\alpha_{\;\beta} u^\beta$, corresponds to the generalized Lorentz force due to the interaction between the particle's magnetic charge and the monopole field of the black hole.
To close the system of equations, we adopt the Tulczyjew spin-supplementary condition~\cite{Iosifidis:2025ano}
\begin{equation}
S^{\alpha \beta} p_\beta = 0,
\label{eq:tssc}
\end{equation}
which ensures that the spin tensor is orthogonal to the four-momentum and defines a unique worldline of the particle's centroid. This condition is particularly appropriate for asymptotically flat spacetimes such as the Bardeen solution.
Differentiating Eq.~\eqref{eq:tssc} covariantly and substituting into the MPD equations yields conservation of the four-momentum magnitude~\cite{Jumaniyozov:2025irx}
\begin{equation}
p^\alpha p_\alpha = -m^2,
\label{eq:mom_norm}
\end{equation}
where $m$ is the particle's rest mass. Similarly, the spin magnitude is conserved~\cite{Chen:2025ncm}
\begin{equation}\label{8}
S^{\mu \nu} S_{\mu \nu} = 2 S^2 = 2 m^2 s^2,
\end{equation}
with $s$ denoting the specific spin parameter. (Note that, following common usage in this literature, the symbol $S$ denotes both the antisymmetric spin tensor $S^{\alpha\beta}$ and, unadorned, its magnitude $S=sm$ as in Eq.~\eqref{8}; the intended meaning is fixed by the number of indices carried.) The stationarity and axial symmetry of the spacetime imply two Killing vectors, $\xi^\alpha_{(t)} = \delta^\alpha_t$ and $\xi^\alpha_{(\phi)} = \delta^\alpha_\phi$, leading to conserved quantities~\cite{Oteev:2025yjh}
\begin{equation}
C_\xi = p^\alpha \xi_\alpha + \frac{1}{2} S^{\alpha \beta} \nabla_\beta \xi_\alpha +  q_m A_\alpha \xi^\alpha- i q_m A^*_\alpha \xi^\alpha,
\label{eq:cons_charge}
\end{equation}

Given the form of the vector potential $A_\phi = g \cos \theta$ and $A^*_t = ig/r$, the real electromagnetic term $q_m A_\alpha \xi^\alpha$ in Eq.~\eqref{eq:cons_charge} contributes only to the conserved angular momentum (through $A_\phi$), while the auxiliary term $-iq_m A^*_\alpha \xi^\alpha$ contributes only to the conserved energy (through $A^*_t$); the explicit factor of $i$ in $A^*_t$ cancels against the explicit $-i$ in Eq.~\eqref{eq:cons_charge}, so that the resulting energy is manifestly real. The conserved energy and angular momentum are as follows:
\begin{eqnarray}\label{s209}\nonumber
-E&=&p_t+\frac{1}{2}g_{t\alpha,\beta}S^{\alpha\beta}-iq_mA^*_t\\
&=& p_t + \frac{1}{2} g_{tt,r} S^{tr} + \frac{q_m g}{r}
\\
\label{s210}\nonumber
J&=&p_\phi+\frac{1}{2}g_{\phi\alpha,\beta}S^{\alpha\beta}+q_m A_{\phi}\\&=& p_\phi + \frac{1}{2} g_{\phi \phi,r} S^{\phi r}+q_m g \cos\theta,
\end{eqnarray}

where $ J = L + S $, with $ S = s m $, $ L = \mathcal{L} m $, and $ E = \mathcal{E} m $. The term $-iq_mA^*_t = q_m g/r$ reflects the electromagnetic interaction and modifies the conserved energy compared to the uncharged case. (Since $\Gamma^\sigma_{\beta\alpha}$ is symmetric in $\alpha,\beta$ while $S^{\alpha\beta}$ is antisymmetric, the Christoffel contribution to $S^{\alpha\beta}\nabla_\beta\xi_\alpha$ vanishes identically, so this term reduces to $S^{\alpha\beta}\partial_\beta\xi_\alpha$; for the diagonal Bardeen metric this gives $g_{tt,r}S^{tr}$ for the time Killing vector and $g_{\phi\phi,r}S^{\phi r}$ for the axial one, both entering $E$ and $J$ with the same overall sign, exactly as Eq.~\eqref{eq:cons_charge} requires.) Restricting the motion to the equatorial plane ($\theta = \pi/2$, $p^\theta = 0$), the spin tensor reduces to fewer independent components, and the Tulczyjew condition yields relations analogous to those in electrically charged cases, with appropriate adjustments for the magnetic interaction.

Here, we introduce the dimensionless specific magnetic charge of the particle, denoted by 

\begin{eqnarray}
    \lambda =\frac{q_m}{m}
    \label{eq:lambda_def}
\end{eqnarray}
where \(q_m\) is the magnetic charge and \(m\) is the rest mass of the particle. This normalization ensures that \(\lambda\) has the same dimensions as the specific spin parameter \(s\) (both are dimensionless in geometrized units), facilitating a consistent comparison of spin and magnetic effects. Similarly, the black hole's magnetic monopole charge is characterized by the dimensionless deviation parameter \(g/M\). The sign of the product \(\lambda g\) determines the nature of the magnetic interaction: \(\lambda g > 0\) corresponds to an attractive (binding) Coulomb-like force, while \(\lambda g < 0\) yields a repulsive force that counteracts the gravitational binding (see the sign convention in Sec.~\ref{sec:spacetime}). This magnetic contribution enters the conserved quantities through the terms proportional to \(q_m = \lambda m\) in Eqs.~\eqref{s209} and \eqref{s210}, modifying both the effective energy and angular momentum of the particle compared to the uncharged case.

In astrophysical contexts, such as accretion disks, we confine the motion to the equatorial plane ($\theta = \pi/2$), where $p_\theta = 0$, and the spin tensor components simplify to $S^{tr}$, $S^{t\phi}$, and $S^{r\phi}$, with $S^{\theta\alpha} = 0$. Applying the Tulczyjew supplementary condition from Eq.~\eqref{eq:tssc}, we obtain:

\begin{eqnarray}
    \label{14}
    && S^{t \phi} p_\phi + S^{tr} p_r = 0 \quad\Rightarrow\quad S^{t \phi} = -\frac{p_r}{p_\phi} S^{tr}, 
    \\\label{15}
    && S^{r \phi} p_t + S^{tr} p_r + S^{t \phi} p_\phi = 0 \quad\Rightarrow\quad S^{r \phi} = \frac{p_t}{p_\phi} S^{tr}, \nonumber
\end{eqnarray}

where the second relation follows after substituting Eq.~\eqref{14} into the first constraint. Using the momentum normalization condition $g_{\alpha \beta} p^\alpha p^\beta = -m^2$ in the equatorial plane, we arrive at~\cite{Benkrane:2025kdq}:
\begin{equation}\label{16}
    (p^r)^2 = -g^{rr} \left[ g^{tt} p_t^2 + g^{\phi \phi} p_\phi^2 + m^2 \right].
\end{equation}

Substituting Eqs.~\eqref{14}-\eqref{16} into the spin conservation law Eq.~\eqref{8}, the non-vanishing spin tensor components become~\cite{Abdukayumova:2025ztr}:
\begin{eqnarray}\label{ssc2}
    S^{tr} = \pm \frac{p_\phi s}{\sqrt{-g_{tt} g_{rr} g_{\phi \phi}}}, \quad
    S^{r \phi} = \pm \frac{p_t s}{\sqrt{-g_{tt} g_{rr} g_{\phi \phi}}},
\end{eqnarray}
where the $\pm$ signs correspond to the direction of the spin relative to $p_\phi$ and are correlated between the two components (both take the upper sign, or both the lower sign, consistently, as required by Eq.~\eqref{15}), and $S = s m$ is the spin magnitude.  The $\pm$ signs in Eq.~\eqref{ssc2} indicate prograde ($+$) or retrograde ($-$) alignment of the particle’s spin with its orbital angular momentum $p_\phi$. Prograde alignment strengthens centrifugal support via spin-curvature coupling, thereby reducing the ISCO radius. Retrograde alignment, conversely, opposes the orbital motion, resulting in a larger $r_{\rm ISCO}$ and a reduced specific angular momentum $\mathcal{L}_{\rm ISCO}$. We have confirmed that stable orbits satisfying $d^2 V_{\rm eff}/dr^2 > 0$ exist for both orientations, though the prograde case is generally more relevant to accretion disk dynamics. (The overall sign of $S^{\alpha\beta}$ is a matter of convention: what is physically meaningful is that $J=\mathcal L+s$, i.e.\ $s>0$ denotes a spin component along $+\phi$, parallel to $\mathcal L>0$. In the numerical verification of Appendix~\ref{app:mpd}, where the MPD system is used exactly as written in Eqs.~\eqref{eq:mpd_mom}--\eqref{eq:mpd_spin} with the standard Riemann convention, the spin tensor must carry the opposite overall sign to Eq.~\eqref{ssc2} for the conserved quantities to take the values $(\mathcal E,J)$ given above.) Finally, solving the conservation laws in Eqs.~\eqref{s209} and \eqref{s210} yields the covariant four-momentum components:
\begin{eqnarray}
    \label{s2e13}
    && p_t = \frac{JSf'-2Er-2g \lambda }{2 r-S^2f'}
    \\\label{s2e14}
    && p_\phi =\frac{2 J r^2-2 r S (g \lambda +E r)}{r \left(2 r-S^2 f'\right)}
\end{eqnarray}
where $f' = \partial f(r) / \partial r$. The contravariant components are obtained by raising indices: $p^\alpha = g^{\alpha \beta} p_\beta$. These expressions provide the foundation for constructing the effective potential and studying orbital dynamics in the following sections.

These equations and conserved quantities form the foundation for deriving the effective potential and analyzing circular orbits, innermost stable circular orbits, superluminal constraints, critical angular momentum, and center-of-mass collision energies in the subsequent sections.
 
\section{Effective Potential for Magnetically Charged Spinning Particles}
\label{sec:effpot}

\begin{figure*}[ht!]
   \centering
\includegraphics[width=0.49\linewidth]{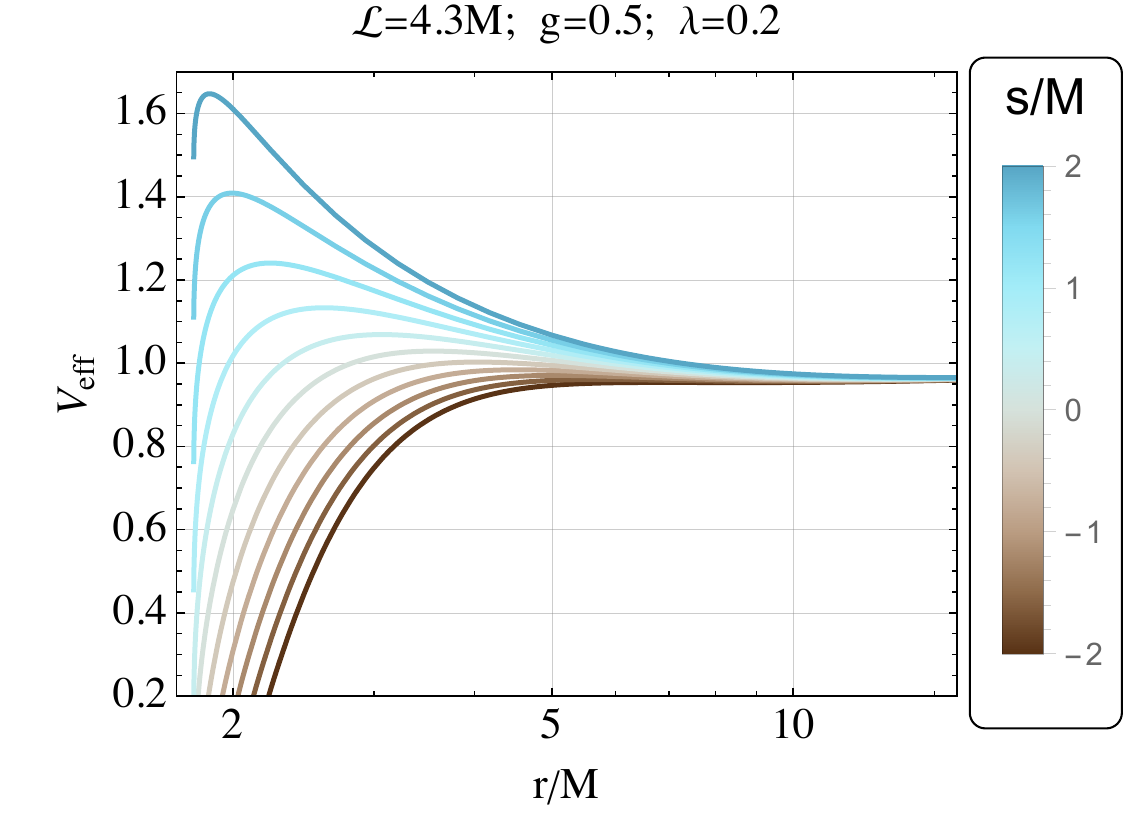}
\includegraphics[width=0.49\linewidth]{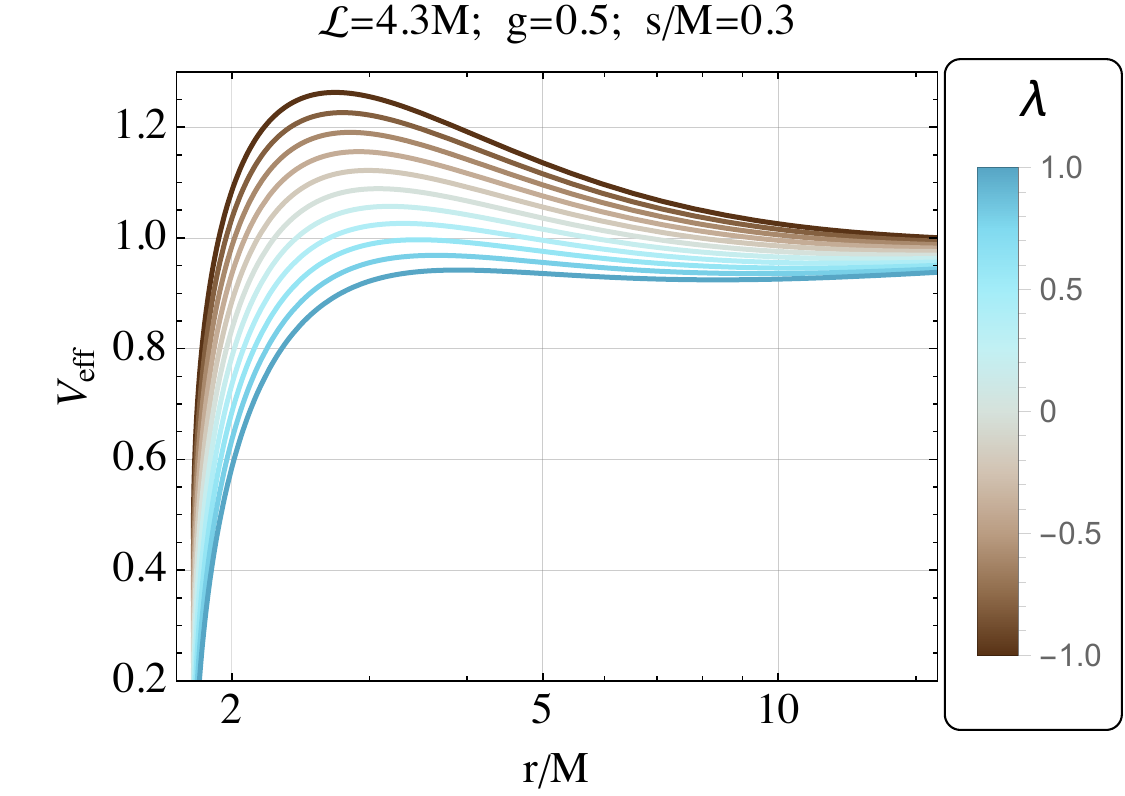}
\includegraphics[width=0.49\linewidth]{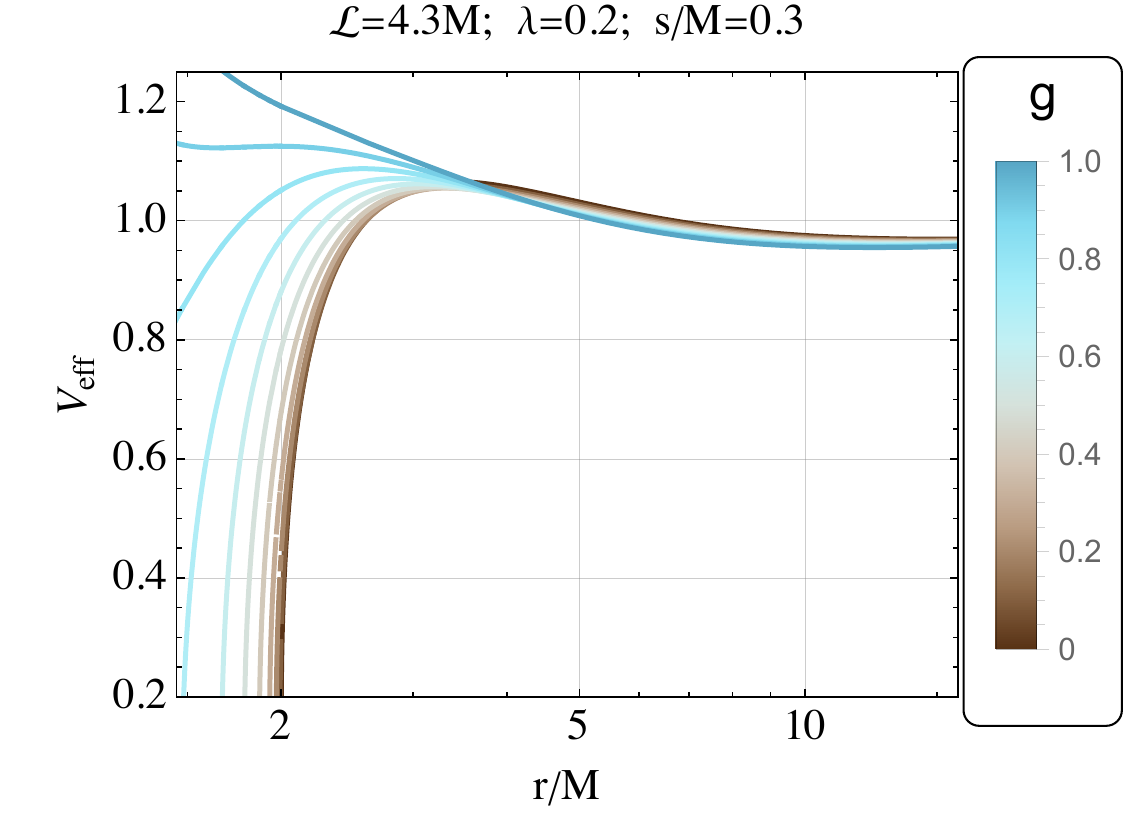} \includegraphics[width=0.49\linewidth]{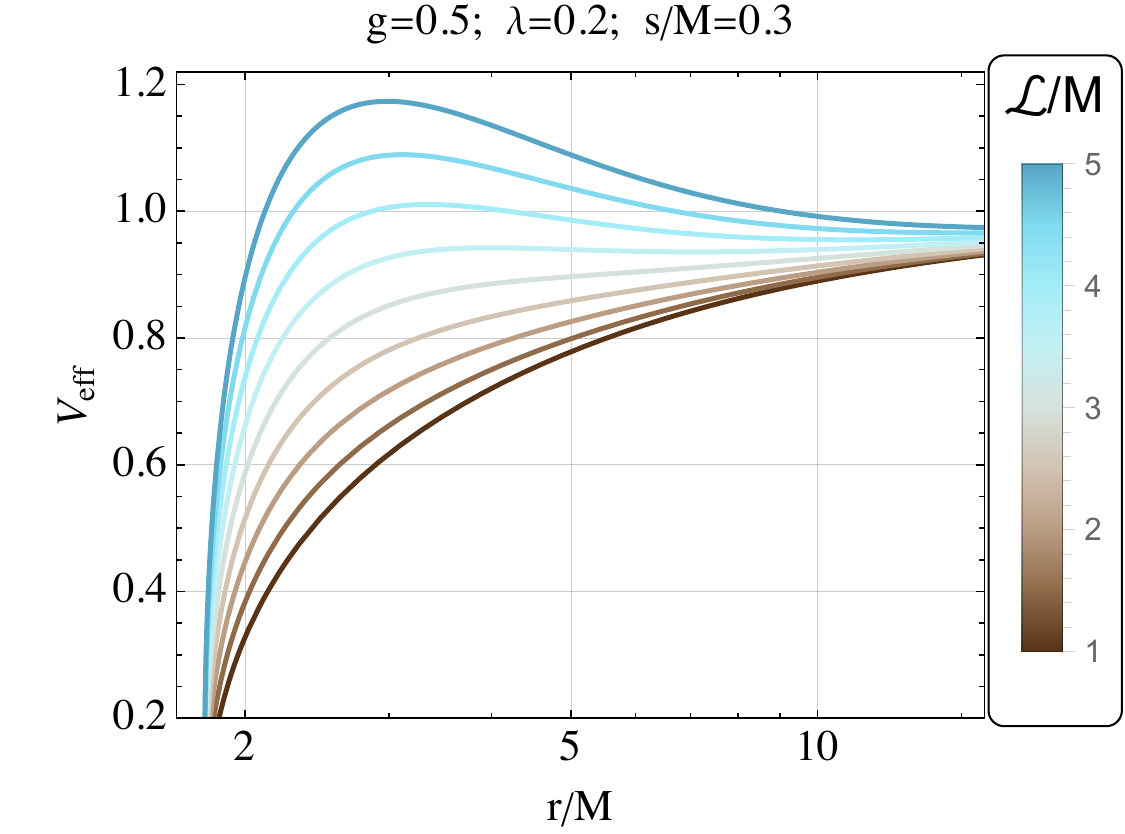}
\caption{Radial profiles of the effective potential $V_{\rm eff}$ for a magnetically charged spinning test particle around the Bardeen black hole with magnetic charge. The plots show variations with the particle's spin parameter $s$, magnetic charge $\lambda$, angular momentum $\mathcal{L}$, and the black hole magnetic charge parameter $g$. } \label{plotVeff}
\end{figure*}

To investigate the orbital dynamics of a spinning test particle with magnetic charge in Bardeen black hole spacetime, we derive the effective potential governing its radial motion. This potential accounts for the interplay of the black hole's mass $M$, its magnetic monopole charge $g$, the particle's spin $s$, and its magnetic charge $q_m$. The effective potential determines the existence, stability, and characteristics of circular orbits, including the innermost stable circular orbit, and reveals regions where spin effects may deviate from geodesic paths, potentially resulting in superluminal velocities.

The effective potential is built from the conserved specific energy $\mathcal{E}$ and specific angular momentum $\mathcal{L}$, in conjunction with the four-momentum normalization condition $p^\alpha p_\alpha = -m^2$. For astrophysical applications, such as accretion disks, we confine the motion to the equatorial plane ($\theta = \pi/2$, $p^\theta = 0$). Under these conditions, the Tulczyjew spin-supplementary condition (Eq.~\eqref{eq:tssc}) simplifies the spin-tensor components. The radial component of the four-momentum, $p^r$, then follows from the normalization condition.

Substituting the expressions for the conserved quantities and the spin tensor into the normalization condition $p^\alpha p_\alpha = -m^2$ yields the radial equation~\cite{Oteev:2025jvf,khan2025circular}
\begin{equation}
d\,\dot{r}^2 = a \mathcal{E}^2 + b \mathcal{E} + c,
\label{eq:radial_eq}
\end{equation}
with coefficients
\begin{eqnarray}
&&d=r^2\left(2 r-S^2f'\right)^2; \nonumber  \\
&&a=4 r^4-4 f r^2 S^2 \nonumber; \\
&&b=-4 J r^3 S f'-8 f g \lambda  r S^2+8 f J r^2 S+8 g \lambda  r^3 ; \nonumber \\
&&c=r^2(-2 g \lambda +JSf')^2-f(4 (g^2 \lambda ^2 S^2-2 g J \lambda  r S+ \nonumber \\
&&+J^2 r^2+r^4)+r^2S^2f'(-4 r+S^2f')).   \nonumber
\end{eqnarray}
which are functions of $r$, $M$, $g$, $s$, $\lambda$, and $\mathcal{L}$ (through $J=\mathcal L+s$).

In the fully geodesic, uncharged limit ($s=0$ and $\lambda=0$, hence $S=0$), the coefficient $b$ vanishes identically and Eq.~\eqref{eq:radial_eq} reduces to the familiar form $\dot r^2 = \mathcal{E}^2 - V_{\rm eff}(r)$ with $V_{\rm eff}=-c/a=f(r)(1+\mathcal{L}^2/r^2)$, the standard Schwarzschild-like expression. More generally, either spin-curvature coupling or the magnetic Lorentz force -- indeed $b\propto g\lambda r^3$ even at $S=0$, and $b\propto JS$ even at $\lambda=0$ introduces gravitomagnetic-like cross terms that are linear in $\mathcal{E}$, so that whenever $s\neq0$ or $\lambda\neq0$ one has $b\neq 0$ and Eq.~\eqref{eq:radial_eq} is a genuine quadratic in $\mathcal{E}$ rather than a simple difference of squares. Completing the square, it can be written as
\begin{equation}
d\,\dot r^2 = a\,(\mathcal{E} - V_{+})(\mathcal{E} - V_{-}), \qquad V_{\pm} = \frac{ -b \pm \sqrt{b^2 - 4ac} }{2a},
\label{eq:Vpm}
\end{equation}
so that a turning point of the radial motion ($\dot r = 0$) occurs when $\mathcal{E}$ coincides with one of the two roots $V_\pm(r)$; these roots therefore play the role of the effective potential governing the allowed range of $\mathcal{E}$ at each radius, generalizing the geodesic $V_{\rm eff}(r)$ to the spinning case. In our analysis, we primarily consider the positive-branch solution $V_{\rm eff}(r) \equiv V_{+}(r)$, as it corresponds to future-directed timelike trajectories with positive energy relevant to astrophysical particles.

Figure~\ref{plotVeff} illustrates the radial dependence of the effective potential $V_{\rm eff}(r)$ for magnetically charged spinning test particles orbiting the Bardeen regular black hole, showing variations with respect to the particle's specific spin parameter $s$, specific magnetic charge $\lambda$, specific angular momentum $\mathcal{L}$, and the black hole's magnetic monopole charge parameter $g$.
The upper-left panel of Figure~\ref{plotVeff} shows the radial profile of the effective potential for different values of the particle's spin parameter $s$, with fixed black hole magnetic charge $g/M = 0.5$, particle magnetic charge $\lambda = 0.2$, and specific angular momentum $\mathcal{L} = 4.3M$. As the spin $s$ increases in the prograde direction (positive $s$, aligned with orbital angular momentum), the potential barrier rises and shifts outward, while retrograde spin (negative $s$) lowers the barrier, reflecting the repulsive gravitomagnetic effect for prograde orbits and the attractive effect for retrograde ones.
The upper-right panel of Figure~\ref{plotVeff} displays the effective potential for varying particle magnetic charge $\lambda$, with fixed $g/M = 0.5$, $s = 0.3$, and $\mathcal{L} = 4.3M$. Positive $\lambda$ (attractive interaction, $\lambda g > 0$) lowers $V_{\rm eff}$, whereas negative $\lambda$ (repulsive interaction, $\lambda g < 0$) raises it, demonstrating how a binding Coulomb-like force deepens the potential well while a repulsive one counteracts gravity; for $s=0$ one finds $V_\pm=-\lambda g/r\pm\sqrt{f(1+\mathcal L^2/r^2)}$ exactly.
The lower-right panel of Figure~\ref{plotVeff} presents the radial dependence of $V_{\rm eff}$ for different values of the specific angular momentum $\mathcal{L}$, with $g/M = 0.5$, $s = 0.3$, and $\lambda = 0.2$. Higher $\mathcal{L}$ significantly raises the centrifugal barrier, as expected, creating deeper potential minima that support stable circular orbits at larger radii.
The lower-left panel of Figure~\ref{plotVeff} illustrates the variation of the effective potential with the black hole's magnetic monopole charge parameter $g$, for fixed particle parameters $s = 0.3$, $\lambda = 0.2$, and $\mathcal{L} = 4.3M$. As $g$ increases, the monopole charge's regularizing effect modifies the potential barrier near the horizon, generally softening the gravitational pull in the near-horizon region compared to the Schwarzschild limit ($g=0$) while preserving the regular de Sitter core.

\subsection{Stable Circular Orbits}
\label{subsec:stable_circ}

Circular orbits are defined by vanishing radial velocity and radial acceleration along the particle's worldline. In terms of the effective potential formalism, these requirements translate to~\cite{Jumaniyozov:2025uwo,Alimova:2025izs,Shermatov:2025qad}
\begin{eqnarray}
    \frac{dr}{d\tau} = 0 &\quad\Rightarrow\quad& \mathcal{E} = V_{\rm eff}(r),
    \label{eq:circ_vel}
    \\
    \frac{d^2 r}{d\tau^2} = 0 &\quad\Rightarrow\quad& \frac{d V_{\rm eff}}{dr} \bigg|_{r = r_{\rm circ}} = 0.
    \label{eq:circ_acc}
\end{eqnarray}
Note that, since $V_{\rm eff}\equiv V_+$ in Eq.~\eqref{eq:Vpm} is itself an energy, the first condition reads $\mathcal E=V_{\rm eff}$ rather than $\mathcal E^2=V_{\rm eff}$. Solving these equations simultaneously provides the specific energy $\mathcal{E}$ and specific angular momentum $\mathcal{L}$ required to sustain a circular orbit at a given radius $r_{\rm circ}$.

For the orbit to be stable against small radial perturbations, the effective potential must exhibit a local minimum at $r_{\rm circ}$:
\begin{equation}
    \frac{d^2 V_{\rm eff}}{dr^2} \bigg|_{r = r_{\rm circ}} > 0.
    \label{eq:stability_condition}
\end{equation}
This second-derivative condition ensures that deviations from $r_{\rm circ}$ produce a restoring force that returns the particle to the equilibrium radius.

The spin-curvature coupling introduces a significant asymmetry between prograde ($s > 0$, spin parallel to orbital angular momentum) and retrograde ($s < 0$) configurations. Prograde spin generates an effective repulsive force analogous to the gravitomagnetic effect in rotating spacetimes, lowering the required $\mathcal{L}$ for a given radius and enabling stable orbits closer to the black hole. Retrograde spin, in contrast, produces an attractive contribution, necessitating higher $\mathcal{L}$ and shifting stable orbits outward.

The magnetic coupling $\lambda g$ further influences orbital parameters through the Coulomb-like term in the MPD equations (see the sign convention in Sec.~\ref{sec:spacetime}). For $\lambda g < 0$ (repulsive) the outward magnetic force partly offsets gravity, so the angular momentum $\mathcal{L}$ required for a circular orbit at fixed $r$ is reduced while the specific energy $\mathcal{E}$ is raised relative to the uncharged case; attractive interactions ($\lambda g > 0$) raise $\mathcal{L}$ and lower $\mathcal{E}$. This pattern is already present for $s=0$.

\begin{figure*}[ht!]
   \centering
\includegraphics[width=0.32\linewidth]{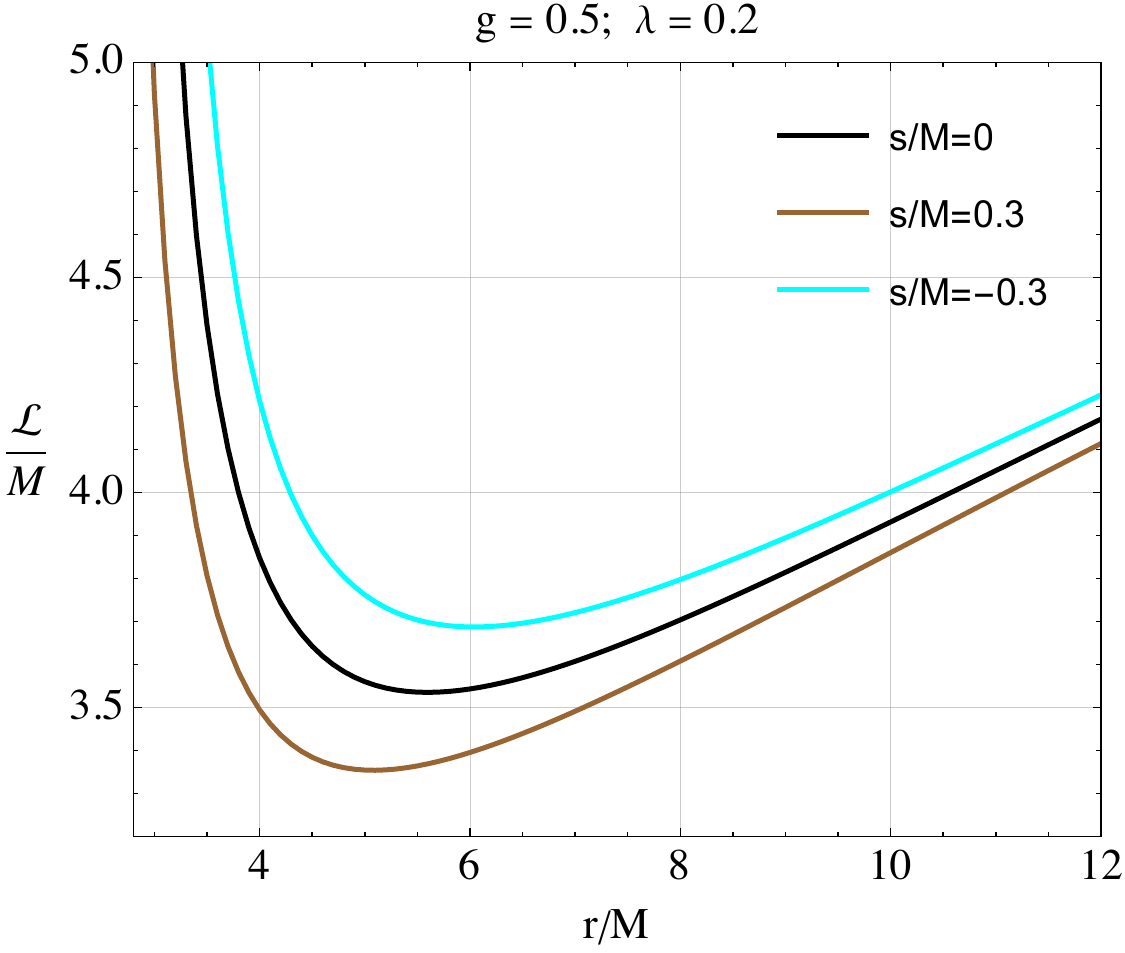}
\includegraphics[width=0.32\linewidth]{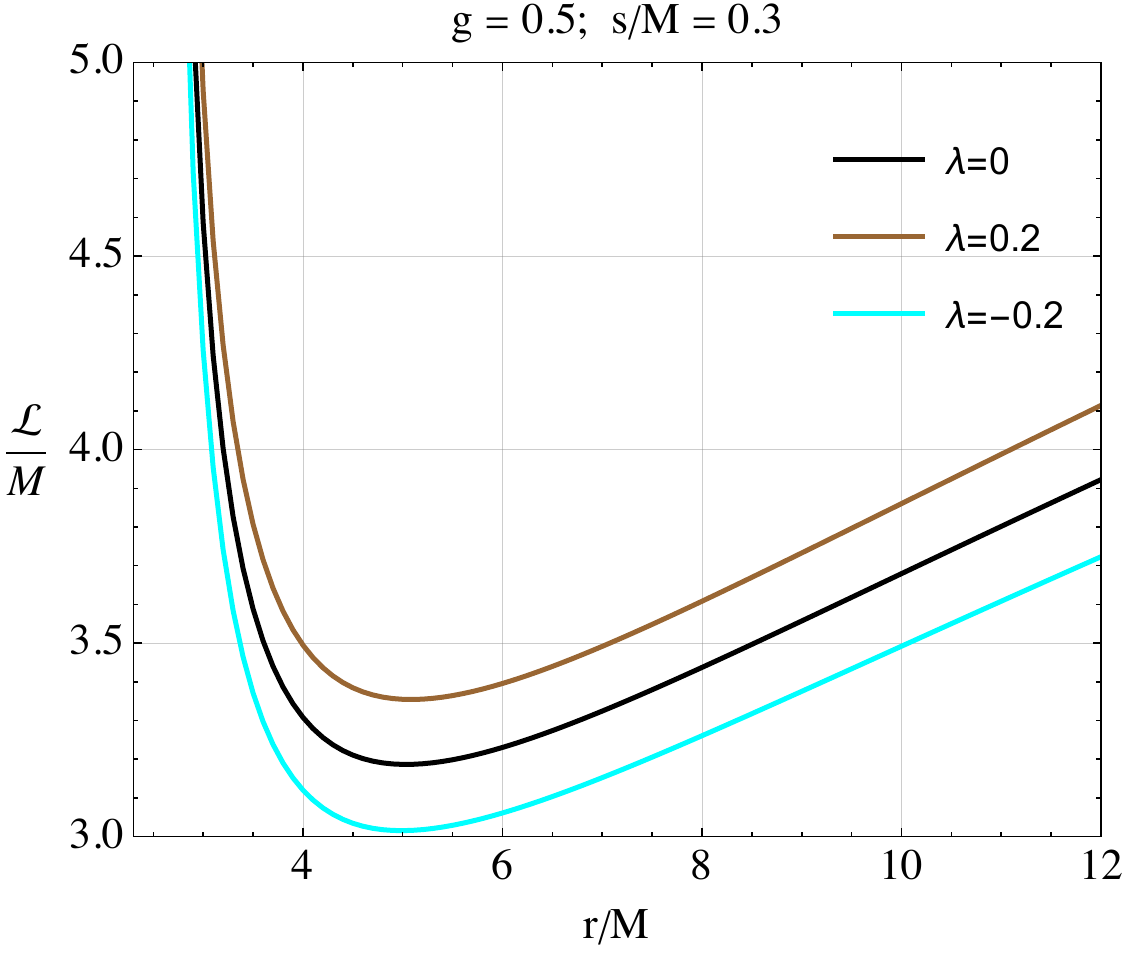}
\includegraphics[width=0.32\linewidth]{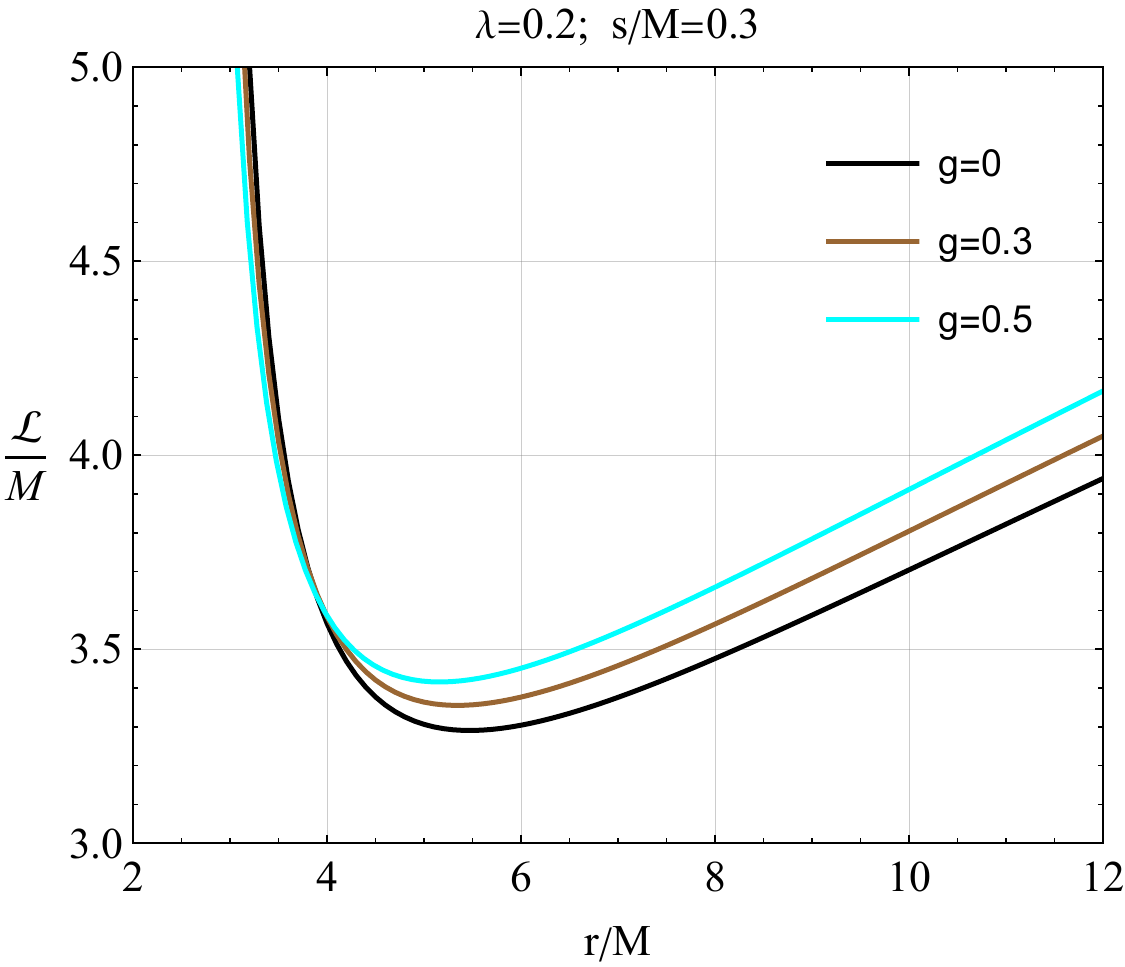}
\caption{Radial dependence of the specific angular momentum $\mathcal{L}$ for stable circular orbits of magnetically charged spinning test particles around the Bardeen black hole. Shown for different values of the particle's spin $s$, its magnetic charge $\lambda$, and black hole magnetic charge $g$.}
   \label{landfign}
\end{figure*}

\begin{figure*}[ht!]
\includegraphics[width=0.32\linewidth]{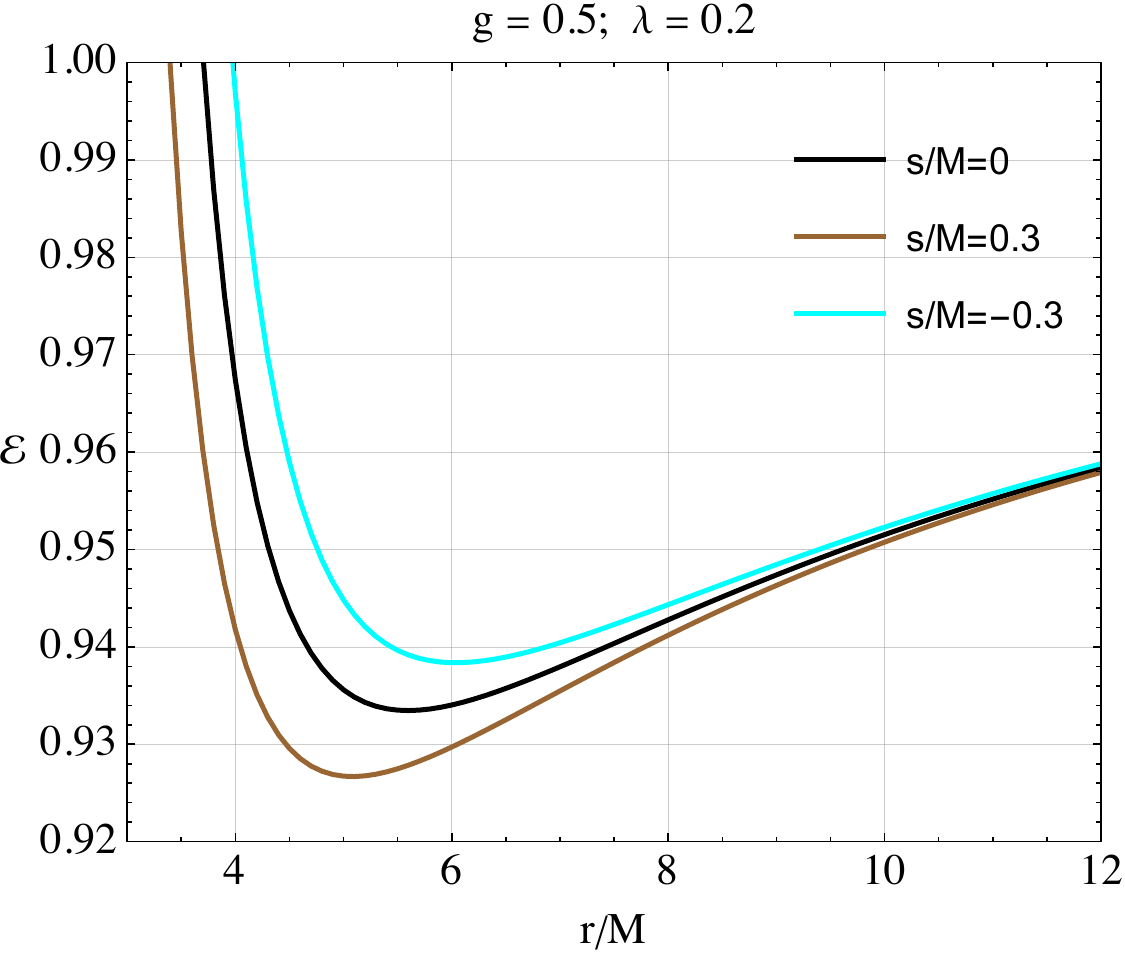}
\includegraphics[width=0.32\linewidth]{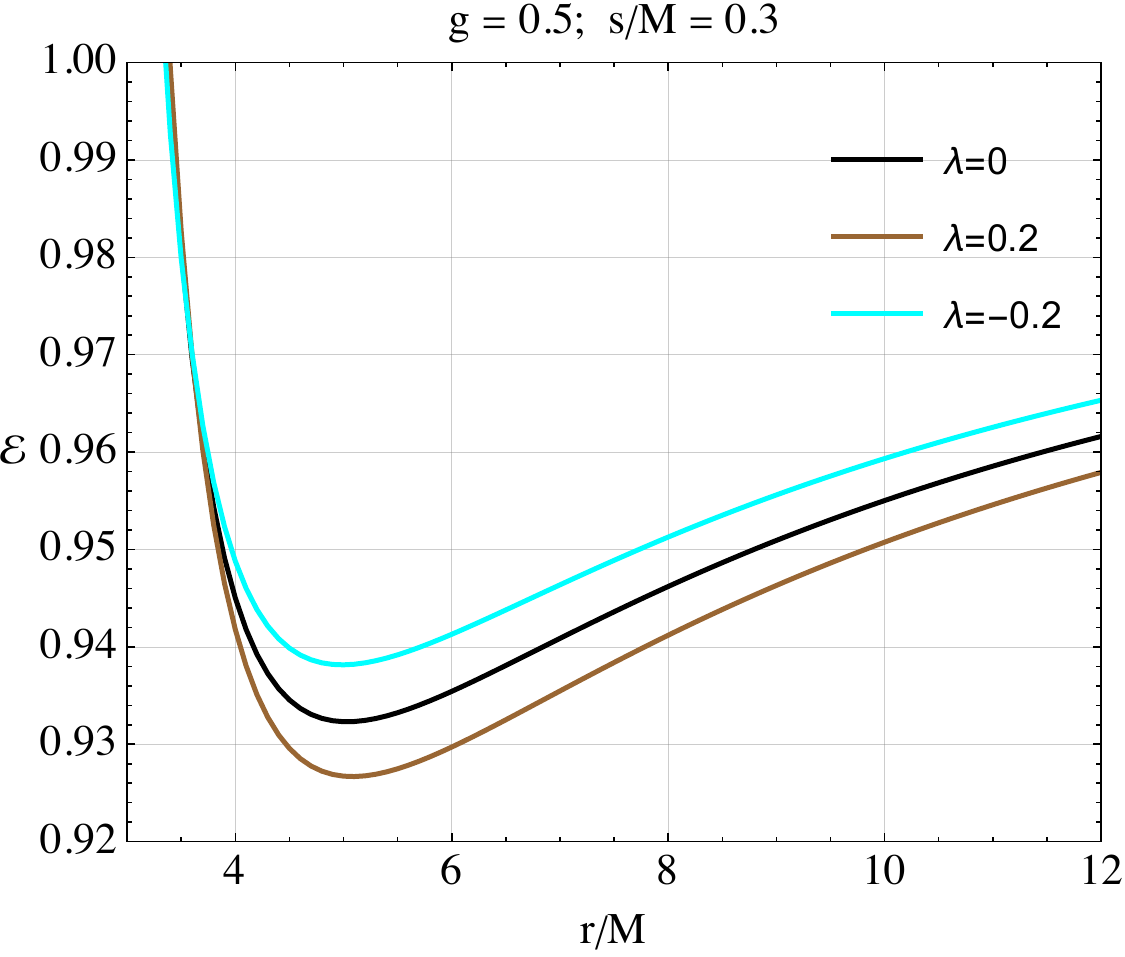}
\includegraphics[width=0.32\linewidth]{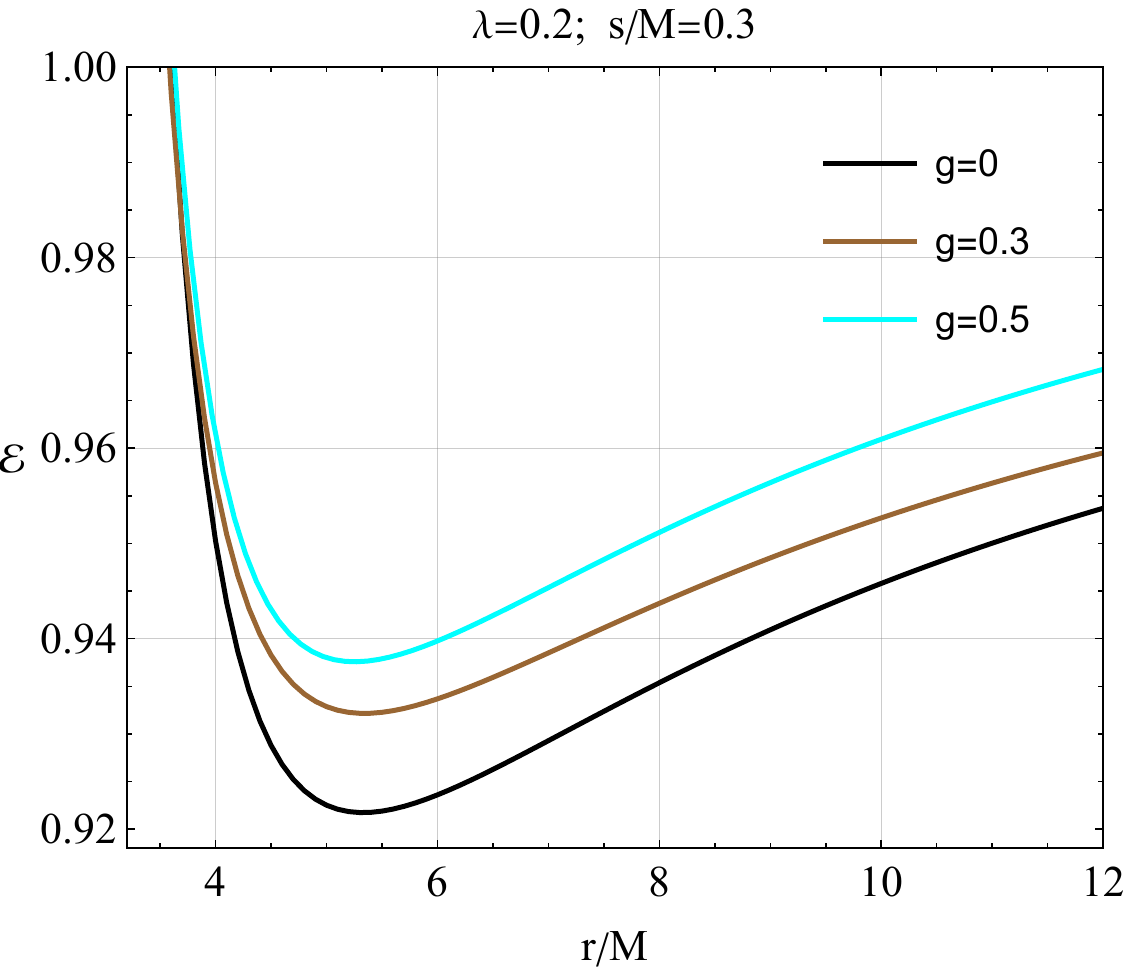} 
\caption{Radial dependence of the specific energy $\mathcal{E}$ for stable circular orbits of magnetically charged spinning test particles around the Bardeen black hole. Shown for different values of the particle's spin $s$, its magnetic charge $\lambda$, and black hole magnetic charge $g$.}
 \label{landefign}
\end{figure*}

\begin{figure*}[ht!]
\begin{center}
    \includegraphics[scale=0.3]{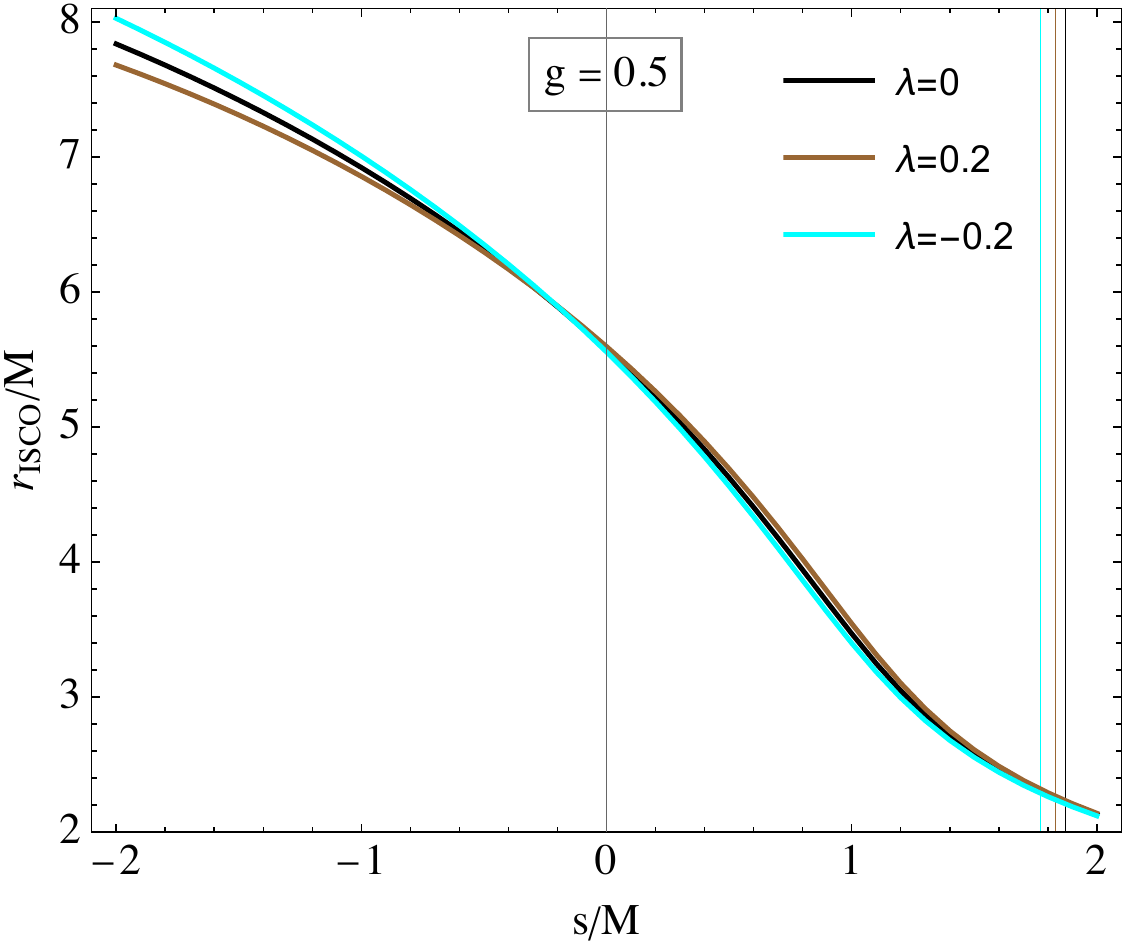}
    \includegraphics[scale=0.3]{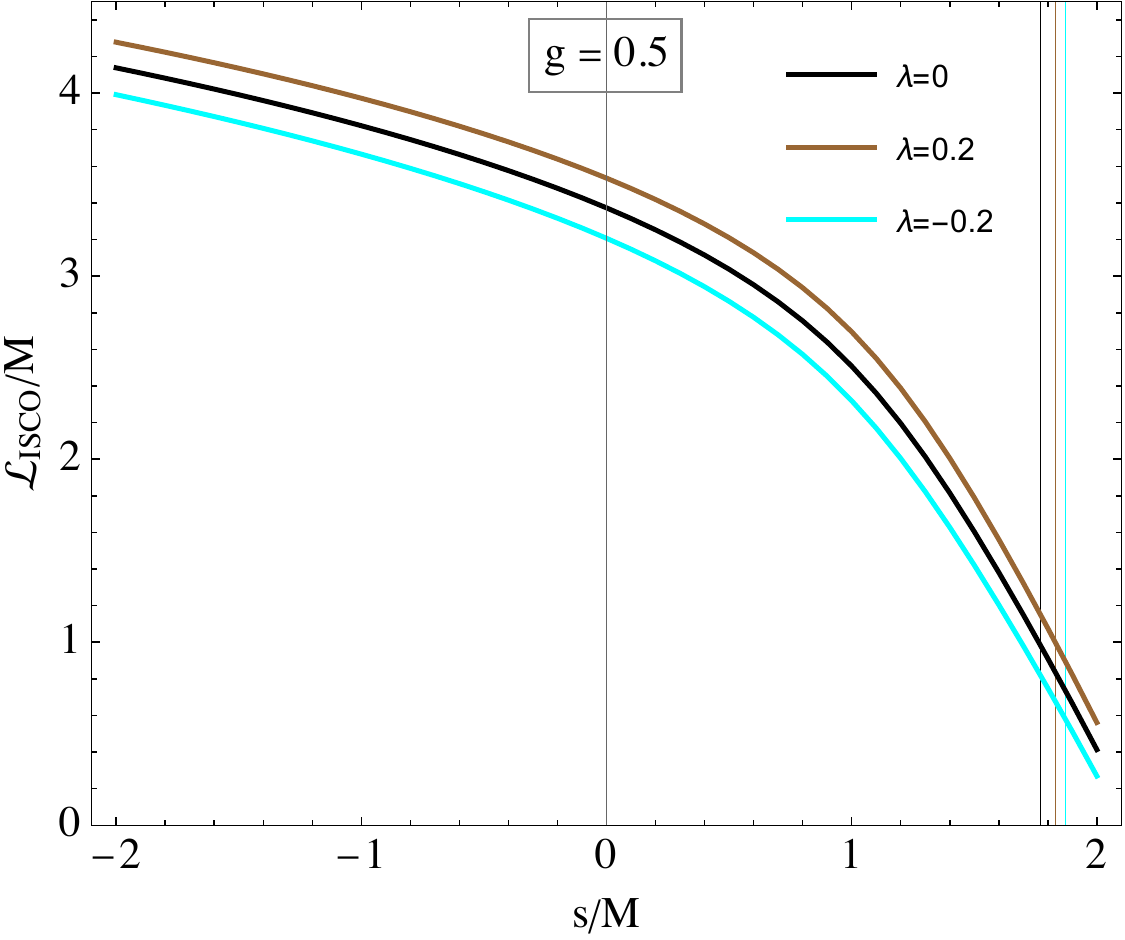}
    \includegraphics[scale=0.32]{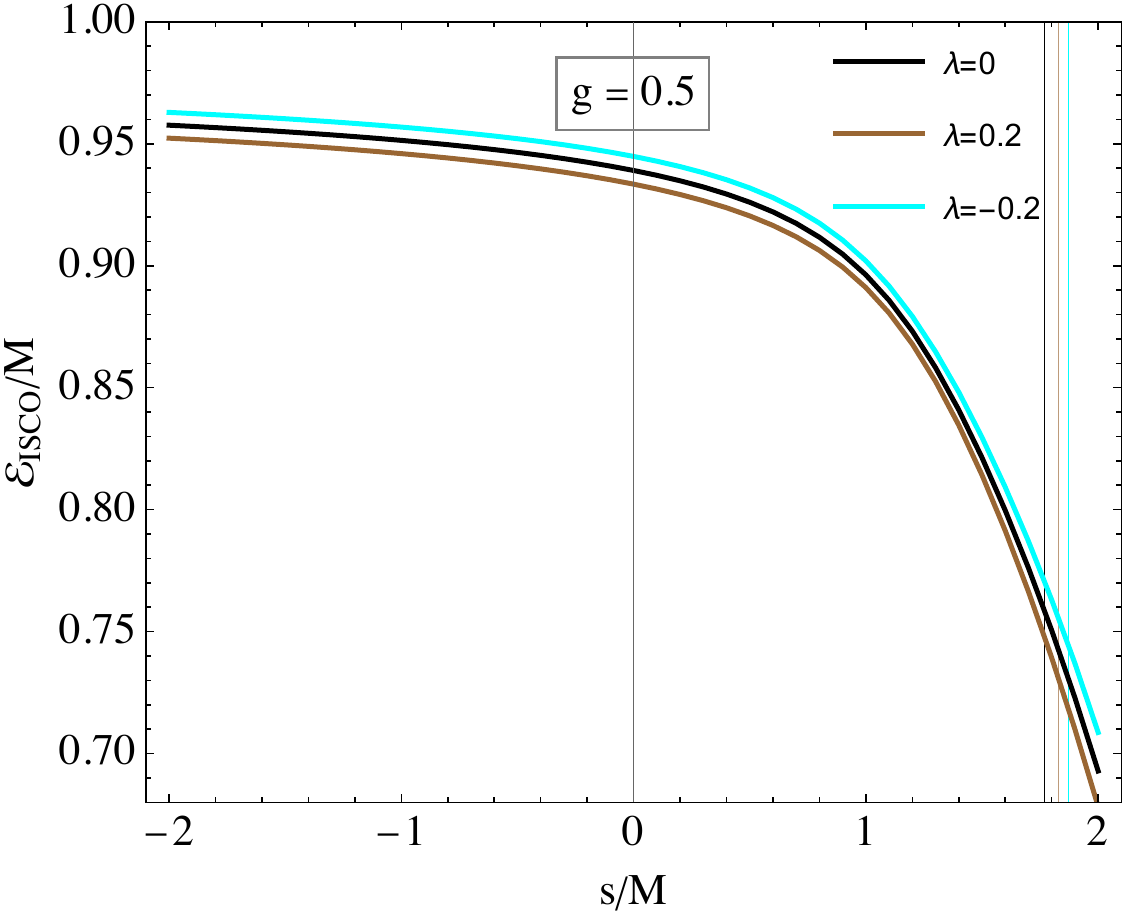}
%%%%
    \includegraphics[scale=0.3]{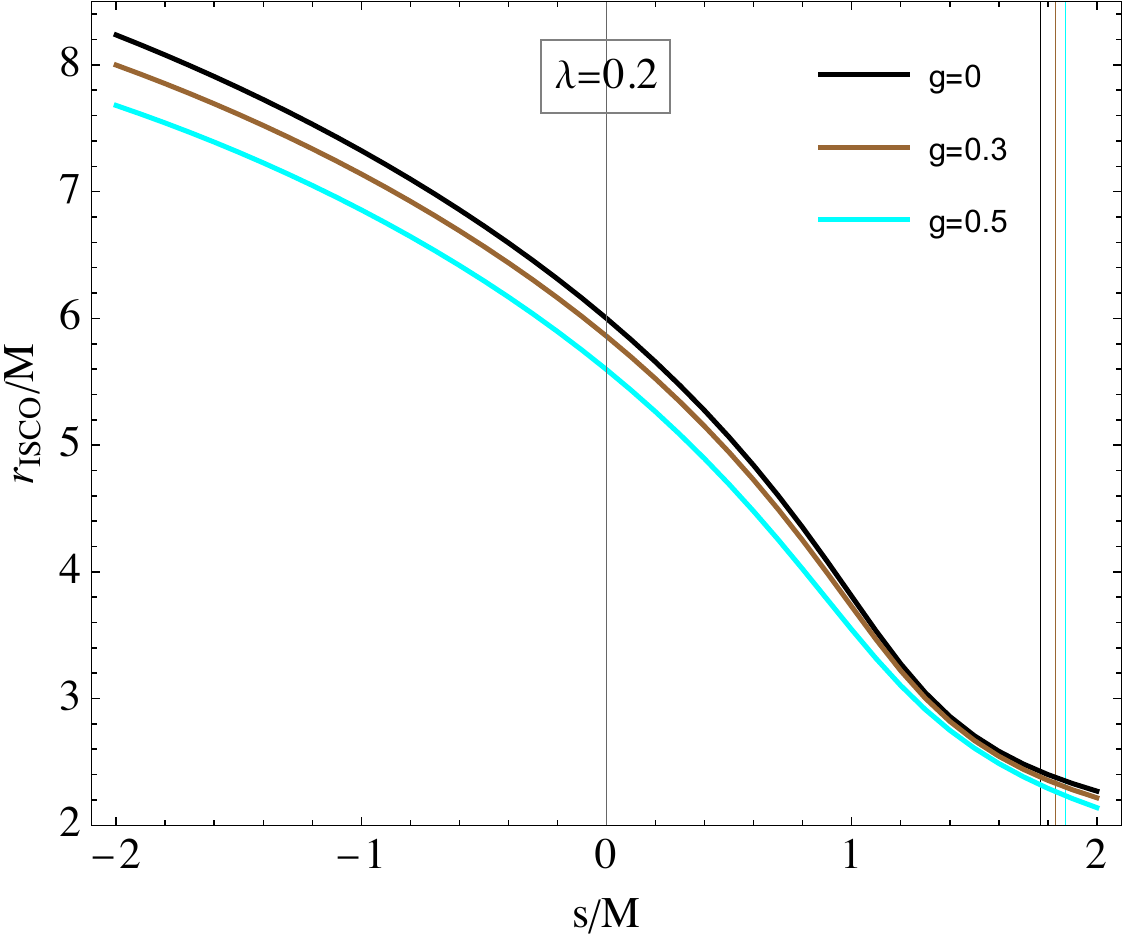}
    \includegraphics[scale=0.3]{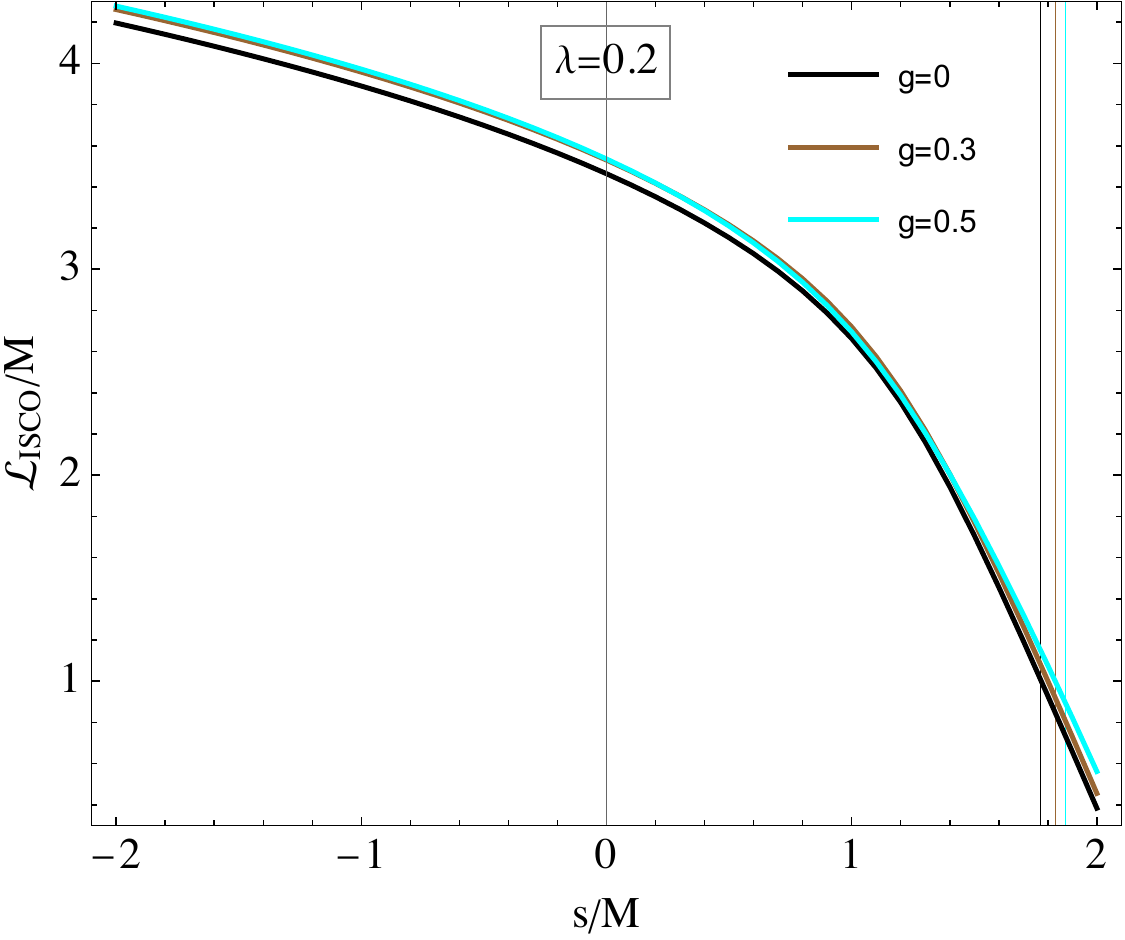}
    \includegraphics[scale=0.32]{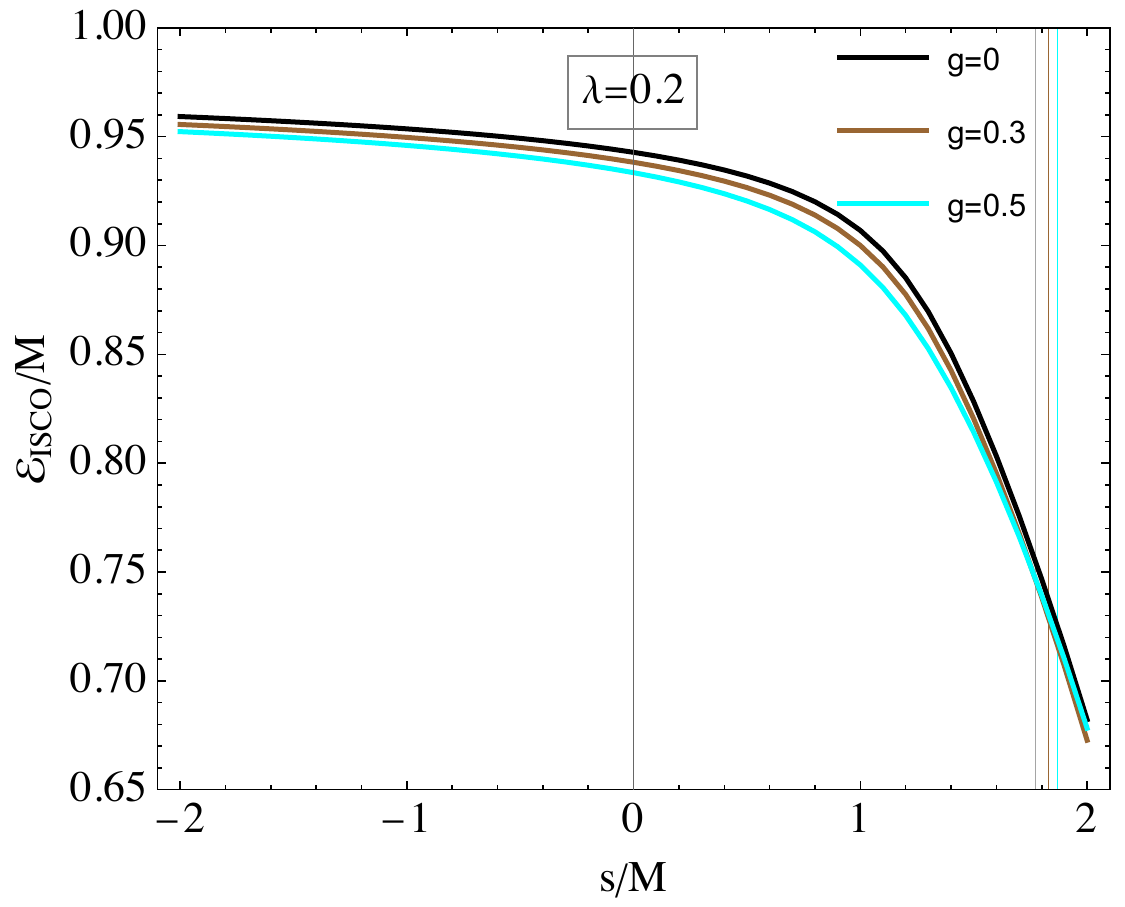}
    \end{center}
\caption{Dependence of the innermost stable circular orbit (ISCO) parameters on the particle's spin parameter $s$: ISCO radius $r_{\rm ISCO}/M$ (left panels), specific angular momentum at ISCO $\mathcal{L}_{\rm ISCO}$ (middle panels), and specific energy at ISCO $\mathcal{E}_{\rm ISCO}$ (right panels). Shown for various values of the particle's magnetic charge $\lambda$ with fixed $g=0.5$ (upper panels), and black hole magnetic charge $g$, with $\lambda=0.2$ (lower panels).}
\label{iscob}
\end{figure*}

Figure~\ref{landfign} illustrates the radial dependence of the specific angular momentum $\mathcal{L}$ required for stable circular orbits of magnetically charged spinning test particles around the Bardeen regular black hole. The plots show variations as functions of the particle's spin parameter $s$, its specific magnetic charge $\lambda$, and the black hole's magnetic monopole charge parameter $g$.

The left panel of Figure~\ref{landfign} shows the radial profile of $\mathcal{L}$ for different values of the particle's spin parameter $s$, with fixed black hole magnetic charge $g/M = 0.5$ and particle magnetic charge $\lambda = 0.2$. Prograde spin ($s > 0$, aligned with orbital angular momentum) significantly reduces the required specific angular momentum $\mathcal{L}$ compared to the spinless case ($s = 0$), reflecting the additional repulsive gravitomagnetic support that allows stable circular orbits at smaller radii with lower $\mathcal{L}$. Conversely, retrograde spin ($s < 0$) increases $\mathcal{L}$, as the attractive spin-curvature force opposes orbital motion and demands greater centrifugal support.

The middle panel of Figure~\ref{landfign} shows $\mathcal{L}$ as a function of particle magnetic charge $\lambda$, with fixed $g=0.5$ and $s=0.3$. Repulsive magnetic interactions ($\lambda < 0$, $\lambda g < 0$) lower the specific angular momentum $\mathcal{L}$, because the outward force partly balances gravity and less centrifugal support is needed. Attractive interactions ($\lambda > 0$, $\lambda g > 0$) raise $\mathcal{L}$, as the additional pull must be balanced by a larger orbital angular momentum.

The right panel of Figure~\ref{landfign} presents the radial dependence of $\mathcal{L}$ for different values of the black hole's magnetic charge parameter $g$, with fixed particle parameters $s = 0.3$ and $\lambda = 0.2$. As $g$ increases from the Schwarzschild limit ($g = 0$), the required specific angular momentum $\mathcal{L}$ increases, particularly at smaller radii. This reflects the regularizing effect of the magnetic charge, which softens the gravitational potential near the horizon and allows stable circular orbits with higher $\mathcal{L}$ compared to the singular Schwarzschild case.

Figure~\ref{landefign} illustrates the radial dependence of the specific energy $\mathcal{E}$ required for stable circular orbits of magnetically charged spinning test particles around the Bardeen regular black hole. The plots show variations as functions of the particle's spin parameter $s$, its specific magnetic charge $\lambda$, and the black hole's magnetic charge parameter $g$.

The left panel of Figure~\ref{landefign} shows the radial profile of $\mathcal{E}$ for different values of the particle's spin parameter $s$, with fixed black hole magnetic charge $g = 0.5$ and particle magnetic charge $\lambda = 0.2$. Prograde spin ($s > 0$, aligned with orbital angular momentum) decreases the specific energy $\mathcal{E}$ compared to the spinless case ($s = 0$), as the repulsive gravitomagnetic effect reduces the binding energy and allows more tightly bound orbits at smaller radii. In contrast, retrograde spin ($s < 0$) increases $\mathcal{E}$, reflecting greater energy required due to the attractive spin-curvature force that weakens orbital binding.

The middle panel of Figure~\ref{landefign} shows $\mathcal{E}$ as a function of particle magnetic charge $\lambda$, with fixed $g=0.5$ and $ s=0.3$. Repulsive interactions ($\lambda < 0$, $\lambda g < 0$) lower the required angular momentum $\mathcal{L}$ (Fig.~\ref{landfign}) but raise the specific energy $\mathcal{E}$ at fixed $r$ relative to the uncharged case, because the Coulomb-like potential energy is positive; attractive interactions ($\lambda > 0$, $\lambda g > 0$) raise $\mathcal{L}$ and lower $\mathcal{E}$. This opposite ordering of $\mathcal{E}$ and $\mathcal{L}$ is the expected behaviour of an additional potential energy $-\lambda g/r$ and is present already for $s=0$.

The right panel of Figure~\ref{landefign} presents the radial dependence of $\mathcal{E}$ for different values of the black hole's magnetic charge parameter $g$, with fixed particle parameters $s = 0.3$ and $\lambda = 0.2$. As $g$ increases from the Schwarzschild limit ($g = 0$), the specific energy $\mathcal{E}$ increases at fixed $r$, especially near the horizon. This behavior arises from the regularizing influence of the magnetic monopole charge, which softens the effective gravitational attraction near the horizon due to the de Sitter-like core, thereby lowering the binding energy and raising $\mathcal{E}$ for stable circular orbits at a given radius compared with the singular Schwarzschild case.

\subsection{Innermost Stable Circular Orbit}
\label{subsec:isco}

\begin{figure*}[t]
\includegraphics[scale=0.301245]{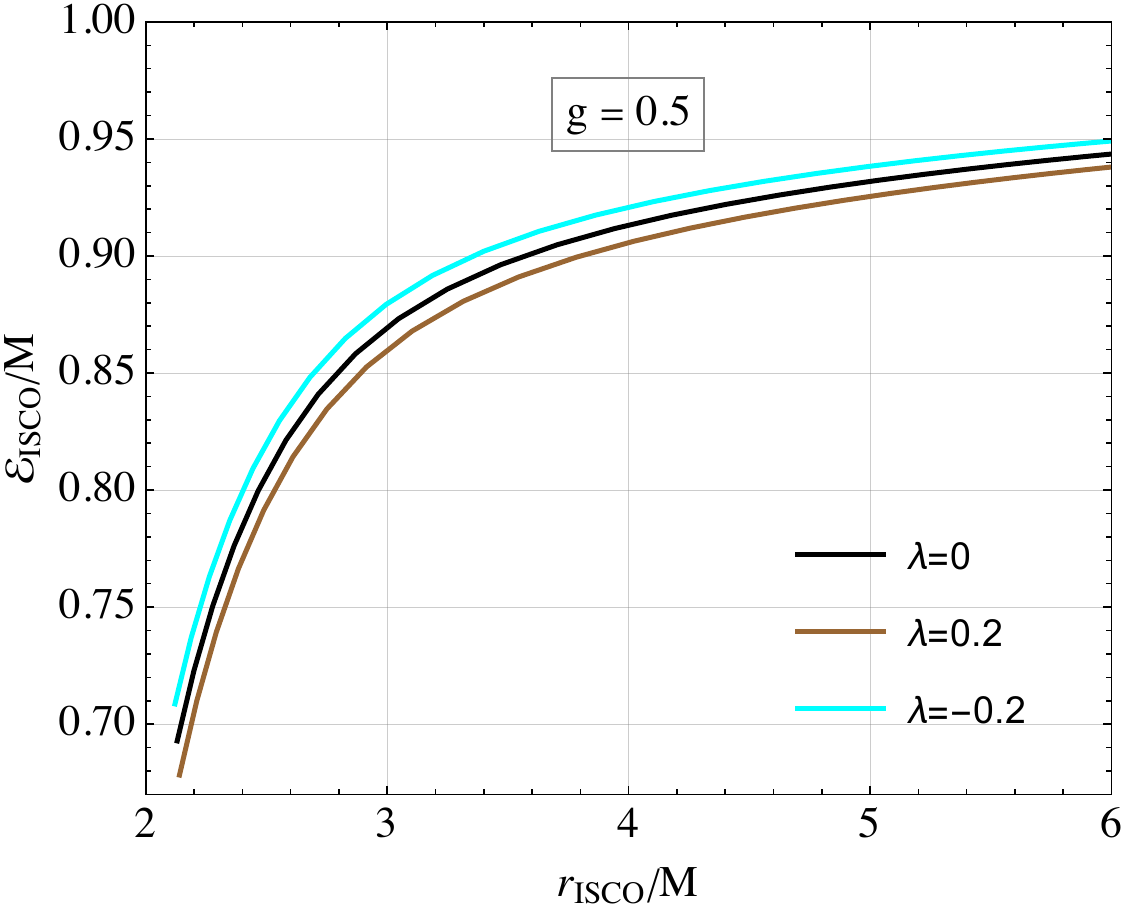}
\includegraphics[scale=0.301245]{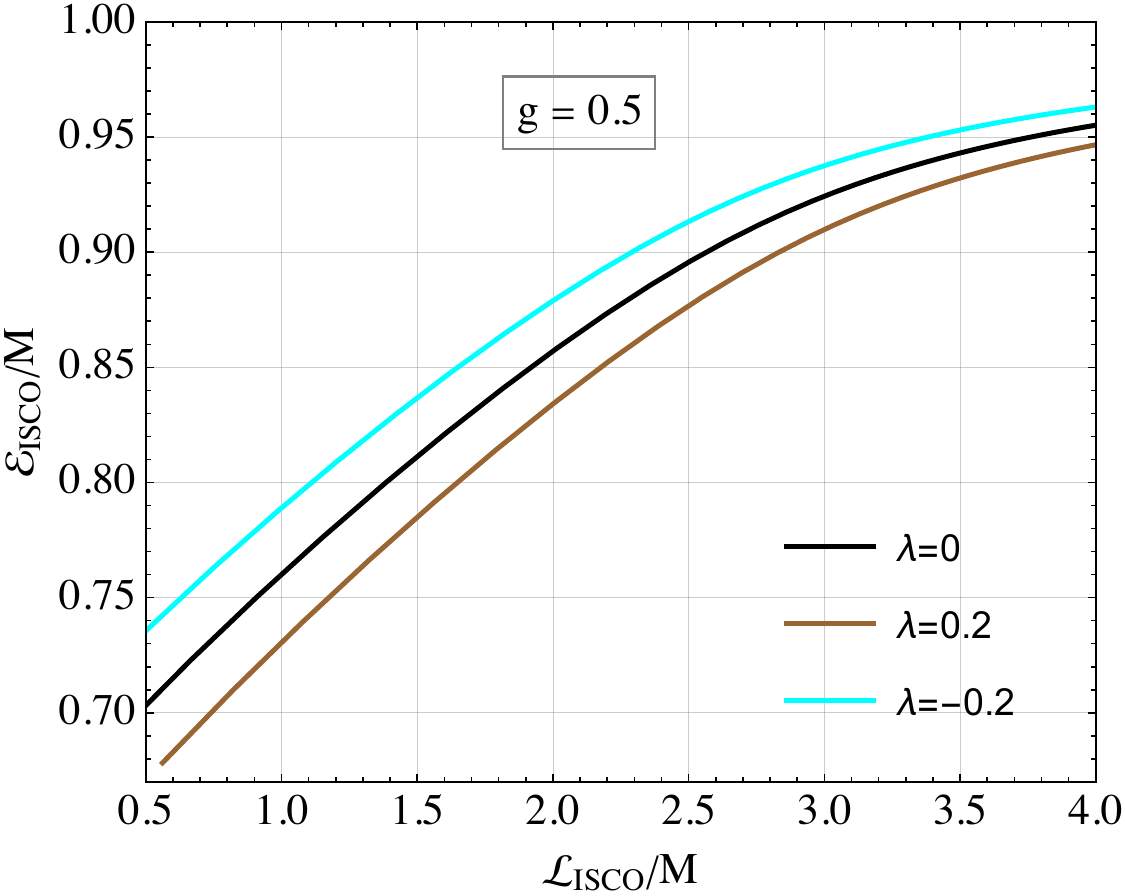}
\includegraphics[scale=0.28]{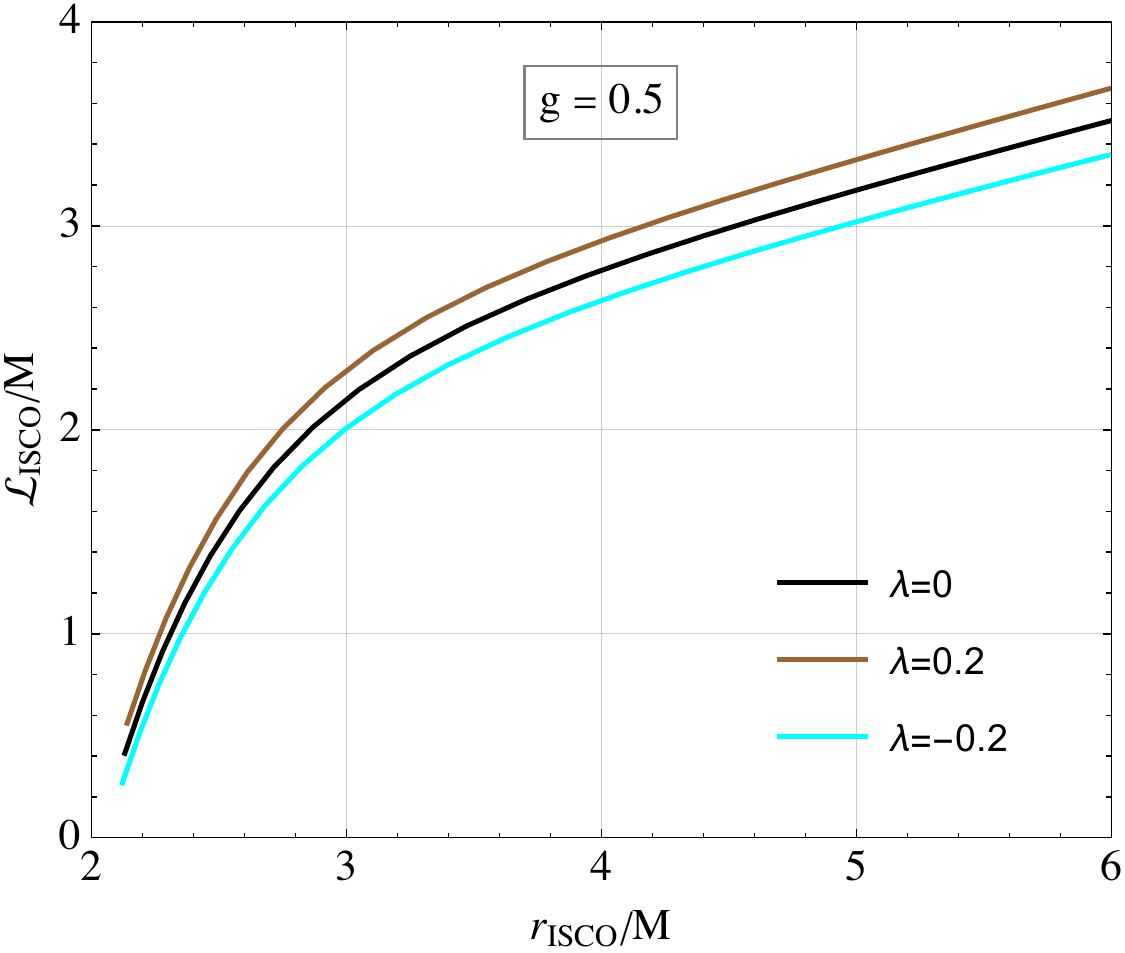}
\includegraphics[scale=0.301245]{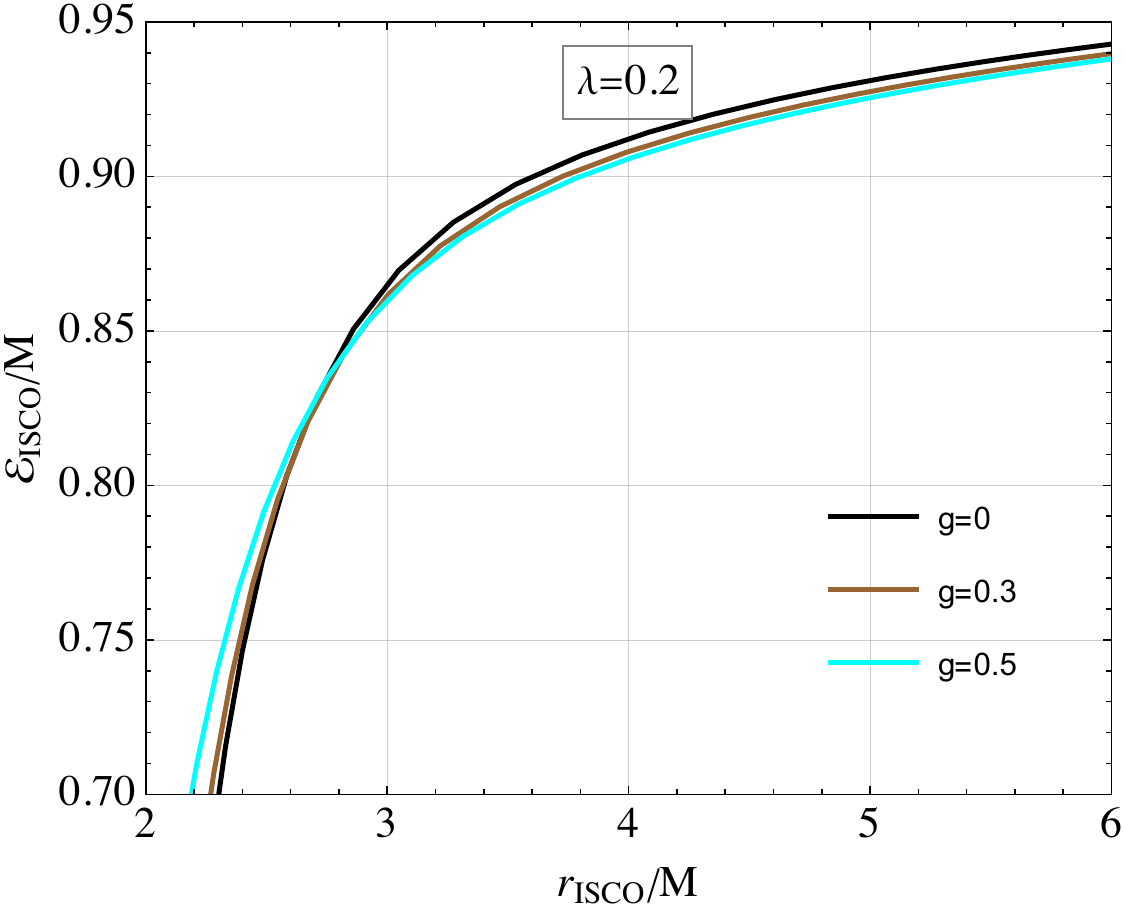}
\includegraphics[scale=0.301245]{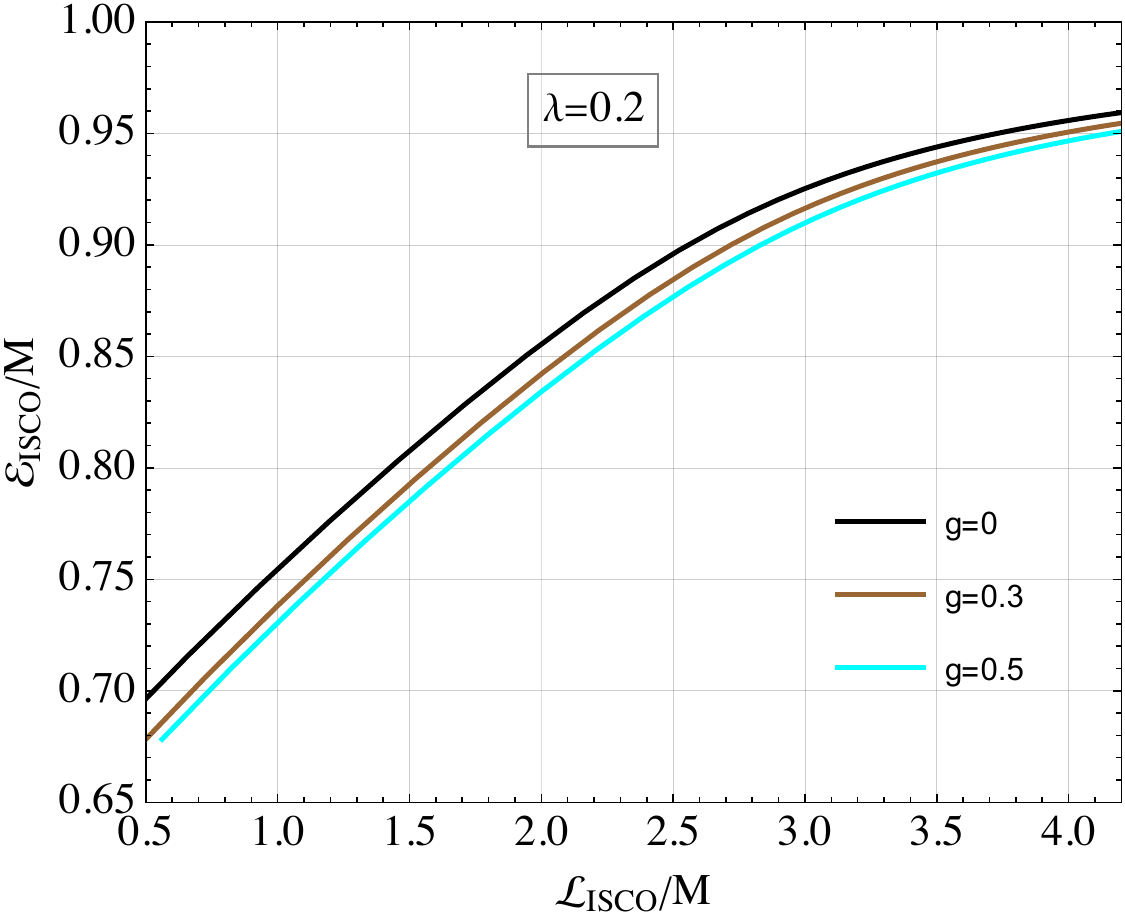}
\includegraphics[scale=0.301245]{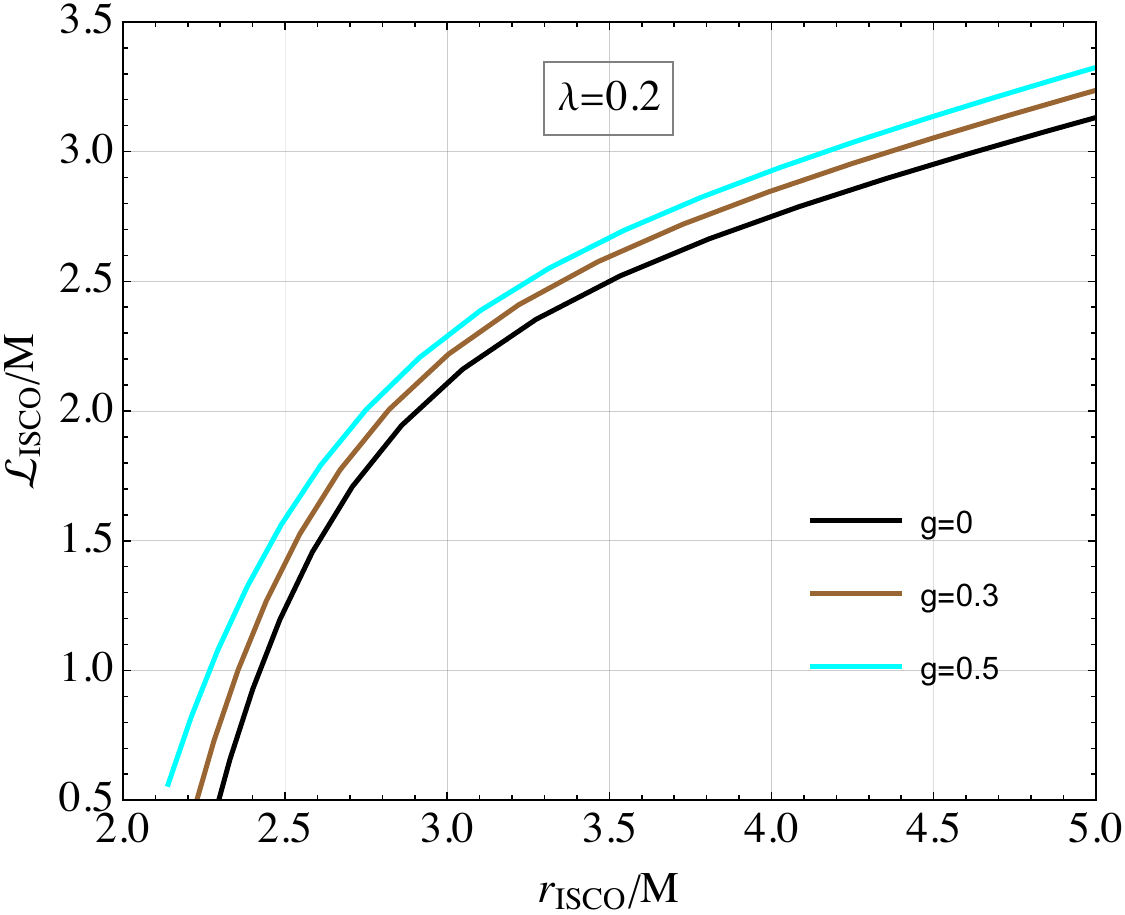}
\caption{Relations between the ISCO radius $r_{\rm ISCO}$, specific energy $\mathcal{E}_{\rm ISCO}$, and specific angular momentum $\mathcal{L}_{\rm ISCO}$ for magnetically charged spinning particles. The upper panels vary the particle's magnetic charge $\lambda$ with fixed $g=0.5$, while the lower panels vary the black hole magnetic charge $g$. \label{iscop}}
\end{figure*}

The ISCO marks the boundary between stable and unstable circular orbits, representing the smallest radius at which a local minimum in the effective potential persists. It is identified by the inflection point where the stability condition becomes marginal~\cite{Umarov:2025wzm,Jumaniyozov:2025xxh,Saydullayev:2025oop,RAHMATOV2025102102}:
\begin{equation}
    \frac{d^2 V_{\rm eff}}{dr^2} \bigg|_{r = r_{\rm ISCO}} = 0,
\end{equation}
while simultaneously satisfying the circular orbit conditions (Eqs.~\eqref{eq:circ_vel}-\eqref{eq:circ_acc}). Beyond this radius (larger $r$), stable circular orbits exist; inward of $r_{\rm ISCO}$, small perturbations lead to plunge trajectories toward the horizon.

The location and properties of the ISCO ($r_{\rm ISCO}$, $\mathcal{L}_{\rm ISCO}$, $\mathcal{E}_{\rm ISCO}$) are highly sensitive to the particle's intrinsic parameters and the background spacetime. Prograde spin ($s > 0$) enhances centrifugal support via spin-curvature coupling, systematically reducing $r_{\rm ISCO}/M$ relative to the spinless limit. This reduction allows particles to maintain stable orbits closer to the horizon, increasing accretion efficiency in disk models. Retrograde spin ($s < 0$), conversely, shifts the ISCO outward, suppressing inward migration and lowering efficiency.

Magnetic interactions introduce additional modulation: repulsive configurations ($\lambda g < 0$) decrease $\mathcal{L}_{\rm ISCO}$ and, more weakly, $r_{\rm ISCO}$, while attractive ones ($\lambda g > 0$) increase both quantities. As shown analytically in Sec.~\ref{subsec:pert}, the effect on $r_{\rm ISCO}$ is of higher order in $g$ and $\lambda$, whereas $\mathcal{L}_{\rm ISCO}$ and $\mathcal{E}_{\rm ISCO}$ respond at first order in $\lambda g$.

Figure~\ref{iscob} illustrates the dependence of the innermost stable circular orbit parameters—ISCO radius $r_{\rm ISCO}/M$ (left panels), specific angular momentum at ISCO $\mathcal{L}_{\rm ISCO}$ (middle panels), and specific energy at ISCO $\mathcal{E}_{\rm ISCO}$ (right panels) on the particle's spin parameter $s$ for magnetically charged spinning test particles around the Bardeen regular black hole. The upper row varies the particle's specific magnetic charge, $\lambda$, while keeping the black hole's magnetic charge fixed at $g=0.5$. In contrast, the lower row varies the black hole magnetic charge parameter $g$ while keeping the particle magnetic charge fixed at $\lambda=0.2$.

The upper-left panel shows $r_{\rm ISCO}/M$ as a function of $s$ for different $\lambda$ values. Prograde spin ($s > 0$) reduces the ISCO radius due to the repulsive gravitomagnetic effect. The dependence on $\lambda$ is weak (Sec.~\ref{subsec:pert}): repulsive interactions ($\lambda < 0$, $\lambda g < 0$) decrease $r_{\rm ISCO}/M$ slightly, while attractive interactions ($\lambda > 0$, $\lambda g > 0$) shift the ISCO slightly outward, especially for prograde spins.

The upper-middle panel displays $\mathcal{L}_{\rm ISCO}$ versus $s$. Prograde spin lowers the angular momentum required at the ISCO. Repulsive $\lambda < 0$ reduces $\mathcal{L}_{\rm ISCO}$, whereas attractive $\lambda > 0$ increases it, consistent with the magnetic force modulating the centrifugal requirement for stability.

The upper-right panel presents $\mathcal{E}_{\rm ISCO}$ as a function of $s$. Prograde spin decreases $\mathcal{E}_{\rm ISCO}$, making orbits more tightly bound. Here the magnetic charge acts oppositely to its effect on $\mathcal{L}_{\rm ISCO}$: repulsive interactions ($\lambda < 0$) raise $\mathcal{E}_{\rm ISCO}$ slightly relative to the uncharged case, while attractive ones ($\lambda > 0$) lower it.

The lower-left panel illustrates $r_{\rm ISCO}/M$ versus $s$ for varying $g$. As the black hole's magnetic charge $g$ increases, the ISCO radius decreases relative to the Schwarzschild limit ($g=0$), reflecting the softening of the gravitational potential near the regular core and permitting closer stable orbits.

The lower-middle panel shows $\mathcal{L}_{\rm ISCO}$ as a function of $s$ for different $g$. Higher $g$ reduces the specific angular momentum needed at ISCO, particularly for prograde orbits, due to the modified near-horizon geometry.

The lower-right panel depicts $\mathcal{E}_{\rm ISCO}$ versus $s$. Increasing $g$ lowers the ISCO energy, indicating stronger binding in the regular spacetime than in the singular Schwarzschild case.

These ISCO properties are crucial for modeling thin accretion disks around regular black holes, determining the inner disk edge, radiative efficiency, and spectral features such as iron line profiles. The parameter-dependent shifts in $r_{\rm ISCO}$, modulated by spin alignment and magnetic interactions, provide distinctive signatures that could distinguish Bardeen black holes from singular models in X-ray observations (e.g., with Athena or Lynx) and gravitational-wave signals from extreme mass-ratio inspirals detectable by LISA.

Figure~\ref{iscop} illustrates the interdependencies between the ISCO radius $r_{\rm ISCO}/M$, specific energy $\mathcal{E}_{\rm ISCO}$, and specific angular momentum $\mathcal{L}_{\rm ISCO}$ for magnetically charged spinning test particles around the Bardeen regular black hole. The upper panels vary the particle's specific magnetic charge $\lambda$ with fixed black hole magnetic charge $g = 0.5$, while the lower panels vary the black hole magnetic monopole charge parameter $g$ with fixed particle magnetic charge $\lambda = 0.2$.

The upper-left panel shows the relationship between $r_{\rm ISCO}/M$ and $\mathcal{E}_{\rm ISCO}$ for different $\lambda$ values. As $r_{\rm ISCO}/M$ increases, $\mathcal{E}_{\rm ISCO}$ rises monotonically for all three values of $\lambda$; at a given $r_{\rm ISCO}$, however, repulsive magnetic interactions ($\lambda < 0$, $\lambda g < 0$) yield a somewhat higher $\mathcal{E}_{\rm ISCO}$ than the uncharged case, while attractive interactions ($\lambda > 0$, $\lambda g > 0$) yield a lower one -- the opposite ordering to that found for $\mathcal{L}_{\rm ISCO}$ (Fig.~\ref{iscob}), which simply reflects the Coulomb-like potential energy $-\lambda g/r$.

The upper-middle panel illustrates the relationship between $\mathcal{E}_{\rm ISCO}$ and $\mathcal{L}_{\rm ISCO}$ for different $\lambda$. Higher angular momentum generally correlates with increased energy; at fixed $\mathcal{L}_{\rm ISCO}$, repulsive magnetic interactions raise $\mathcal{E}_{\rm ISCO}$ slightly relative to the uncharged case, while attractive interactions lower it.

The upper-right panel depicts the dependence of $r_{\rm ISCO}/M$ on $\mathcal{L}_{\rm ISCO}$ for varying $\lambda$. Larger ISCO radii correspond to higher specific angular momentum, where repulsive $\lambda < 0$ reduces $\mathcal{L}_{\rm ISCO}$ at a given radius because the outward force partly offsets gravity, whereas attractive $\lambda > 0$ demands greater $\mathcal{L}_{\rm ISCO}$ to maintain stability.

The lower-left panel shows the relationship between $r_{\rm ISCO}/M$ and $\mathcal{E}_{\rm ISCO}$ for different values of $g$. As the black hole magnetic charge increases, smaller ISCO radii are associated with lower energies than in the Schwarzschild limit ($g=0$), because the regular core softens the near-horizon gravitational potential.

The lower-middle panel illustrates the relationship between $\mathcal{E}_{\rm ISCO}$ and $\mathcal{L}_{\rm ISCO}$ for different $g$. Increasing $g$ lowers both energy and angular momentum at ISCO, indicating stronger binding in the regular spacetime relative to singular cases.

The lower-right panel depicts the dependence of $r_{\rm ISCO}/M$ on $\mathcal{L}_{\rm ISCO}$ for varying $g$. Higher $g$ shifts the curve toward smaller ISCO radii, while for a given $r_{\rm ISCO}$ the required angular momentum is slightly higher than in the Schwarzschild case, highlighting the regularizing effect that permits tighter, more centrally concentrated orbits despite the modestly larger centrifugal support they require.

These interdependencies show how magnetic interactions and spacetime regularity shape ISCO characteristics, which are crucial for understanding accretion disk structure and efficiency around regular black holes. Repulsive particle-black hole magnetic configurations ($\lambda g<0$) reduce the angular-momentum requirement of the ISCO (at the price of a higher orbital energy), attractive ones do the opposite, while a larger monopole charge $g$ promotes closer, more tightly bound orbits with both reduced energy and reduced angular-momentum requirements. Together with prograde spin and attractive coupling, this increases the radiative efficiency and alters spectral signatures observable in X-ray emission; quantitative values are given in Sec.~\ref{subsec:eff}.

\subsection{Analytic expansion of the ISCO}
\label{subsec:pert}

\begin{figure*}[t]
\includegraphics[width=0.97\linewidth]{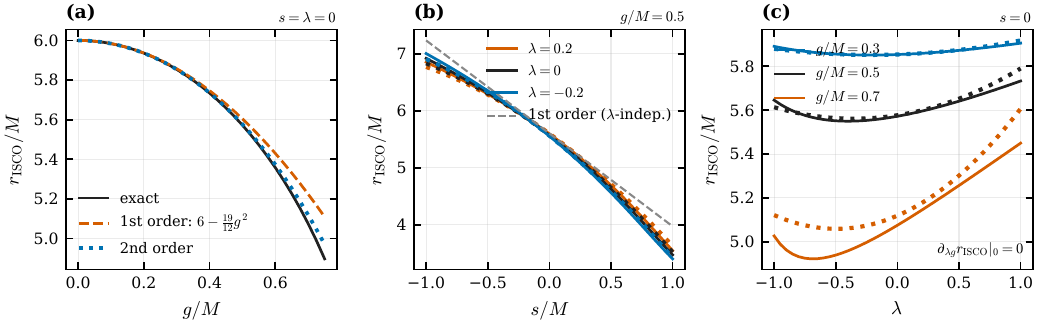}
\caption{Exact ISCO radius (solid) versus its perturbative expansion, Eq.~\eqref{eq:risco_pert} (first order dashed, second order dotted). (a) $r_{\rm ISCO}$ versus the monopole charge for $s=\lambda=0$. (b) Versus the particle spin for $g/M=0.5$ and $\lambda=0,\pm0.2$; the grey dashed line is the first-order result, which does not depend on $\lambda$. (c) Versus $\lambda$ for $s=0$ and $g/M=0.3,0.5,0.7$; the absence of a term linear in $\lambda g$ makes the dependence on $\lambda$ weak.}
\label{fig:pert}
\end{figure*}

The magnetic coupling enters Eq.~\eqref{eq:radial_eq} only through the combinations $\mu\equiv\lambda g$ (the quantities $g\lambda$ and $g^2\lambda^2$) and, via the lapse function, $\kappa\equiv g^2$. Near the Schwarzschild point $(r,\mathcal L)=(6M,2\sqrt3\,M)$ the ISCO conditions $V_{\rm eff}'=V_{\rm eff}''=0$ can therefore be solved by implicit differentiation in the small quantities $(s,\mu,\kappa)$. The exact (computer-algebra) first-order coefficients are collected in Table~\ref{tab:coeff}. To second order, in units $M=1$,
\begin{eqnarray}
r_{\rm ISCO}&=&6-\tfrac{2\sqrt6}{3}\,s-\tfrac{19}{12}\,g^2-\tfrac{29}{72}\,s^2+\tfrac{\sqrt3}{2}\,\lambda g\,s+\tfrac12\,\lambda^2g^2\nonumber\\
&&-\tfrac{5\sqrt6}{108}\,g^2s+\tfrac{\sqrt2}{2}\,\lambda g^3-\tfrac{757}{1728}\,g^4+\mathcal O(3).
\label{eq:risco_pert}
\end{eqnarray}
Three features are worth stressing. (i) For $g=\lambda=0$ the term linear in $s$ reproduces the known shift $r_{\rm ISCO}=6M-2\sqrt{2/3}\,s$ of a spinning test particle in Schwarzschild~\cite{Jefremov:2015gza}, which is a non-trivial check of Eqs.~\eqref{eq:radial_eq}--\eqref{eq:Vpm}. (ii) The ISCO radius has \emph{no} term linear in $\lambda g$: the magnetic coupling shifts $r_{\rm ISCO}$ only through $\lambda^2g^2$ and $\lambda g^3$, i.e.\ at the percent level or below for $|\lambda|\lesssim0.2$ (cf.\ Fig.~\ref{iscob}), whereas $\mathcal L_{\rm ISCO}$ and $\mathcal E_{\rm ISCO}$ do respond at first order, $\partial\mathcal L_{\rm ISCO}/\partial(\lambda g)=2\sqrt6/3$ and $\partial\mathcal E_{\rm ISCO}/\partial(\lambda g)=-1/18$. (iii) The monopole charge lowers the ISCO already at order $g^2$, $\delta r_{\rm ISCO}=-\tfrac{19}{12}g^2$. Figure~\ref{fig:pert} compares the expansion with the exact solution: Eq.~\eqref{eq:risco_pert} is accurate to better than $0.4\%$ for all entries of Table~\ref{tab:isco} with $g/M\le0.5$ and to about $2\%$ at $g/M=0.7$, while the first-order result alone is accurate to $0.6\%$ at $g/M=0.5$.

\begin{table*}[t]
\caption{Exact first-order coefficients of the ISCO quantities around the Schwarzschild values, $Q=Q_0+Q_s\,s+Q_\mu\,\lambda g+Q_\kappa\,g^2+\dots$ (units $M=1$; $\mu\equiv\lambda g$, $\kappa\equiv g^2$). The numerical value of each coefficient is given in parentheses. $J_{\rm ISCO}=\mathcal L_{\rm ISCO}+s$ is the total angular momentum.}
\label{tab:coeff}
\begin{ruledtabular}
\begin{tabular}{lcccc}
$Q$ & $Q_0$ & $Q_s$ & $Q_\mu$ & $Q_\kappa$\\
\hline
$r_{\rm ISCO}$ & $6$ & $-\frac{2\sqrt6}{3}\;(-1.633)$ & $0$ & $-\frac{19}{12}\;(-1.583)$\\[2pt]
$\mathcal L_{\rm ISCO}$ & $2\sqrt3\;(3.464)$ & $-1+\frac{\sqrt2}{3}\;(-0.529)$ & $\frac{2\sqrt6}{3}\;(1.633)$ & $-\frac{7\sqrt3}{36}\;(-0.337)$\\[2pt]
$J_{\rm ISCO}$ & $2\sqrt3\;(3.464)$ & $\frac{\sqrt2}{3}\;(0.471)$ & $\frac{2\sqrt6}{3}\;(1.633)$ & $-\frac{7\sqrt3}{36}\;(-0.337)$\\[2pt]
$\mathcal E_{\rm ISCO}$ & $\frac{2\sqrt2}{3}\;(0.9428)$ & $-\frac{\sqrt3}{108}\;(-0.0160)$ & $-\frac{1}{18}\;(-0.0556)$ & $-\frac{\sqrt2}{108}\;(-0.0131)$\\
\end{tabular}
\end{ruledtabular}
\end{table*}

\subsection{Radiative efficiency and characteristic frequencies}
\label{subsec:eff}

\begin{table*}[t]
\caption{ISCO parameters from the exact solution of $V_{\rm eff}'=V_{\rm eff}''=0$ ($M=1$): radius $r_{\rm ISCO}$, specific angular momentum $\mathcal L_{\rm ISCO}$, specific energy $\mathcal E_{\rm ISCO}$ and radiative efficiency $\eta=(1-\mathcal E_{\rm ISCO})\times100\%$. At $g=0$ the ISCO does not depend on $\lambda$ (only the combination $\lambda g$ enters), and for $s=\lambda=0$ one recovers $r_{\rm ISCO}=6M$, $\eta=5.72\%$.}
\label{tab:isco}
\begin{ruledtabular}
\begin{tabular}{cc cccc cc cccc cc cccc}
\multicolumn{6}{c}{$g/M=0$} & \multicolumn{6}{c}{$g/M=0.5$} & \multicolumn{6}{c}{$g/M=0.7$}\\
$s$ & $\lambda$ & $r_{\rm ISCO}$ & $\mathcal L_{\rm ISCO}$ & $\mathcal E_{\rm ISCO}$ & $\eta$ & $s$ & $\lambda$ & $r_{\rm ISCO}$ & $\mathcal L_{\rm ISCO}$ & $\mathcal E_{\rm ISCO}$ & $\eta$ & $s$ & $\lambda$ & $r_{\rm ISCO}$ & $\mathcal L_{\rm ISCO}$ & $\mathcal E_{\rm ISCO}$ & $\eta$\\
\hline
+0.0 & +0.0 & 6.000 & 3.464 & 0.9428 & 5.72 & +0.0 & +0.0 & 5.572 & 3.373 & 0.9391 & 6.09 & +0.0 & +0.0 & 5.075 & 3.268 & 0.9342 & 6.58 \\
+0.3 & +0.0 & 5.470 & 3.291 & 0.9371 & 6.29 & +0.3 & +0.0 & 5.033 & 3.187 & 0.9323 & 6.77 & +0.3 & +0.0 & 4.523 & 3.064 & 0.9258 & 7.42 \\
-0.3 & +0.0 & 6.457 & 3.611 & 0.9470 & 5.30 & -0.3 & +0.0 & 6.039 & 3.529 & 0.9439 & 5.61 & -0.3 & +0.0 & 5.558 & 3.437 & 0.9400 & 6.00 \\
+0.0 & +0.2 & 6.000 & 3.464 & 0.9428 & 5.72 & +0.0 & +0.2 & 5.596 & 3.536 & 0.9335 & 6.65 & +0.0 & +0.2 & 5.145 & 3.497 & 0.9263 & 7.37 \\
+0.0 & -0.2 & 6.000 & 3.464 & 0.9428 & 5.72 & +0.0 & -0.2 & 5.556 & 3.207 & 0.9448 & 5.52 & +0.0 & -0.2 & 5.011 & 3.031 & 0.9424 & 5.76 \\
+0.3 & +0.2 & 5.470 & 3.291 & 0.9371 & 6.29 & +0.3 & +0.2 & 5.083 & 3.354 & 0.9267 & 7.33 & +0.3 & +0.2 & 4.633 & 3.304 & 0.9181 & 8.19 \\
\end{tabular}
\end{ruledtabular}
\end{table*}

\emph{Radiative efficiency.} For a thin accretion disk truncated at the ISCO the radiative efficiency is $\eta=1-\mathcal E_{\rm ISCO}$~\cite{Novikov:1973,Bardeen:1972fi}. From Table~\ref{tab:coeff},
\begin{equation}
\eta\,[\%]\simeq5.72+1.60\,s+5.56\,\lambda g+1.31\,g^2+\dots ,
\label{eq:eta_pert}
\end{equation}
so that the Bardeen geometry alone ($s=\lambda=0$) raises the efficiency monotonically from $5.72\%$ (Schwarzschild) to $6.09\%$ at $g/M=0.5$, $6.58\%$ at $g/M=0.7$ and $6.86\%$ at the extremal charge, while prograde spin and an attractive coupling ($\lambda g>0$) increase it further, up to $\eta=8.19\%$ for $(g,s,\lambda)=(0.7,0.3,0.2)$ (Table~\ref{tab:isco}).

\begin{figure*}[t]
\includegraphics[width=0.97\linewidth]{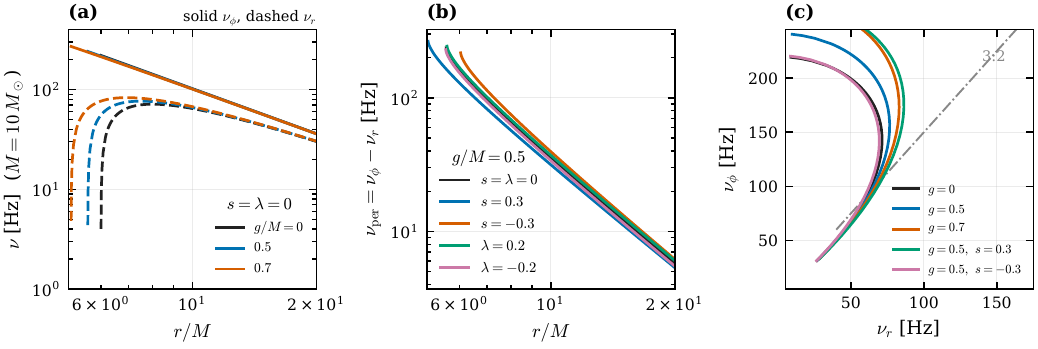}
\caption{Orbital and radial epicyclic frequencies of circular orbits for $M=10\,M_\odot$ ($\nu\propto1/M$). (a) $\nu_\phi$ (solid) and $\nu_r$ (dashed) versus radius for $s=\lambda=0$ and $g/M=0,0.5,0.7$; $\nu_r$ vanishes at the ISCO. (b) Periastron precession frequency $\nu_{\rm per}=\nu_\phi-\nu_r$ for $g/M=0.5$ and several $(s,\lambda)$. (c) Frequency--frequency diagram $(\nu_r,\nu_\phi)$ with the 3:2 line.}
\label{fig:freq}
\end{figure*}

\emph{Orbital and epicyclic frequencies.} For a circular orbit the orbital frequency is $\Omega=u^\phi/u^t$, where the four-velocity $u^\mu$ of the Tulczyjew centroid is obtained from the MPD equations as described in Appendix~\ref{app:mpd} (it differs from $p^\mu/m$ at order $s\lambda g$ and $s^2$); for $s=\lambda=0$ it reduces to $\Omega^2=f'(r)/2r$. The radial epicyclic frequency $\Omega_r$ is the frequency of small radial oscillations in coordinate time, obtained from the linearised MPD flow around the circular orbit (Appendix~\ref{app:mpd}); it vanishes at the ISCO. Restoring units, $\nu=\Omega/(2\pi M)$ with $GM_\odot/c^3=4.925\,\mu{\rm s}$, i.e.\ $\nu_{\rm ISCO}=219.9\,{\rm Hz}\,(10\,M_\odot/M)$ for Schwarzschild. The results are shown in Fig.~\ref{fig:freq} and Table~\ref{tab:freq}. At fixed mass the ISCO frequency increases by $9.7\%$ ($23\%$) for $g/M=0.5$ ($0.7$), and the frequency of the 3:2 resonance, $\nu_\phi/\nu_r=3/2$, which in Schwarzschild occurs at $r=10.8M$, moves inwards ($r_{3:2}=10.40M$ for $g/M=0.5$) and its frequency increases by $5.3\%$ ($11.4\%$). Prograde spin $s=0.3M$ shifts it by a further $+13\%$ and an attractive coupling $\lambda=0.2$ by $+4\%$ at $g/M=0.5$. Because $\nu\propto1/M$, these shifts are degenerate with a rescaling of the mass: fitting an observed 3:2 pair with the Schwarzschild relation would underestimate the mass of a Bardeen black hole with $g/M=0.5$ by about $5\%$. Breaking this degeneracy requires an independent mass measurement or the simultaneous use of several frequencies, e.g.\ the periastron precession $\nu_{\rm per}=\nu_\phi-\nu_r$ shown in Fig.~\ref{fig:freq}(b).

\begin{table*}[t]
\caption{Orbital frequencies for $M=10\,M_\odot$ ($\nu\propto1/M$). $\nu_{\rm ISCO}=\Omega_{\rm ISCO}/(2\pi M)$ is the orbital frequency at the ISCO; the 3:2 resonance is defined by $\nu_\phi/\nu_r=3/2$, where $\nu_r$ is the radial epicyclic frequency measured in coordinate time. The last two columns give the relative shift (in \%) of $\nu_{\rm ISCO}$ and of $\nu_\phi^{3:2}$ with respect to Schwarzschild ($g=0$).}
\label{tab:freq}
\begin{ruledtabular}
\begin{tabular}{ccc ccccc cc}
$g/M$ & $s/M$ & $\lambda$ & $r_{\rm ISCO}/M$ & $\nu_{\rm ISCO}$ [Hz] & $r_{3:2}/M$ & $\nu_\phi^{3:2}$ [Hz] & $\nu_r^{3:2}$ [Hz] & $\Delta\nu_{\rm ISCO}$ & $\Delta\nu^{3:2}_\phi$\\
\hline
$0$ & $0$ & $0$ & $6.000$ & $219.9$ & $10.80$ & $91.0$ & $60.7$ & $0$ & $0$ \\
$0.3$ & $0$ & $0$ & $5.854$ & $226.8$ & $10.66$ & $92.7$ & $61.8$ & $+3.2$ & $+1.8$ \\
$0.5$ & $0$ & $0$ & $5.572$ & $241.2$ & $10.40$ & $95.9$ & $63.9$ & $+9.7$ & $+5.3$ \\
$0.7$ & $0$ & $0$ & $5.075$ & $270.7$ & $9.98$ & $101.4$ & $67.6$ & $+23.1$ & $+11.4$ \\
$0.5$ & $+0.3$ & $0$ & $5.033$ & $269.8$ & $9.52$ & $107.7$ & $71.8$ & $+22.7$ & $+18.3$ \\
$0.5$ & $-0.3$ & $0$ & $6.039$ & $220.9$ & $11.19$ & $87.0$ & $58.0$ & $+0.5$ & $-4.4$ \\
$0.5$ & $0$ & $+0.2$ & $5.596$ & $248.1$ & $10.44$ & $99.3$ & $66.2$ & $+12.8$ & $+9.1$ \\
$0.5$ & $0$ & $-0.2$ & $5.556$ & $233.4$ & $10.38$ & $91.9$ & $61.3$ & $+6.1$ & $+0.9$ \\
\end{tabular}
\end{ruledtabular}
\end{table*}

\section{Superluminal Motion Constraints}
\label{sec:superluminal}

The motion of spinning test particles described by the MPD equations~\eqref{eq:mpd_mom}--\eqref{eq:mpd_spin}, supplemented by the Tulczyjew condition~\eqref{eq:tssc}, is fundamentally non-geodesic because of spin--curvature coupling. While the four-momentum $p^\mu$ remains timelike ($p^\mu p_\mu = -m^2 < 0$) and its magnitude is conserved, the four-velocity $u^\mu = dx^\mu / d\tau$ is generally not parallel to $p^\mu$. This misalignment implies that the four-velocity normalization, $u^\mu u_\mu$, is not necessarily $-1$ and can deviate from this value along the trajectory.

In certain regimes, particularly for large spin magnitudes, high orbital angular momenta, or near the black hole horizon, the spin curvature force can accelerate the particle's centroid such that $u^\mu u_\mu  > 0$, rendering the worldline spacelike. Such superluminal motion violates causality and is physically inadmissible for massive particles, as it would imply velocities exceeding the local speed of light relative to observers comoving with the centroid.
To ensure physical viability, we impose the timelike constraint~\cite{Mannobova:2025uqf}
\begin{equation}
u^\mu u_\mu = -1 
\label{eq:timelike_constraint}
\end{equation}

throughout the trajectory. This condition restricts the admissible parameter space spanned by the particle's specific spin $s$, magnetic charge $\lambda$, specific angular momentum $\mathcal{L}$, and the black hole's magnetic monopole charge $g$. Here we have to take 

\begin{eqnarray}
    u^{a} u_{a}
    &&= g_{tt} \left( \frac{dt}{d\tau} \right)^{2}
      + g_{rr} \left( \frac{dr}{d\tau} \right)^{2}
      + g_{\phi\phi} \left( \frac{d\phi}{d\tau} \right)^{2} \nonumber \\
   && = g_{tt} + g_{rr} \left( u^{r} \right)^{2}
      + g_{\phi\phi} \left( u^{\phi} \right)^{2}
      \leq 0.
    \label{eq:placeholder_label}
\end{eqnarray}

\begin{figure*}[ht!]
  \centering
   \includegraphics[width=0.32\linewidth]{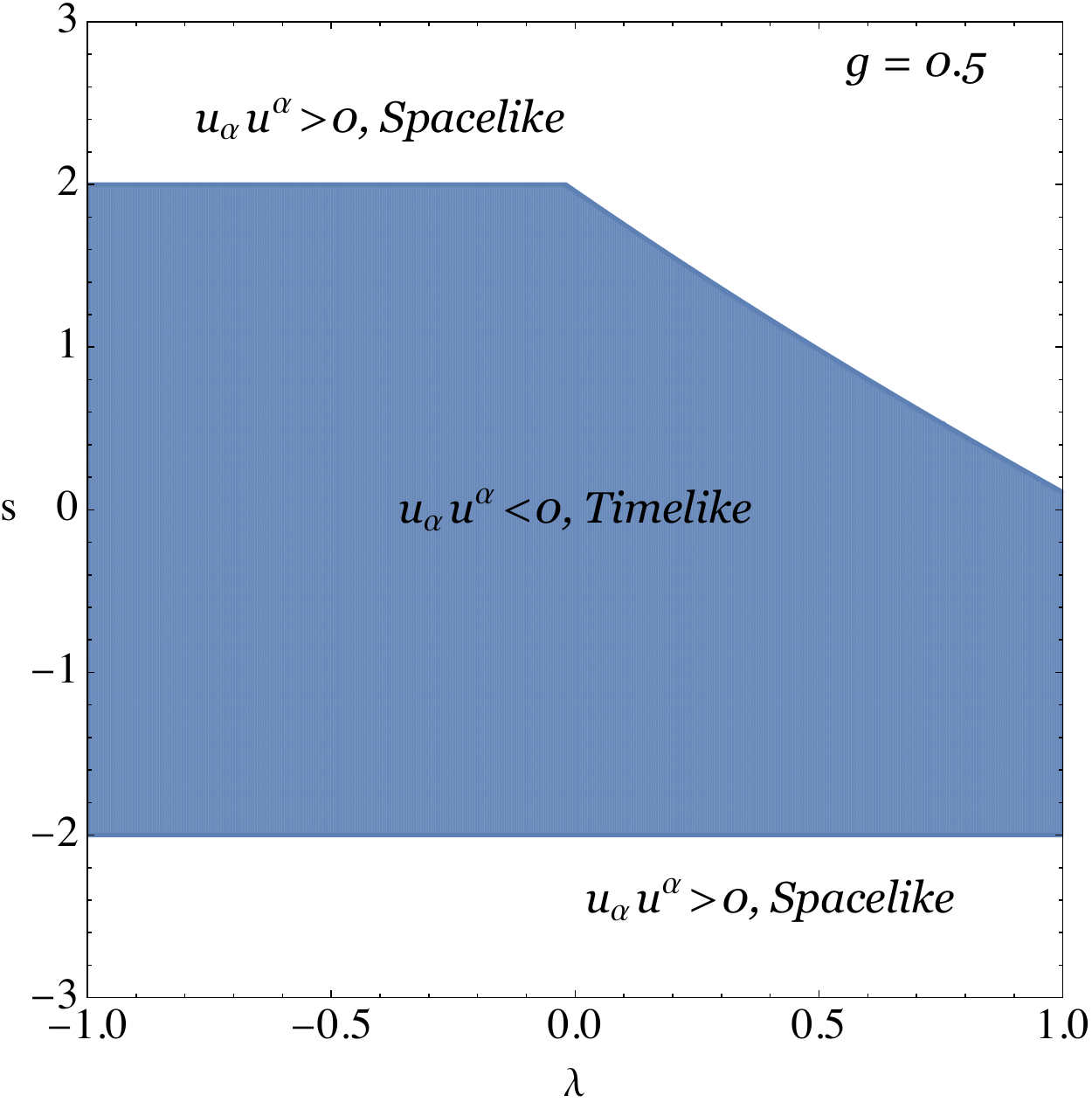}
   \includegraphics[width=0.32\linewidth]{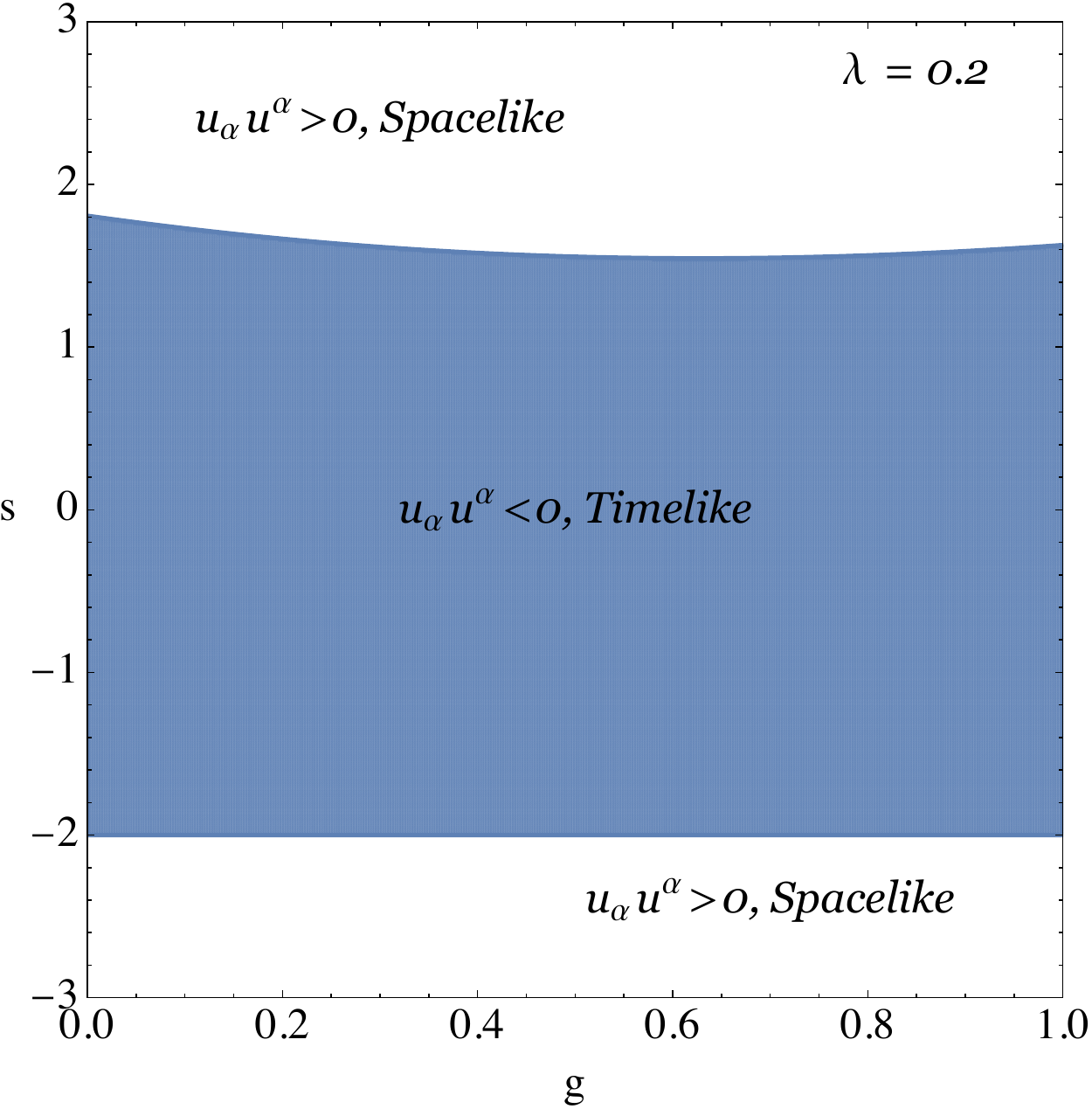}
    \includegraphics[width=0.32\linewidth]{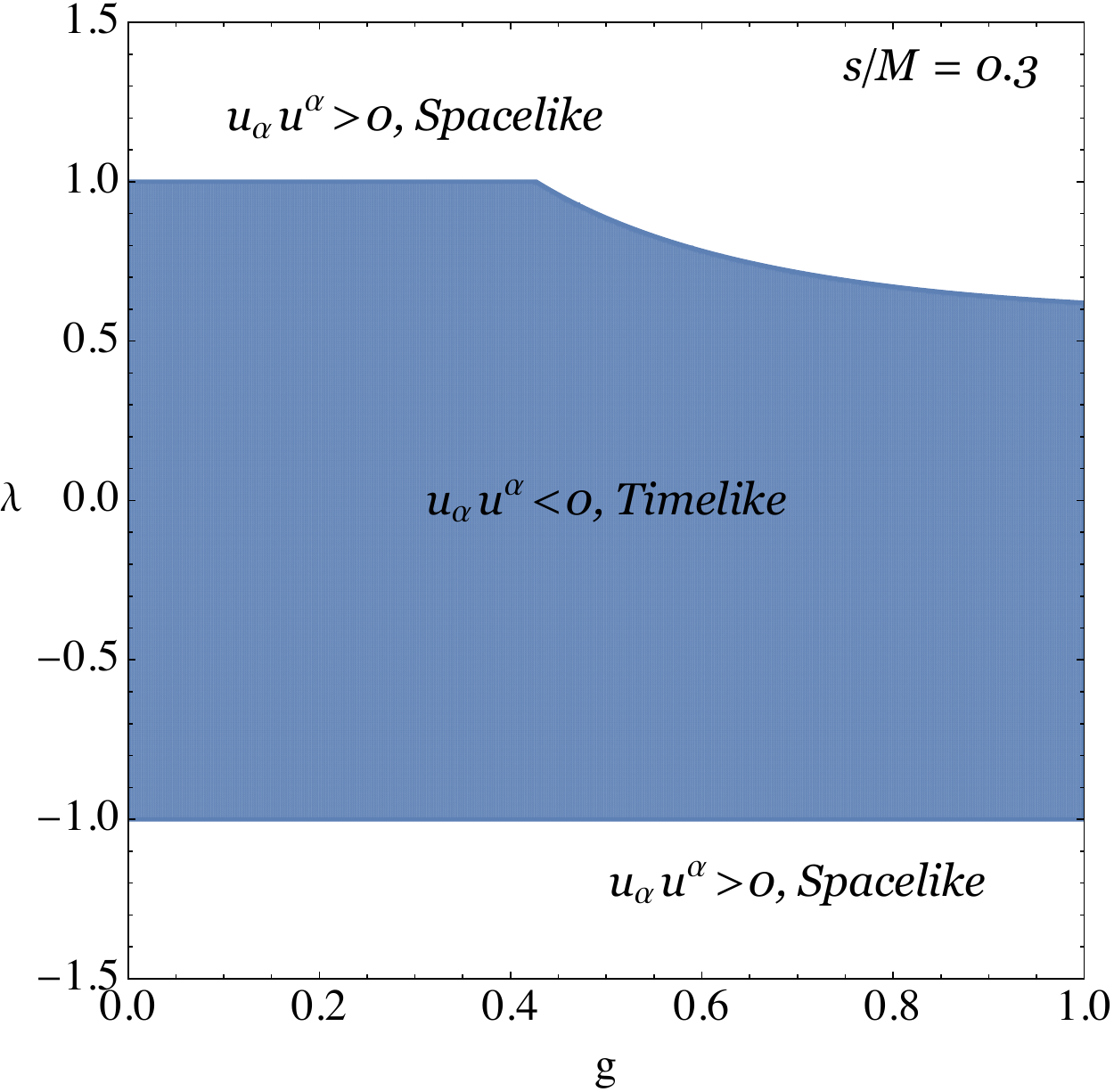}
  \caption{Parameter space regions where the function $u^\mu u_\mu$ (related to timelike/spacelike conditions or superluminal constraints) takes specific values, shown in different planes for timelike and spacelike regimes. We plot it for various parameter combinations with $M=1$.}
   \label{fig:superluminal}
\end{figure*}

which measures the deviation of the four-velocity from unitary normalization. We compute this function from $u^\mu u_\mu$ using the expressions for $p_t$ and $p_\phi$ in Eqs.~\eqref{s2e13}--\eqref{s2e14} and the radial velocity derived from the effective potential. Physically viable trajectories satisfy $u^\mu u_\mu \leq 0$ for all $r$ along the orbit, while $u^\mu u_\mu > 0$ indicates forbidden superluminal motion. The zero contour $u^\mu u_\mu=0$ defines the boundary of the allowed region in the $(s, \lambda, g)$ parameter space.

For prograde orbits ($s > 0$, spin aligned with orbital angular momentum), the spin-curvature force acts repulsively, pushing the particle outward and facilitating closer approaches to the horizon; this can more readily induce superluminal velocities for sufficiently large $s$. Conversely, retrograde spins ($s < 0$) generate an attractive force, stabilizing orbits but restricting the maximum $s$ before superluminal motion ensues.

The interplay with the magnetic charge is particularly rich: in Fig.~\ref{fig:superluminal} repulsive interactions ($\lambda g < 0$) leave the admissible domain unchanged, while attractive interactions ($\lambda g > 0$) strongly reduce the allowed prograde spin.

Figure~\ref{fig:superluminal} illustrates the admissible regions in the parameter space where the timelike condition $u^\mu u_\mu = -1$ is satisfied, ensuring physically viable (non-superluminal) trajectories for magnetically charged spinning test particles around the Bardeen regular black hole. The plots show the regions of the function $u^\mu u_\mu$ (quantifying the deviation from the timelike constraint, with $u^\mu u_\mu \leq 0$ corresponding to allowed timelike worldlines and $u^\mu u_\mu > 0$ to forbidden spacelike/superluminal regimes) in different parameter planes, for $M=1$.

The left panel of Figure~\ref{fig:superluminal} displays the allowed region in the $(s, \lambda)$ plane for fixed specific angular momentum $\mathcal{L}$ and black hole magnetic charge $g$. For $\lambda\lesssim0$ the allowed region is bounded by $|s|\simeq2M$, while for positive $\lambda$ (attractive, $\lambda g>0$) the upper (prograde) bound decreases roughly linearly with $\lambda$; the retrograde bound is unchanged.

The middle panel of Figure~\ref{fig:superluminal} presents the parameter space in the $(s, g)$ plane for fixed $\mathcal{L}$ and particle magnetic charge $\lambda$. The allowed region is asymmetric with respect to the sign of $s$ and depends only weakly on $g$ over the range shown.

The right panel of Figure~\ref{fig:superluminal} shows the constraints in the $(\lambda, g)$ plane for fixed spin $s$ and angular momentum $\mathcal{L}$. The allowed range of $\lambda$ is bounded by $|\lambda|\simeq1$; the upper bound on $\lambda$ (attractive coupling) decreases with increasing $g$, whereas the lower one is not affected.

These superluminal constraints define the physically valid domain within the pole-dipole approximation of the MPD formalism. Analyses of circular orbits, ISCO properties, and high-energy collisions should respect these regions to ensure all considered trajectories remain causal and timelike. These bounds are particularly relevant for astrophysical applications, such as extreme mass-ratio inspirals involving magnetized compact objects, where exceeding them would indicate the need for higher-order multipole corrections or the breakdown of the test-particle limit.

In summary, the superluminal constraints delineate the physically allowable regime for magnetically charged spinning particles in the Bardeen spacetime. These bounds must be respected when analyzing circular orbits, ISCO properties, and high-energy collision processes discussed in subsequent sections.

\subsection{Superluminality bound for circular orbits}
\label{subsec:sl_circ}

\begin{figure*}[ht]
\includegraphics[width=\linewidth]{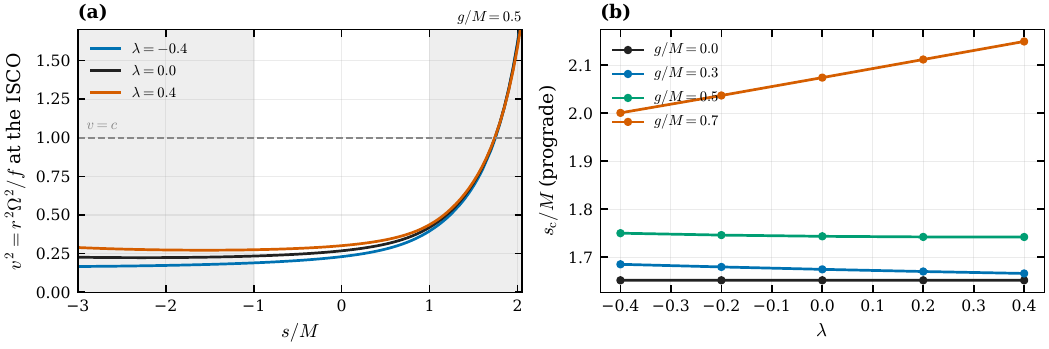}
\caption{Superluminality of circular orbits. (a) Squared velocity $v^2=r^2\Omega^2/f$ (relative to static observers) at the ISCO as a function of $s$ for $g/M=0.5$; $v^2=1$ marks the null limit and the shaded bands $|s|>M$ lie outside the pole-dipole validity. (b) Critical prograde spin $s_{\rm c}$ versus $\lambda$ for several $g$.}
\label{fig:sl}
\end{figure*}

As an independent and quantitative complement we examine the timelike condition for circular orbits. For a circular orbit $u^\mu\propto(1,0,0,\Omega)$ with $\Omega=u^\phi/u^t$, so that $u^\mu u_\mu\le0$ is equivalent to $v^2\equiv r^2\Omega^2/f\le1$, i.e.\ the velocity relative to a static observer must not exceed the speed of light. The direction of $u^\mu$ is obtained from the consistency of the Tulczyjew condition with the MPD equations, Eq.~\eqref{eq:A1} of Appendix~\ref{app:mpd}. The velocity $v$ is largest at the ISCO, and we therefore find, for each $(g,\lambda)$, the critical prograde spin $s_{\rm c}$ at which $v^2=1$ at the ISCO (Fig.~\ref{fig:sl}, Table~\ref{tab:sc}). Retrograde spin never leads to superluminal circular orbits ($v^2\simeq0.2$--$0.3$). For $g/M=0.5$ one finds $s_{\rm c}\simeq1.74M$ with a very weak dependence on $\lambda$ ($1.742$ to $1.750$ for $\lambda\in[-0.4,0.4]$); $s_{\rm c}$ increases with the monopole charge, from $1.652M$ (Schwarzschild) to $2.07M$ at $g/M=0.7$. This criterion applies to circular orbits only, whereas Fig.~\ref{fig:superluminal} is computed for a different family of orbits (the values of $r$ at which Eq.~\eqref{eq:placeholder_label} is evaluated are not fixed in the figure), so a point-by-point comparison is not possible; both give a critical spin of order $2M$ at $g/M=0.5$. In any case the pole-dipole description requires $s\ll M$ (a body of radius $R$ cannot carry a specific spin larger than $\sim R$ in geometrized units), so that the superluminal bound lies well outside the domain of validity of the test-particle approximation; the ranges $|s|>M$ in Figs.~\ref{iscob}--\ref{iscop} and~\ref{fig:superluminal} should be understood as formal.

\begin{table}[b]
\caption{Critical prograde spin $s_{\rm c}/M$ above which circular orbits at the ISCO become superluminal ($v^2=r^2\Omega^2/f=1$) for the Tulczyjew spin condition.}
\label{tab:sc}
\begin{ruledtabular}
\begin{tabular}{c ccccc}
 & \multicolumn{5}{c}{$\lambda$}\\
$g/M$ & $-0.4$ & $-0.2$ & $0$ & $0.2$ & $0.4$\\
\hline
$0$ & $1.652$ & $1.652$ & $1.652$ & $1.652$ & $1.652$ \\
$0.3$ & $1.685$ & $1.680$ & $1.675$ & $1.670$ & $1.666$ \\
$0.5$ & $1.750$ & $1.746$ & $1.743$ & $1.742$ & $1.742$ \\
$0.7$ & $2.000$ & $2.037$ & $2.074$ & $2.112$ & $2.149$ \\
\end{tabular}
\end{ruledtabular}
\end{table}

\section{Collisions of Magnetically Charged Spinning Particles around Bardeen black holes}
\label{sec:collisions}

\begin{figure*}[ht!]
\centering
\includegraphics[width=0.85\linewidth]{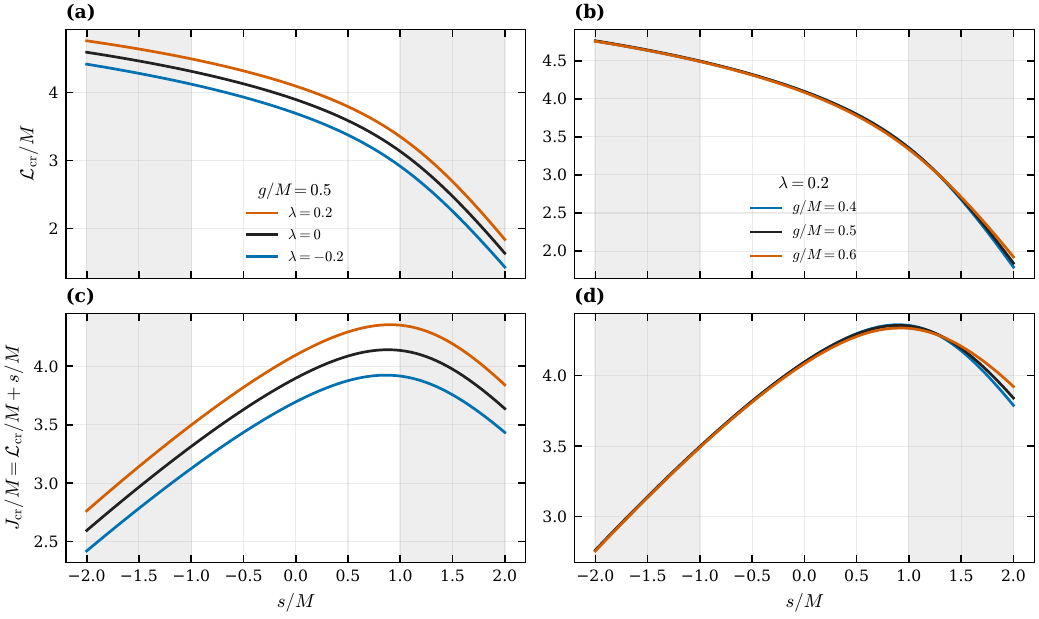}
\caption{Critical (marginally bound) angular momentum of a magnetically charged spinning particle, defined by $V_{\rm eff}=1$, $V_{\rm eff}'=0$ outside the horizon. Upper row: $\mathcal L_{\rm cr}/M$; lower row: total angular momentum $J_{\rm cr}/M=\mathcal L_{\rm cr}/M+s/M$. Left column: $g/M=0.5$, $\lambda=0,\pm0.2$; right column: $\lambda=0.2$, $g/M=0.4,0.5,0.6$. The grey bands $|s|>M$ lie outside the validity range of the pole--dipole approximation (Sec.~\ref{sec:discussion}).}
\label{fig:jcrit}
\end{figure*}

In extremal or near-extremal black hole spacetimes, high-energy collisions near the horizon can reach arbitrarily large centre-of-mass energies, the Ba\~nados--Silk--West (BSW) effect~\cite{Banados:2009pr}. The mechanism requires (i) a degenerate (extremal) horizon and (ii) one of the colliding particles to be \emph{critical}, i.e.\ to reach the horizon with a vanishing radial kinetic factor~\cite{Zaslavskii:2010jd,Berti:2009bk}. Both conditions are sensitive to the near-horizon geometry. A regular black hole is not excluded a priori: the Bardeen solution has an extremal limit $g=g_{\rm ext}=4/(3\sqrt3)\,M$ with $r_h=\sqrt{32/27}\,M$, and the monopole coupling gives a charged particle a Coulomb-like term that can play the role of the critical charge. We therefore treat two distinct questions: how the \emph{capture threshold} $\mathcal L_{\rm cr}$ depends on $(s,\lambda,g)$ (Sec.~\ref{subsec:critical}), and whether $\mathcal E_{\rm CM}$ is finite or divergent at the horizon (Secs.~\ref{subsec:center_of_mass_energy} and \ref{subsec:horizon}).

\subsection{Critical angular momentum}\label{subsec:critical}

For a particle with $\mathcal E=1$ (released from rest at infinity) we define the critical angular momentum $\mathcal L_{\rm cr}$ as the \emph{marginally bound} value: the unstable circular orbit at which $V_{\rm eff}=\mathcal E$ and $V_{\rm eff}'=0$ simultaneously, so that the particle takes infinite proper time to leave and a slight reduction of $\mathcal L$ sends it over the barrier into the black hole. Equivalently, $\mathcal L_{\rm cr}$ is the capture threshold: particles with $\mathcal L<\mathcal L_{\rm cr}$ fall into the horizon, those with $\mathcal L>\mathcal L_{\rm cr}$ are scattered. For $g=s=\lambda=0$, this gives $\mathcal L_{\rm cr}=4M$, the familiar Schwarzschild value. Note that this is a statement about the radial turning points outside the horizon; it is the \emph{critical-particle} condition at the horizon, $\mathcal E=-\lambda g/r_h$ in the present sign convention, that governs a possible BSW divergence (Sec.~\ref{subsec:horizon}), and the two should not be confused.

Figure~\ref{fig:jcrit} shows $\mathcal L_{\rm cr}$ and the total angular momentum $J_{\rm cr}=\mathcal L_{\rm cr}+s$ (the conserved total angular momentum) from the numerical solution of $V_{\rm eff}=1$, $V_{\rm eff}'=0$. The following trends are found.

\emph{Spin.} Prograde spin lowers the threshold: for $g/M=0.5$, $\lambda=0$, $\mathcal L_{\rm cr}=4.31,\,3.90,\,3.14$ at $s=-1,\,0,\,+1$ (in units of $M$), in line with the spin--curvature force being repulsive for $s>0$. The same ordering holds for $J_{\rm cr}$ although with a weaker slope.

\emph{Magnetic coupling.} At fixed spin $\mathcal L_{\rm cr}$ grows with $\lambda$: at $s=0$, $g/M=0.5$ it is $3.70$, $3.90$ and $4.10$ for $\lambda=-0.2,0,+0.2$. In our sign convention $\lambda g>0$ is attractive, so an attractive coupling increases the angular momentum needed to be captured, because the additional binding must be compensated by a larger centrifugal barrier at the marginally bound orbit; the repulsive case ($\lambda g<0$) lowers it. The magnitude of the effect, about $\pm5\%$ for $|\lambda|=0.2$, is comparable to that of a spin $|s|\simeq0.3M$.

\emph{Monopole charge.} The dependence on $g$ is weak. At $s=\lambda=0$ one finds $\mathcal L_{\rm cr}/M=4$, $3.937$, $3.898$ and $3.847$ for $g/M=0,\,0.4,\,0.5,\,0.6$, i.e.\ the threshold \emph{decreases} slowly with $g$; for $\lambda=0.2$ the spread between $g/M=0.4$ and $0.6$ is below $1\%$ for $|s|\lesssim1$ (right column of Fig.~\ref{fig:jcrit}). This corrects the impression, conveyed by earlier versions of this analysis, that a larger $g$ raises $\mathcal L_{\rm cr}$.

In summary, $\mathcal L_{\rm cr}$ is controlled mainly by $s$ and by $\lambda$ and is rather insensitive to the regular-core parameter $g$. Because the threshold does not by itself imply that $\mathcal E_{\rm CM}$ is bounded or unbounded, we examine the collision energy separately.

\subsection{The center-of-mass energy of magnetically charged spinning particles}
\label{subsec:center_of_mass_energy}

\begin{figure*}[ht!]
   \centering
 \includegraphics[width=0.40\linewidth]{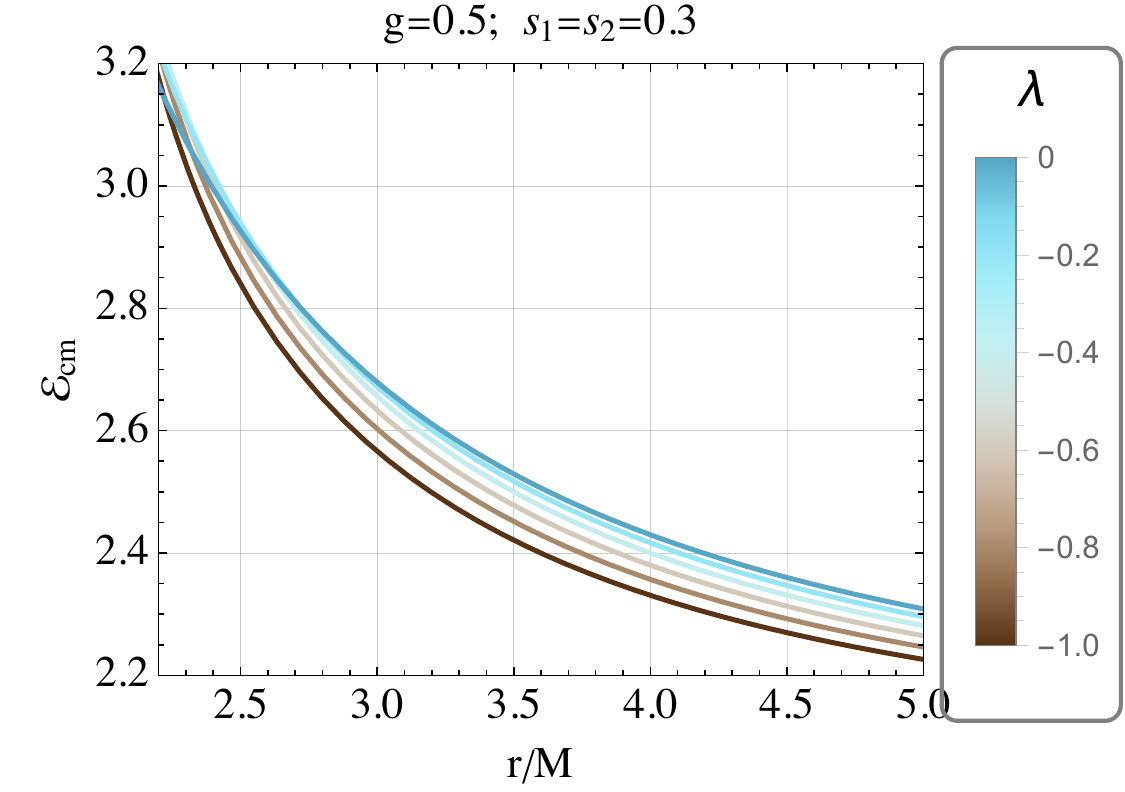}
 \includegraphics[width=0.40\linewidth]{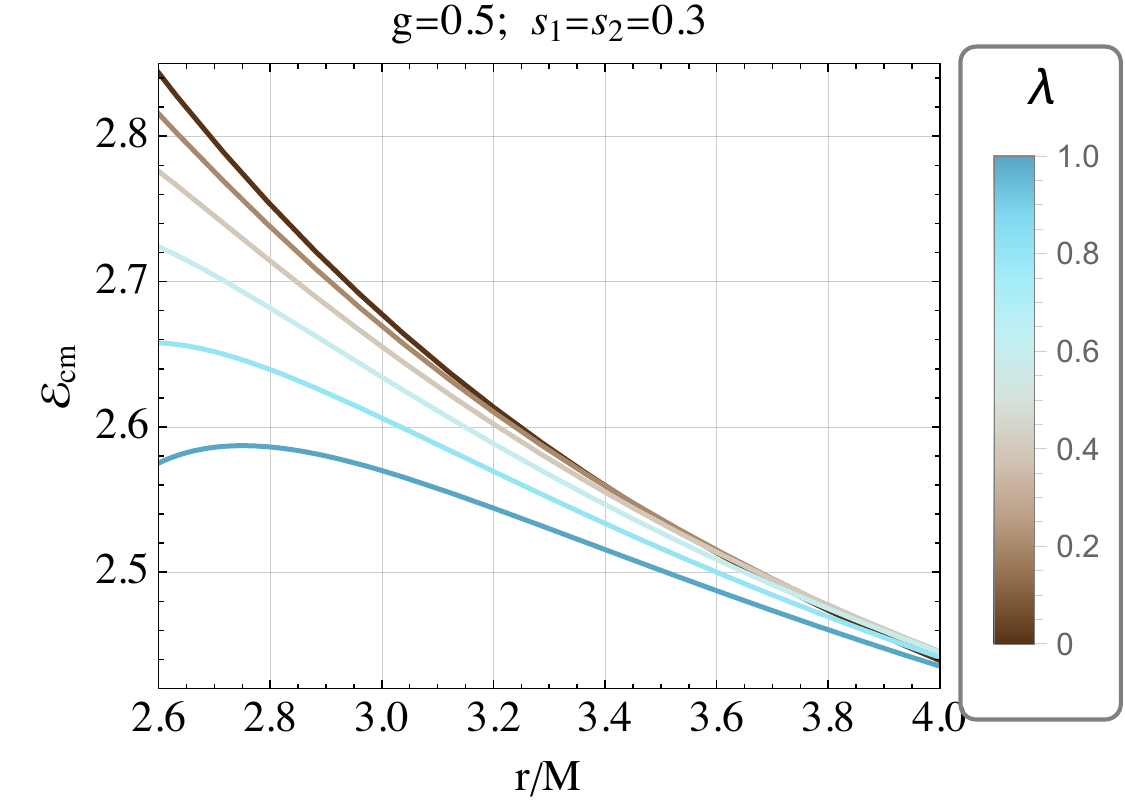}
\includegraphics[width=0.40\linewidth]{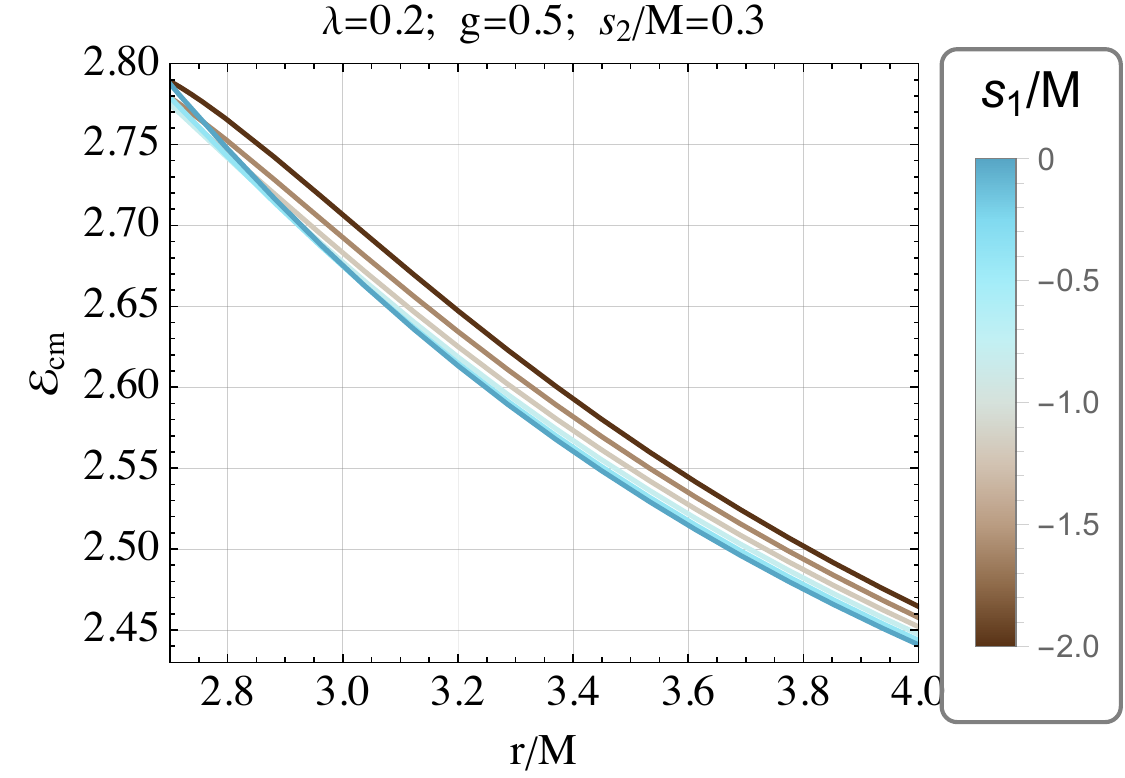}
\includegraphics[width=0.40\linewidth]{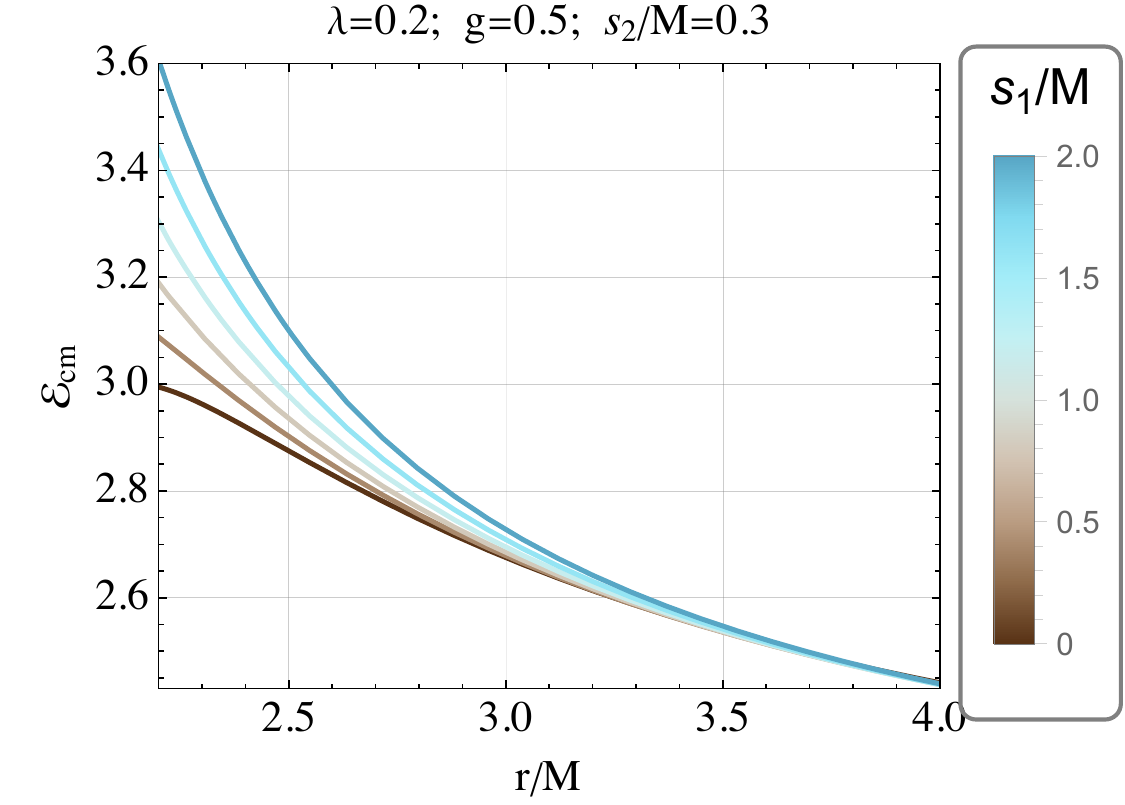}
\includegraphics[width=0.40\linewidth]{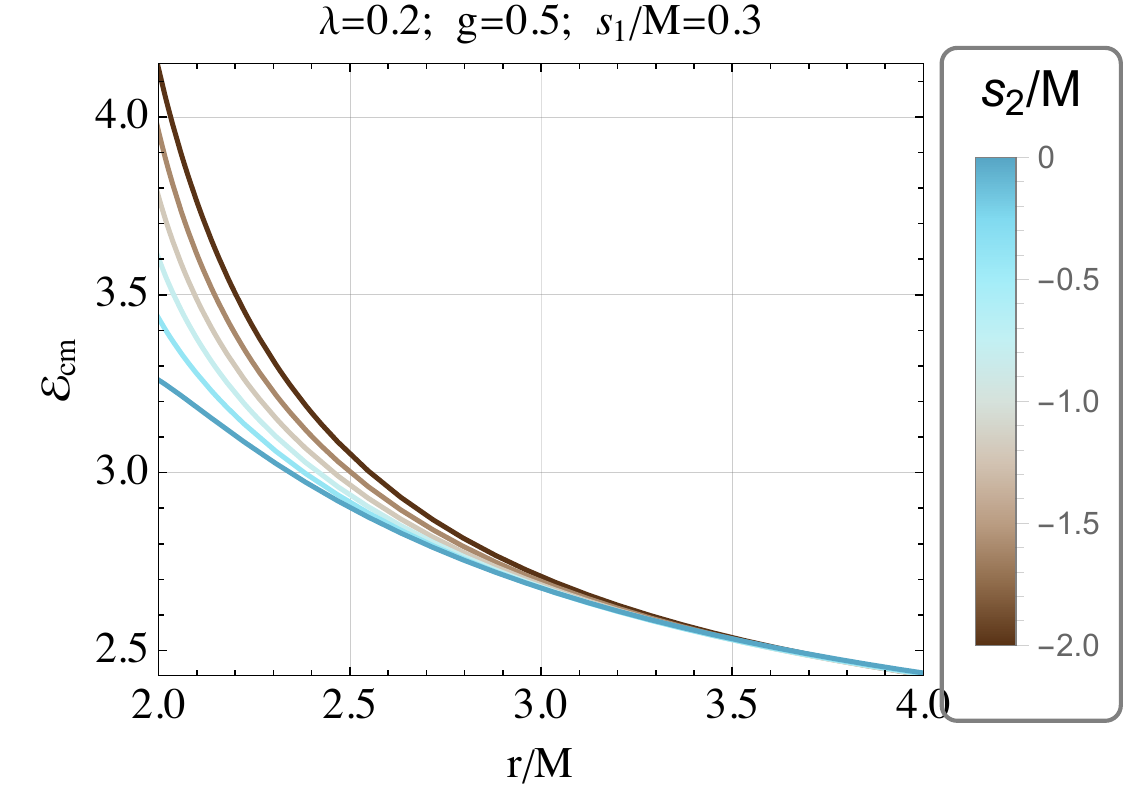}
\includegraphics[width=0.40\linewidth]{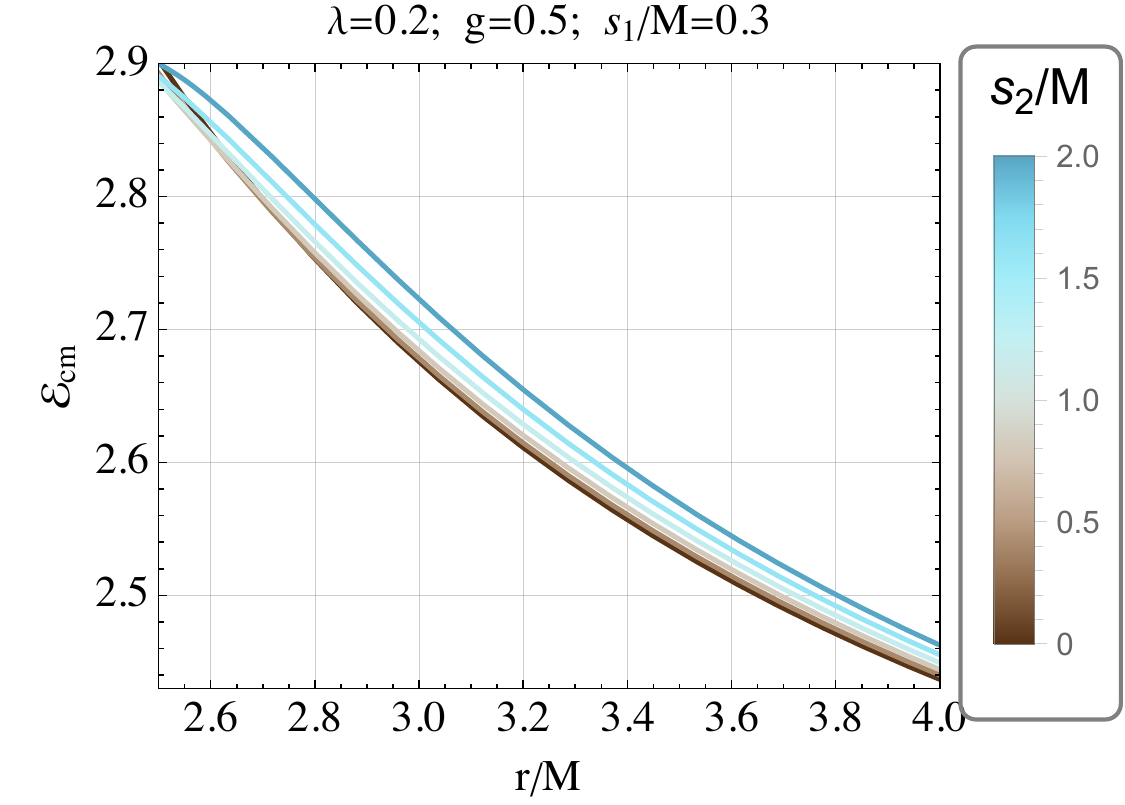}
\includegraphics[width=0.40\linewidth]{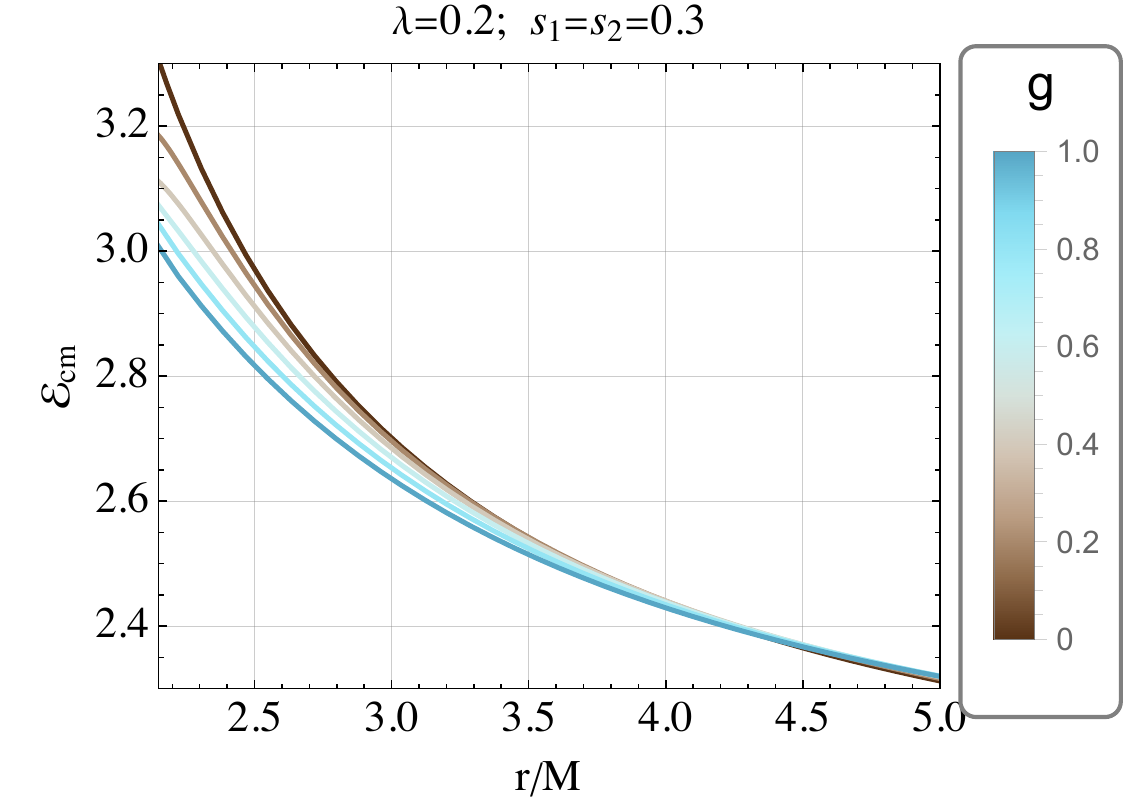}
\caption{Center-of-mass energy $\mathcal{E}_{\rm CM}$ of two colliding magnetically charged spinning particles as a function of radial coordinate $r/M$ near the Bardeen black hole. Illustrated for various combinations of particle spins $s_1$ and $s_2$, the particle's magnetic charge $\lambda$, and black hole magnetic charge $g$, with $M=1$.}
   \label{fig:ecm}
\end{figure*}

Building on the critical angular momentum analysis, we now examine the center-of-mass energy of collisions between two magnetically charged spinning test particles near the Bardeen black hole horizon. The center-of-mass energy for two particles is given by~\cite{Jumaniyozov:2025lox}
\begin{eqnarray}\label{eq:cm_energy}\nonumber 
   E_{CM}^2&=&-g^{\mu\nu}(p^{(1)}_\mu+p^{(2)}_\mu)(p^{(1)}_\nu+p^{(2)}_\nu)\\
   &=& m_1^2+m_2^2-2g^{\mu\nu}p^{(1)}_\mu p^{(2)}_\nu\,.
\end{eqnarray}

Here, $p_{\mu}^{(1)}$ and $p_{\mu}^{(2)}$ are the momenta of the first and second particles, respectively, given in Eqs. (\ref{s2e13}) and (\ref{s2e14}).

\begin{eqnarray}\label{ecmps}
   {\cal E}_{CM}\equiv\frac{E_{CM}^2}{2m^2} =1-g^{tt}p^{(1)}_t p^{(2)}_t-g^{rr}p^{(1)}_r p^{(2)}_r-g^{\phi \phi}p^{(1)}_{\phi} p^{(2)}_{\phi},
\end{eqnarray}
Here $p_\mu$ denotes the momentum per unit mass and $\mathcal E_{\rm CM}=E_{\rm CM}^2/(2m^2)$, so that $\mathcal E_{\rm CM}=1$ for two particles at rest relative to each other; this is the normalisation used in all figures (note that it is the \emph{square} of the centre-of-mass energy that is plotted). The finiteness of $\mathcal E_{\rm CM}$ at the horizon is decided by the degeneracy of the horizon and not by the regularity of the core; this is analysed in Sec.~\ref{subsec:horizon}.

To compute $\mathcal{E}_{\rm CM}(r)$, we consider multiple collision scenarios: (i) both particles on near-critical orbits with opposite directions, (ii) one critical and one non-critical particle falling from infinity, and (iii) symmetric collisions of identical particles. We obtain the four-velocities $u^\alpha_{(i)}$ from the conserved quantities $\mathcal{E}_i$, $\mathcal{L}_i$, the spin tensor components (under the Tulczyjew condition), and the radial momentum derived from the effective potential. Magnetic interactions enter indirectly through trajectory modifications and directly via the Lorentz term in the MPD equations.
% TODO(authors): state the (E_i, L_i) used for each curve of Fig. 8 (the original figure data are not part of the source).

Figure~\ref{fig:ecm} illustrates the center-of-mass energy $\mathcal{E}_{\rm CM}$ of two colliding magnetically charged spinning test particles as a function of the radial coordinate $r/M$ near the horizon of the Bardeen regular black hole, with $M=1$. The plots show variations with respect to the spin parameters $s_1$ and $s_2$ of the first and second particles, the particle's specific magnetic charge $\lambda$, and the black hole's magnetic monopole charge parameter $g$.
The top panel shows $\mathcal{E}_{\text{CM}}$ versus $r$ for a specific set of parameters, such as $s_1/M = 0.3$, $s_2/M = 0.3$, $g = 0.5$. Over the negative range shown (top-left), $\mathcal{E}_{\text{CM}}$ decreases monotonically as $\lambda$ is made more negative (for $g>0$, the coupling $\lambda g<0$ is repulsive, and increasingly so), while over the positive range (top-right) $\mathcal{E}_{\text{CM}}$ decreases as $\lambda$ increases (the attractive coupling $\lambda g>0$ becoming stronger), so that the uncharged case $\lambda=0$ marks a local maximum of $\mathcal{E}_{\text{CM}}$ with respect to $\lambda$.
The middle panel illustrates $\mathcal{E}_{\text{CM}}$ versus $r$ for various values of the first particle's spin parameter $s_1/M$, with $s_2/M = 0.3$, $g = 0.5$, and $\lambda=0.2$ fixed. The dependence on $s_1$ is non-monotonic: $\mathcal{E}_{\text{CM}}$ is smallest near $s_1 \simeq 0$ and increases as $s_1$ moves toward either strongly retrograde (middle-left) or strongly prograde (middle-right) values.
The third panel shows $\mathcal{E}_{\text{CM}}$ versus $r$ for the analogous variation of the second particle's spin parameter $s_2/M$, with $s_1/M = 0.3$, $g = 0.5$, and $\lambda=0.2$ fixed. The same non-monotonic pattern is found: $\mathcal{E}_{\text{CM}}$ increases as $s_2$ becomes more strongly retrograde (left panel) or more strongly prograde (right panel), relative to the $s_2\simeq0$ case.
The bottom panel  depicts $\mathcal{E}_{\text{CM}}$ versus $r$ for a different combination, such as $s_1/M = 0.3$, $s_2/M = 0.3$, and $\lambda = 0.2$. From the figure, we can conclude that as the BH magnetic charge increases, the center-of-mass energy decreases.

Overall, Fig.~\ref{fig:ecm} shows that $\mathcal E_{\rm CM}$ near the horizon is large but finite for the parameters considered and is modulated by spin and by the magnetic couplings. Whether this finiteness is a consequence of regularity is a separate question that we address next; it should not be inferred from the curves of Fig.~\ref{fig:ecm} alone.

\subsection{Near-horizon limit and the degenerate horizon}\label{subsec:horizon}

\begin{figure*}[t]
\centering
\includegraphics[width=0.97\linewidth]{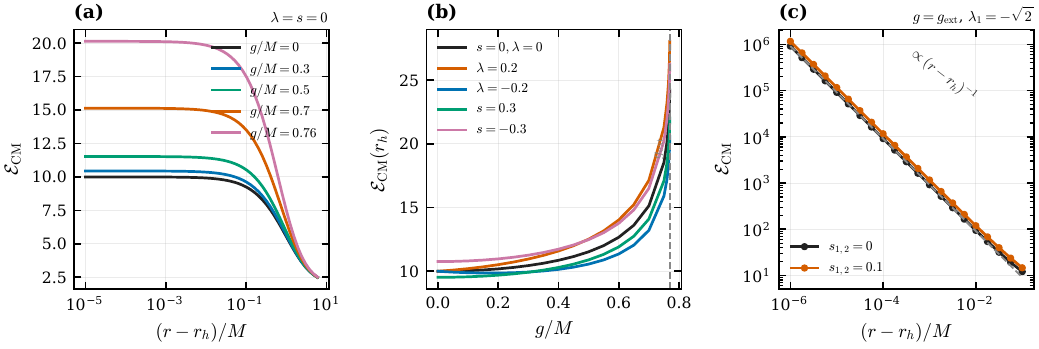}
\caption{Collision energy near the horizon. (a) $\mathcal E_{\rm CM}$ versus distance from the horizon for two marginally bound spinless, uncharged particles with $\mathcal L=\pm\mathcal L_{\rm cr}$, for $g/M=0,0.3,0.5,0.7,0.76$: it saturates at the finite value in Eq.~\eqref{eq:ecm_h}. (b) The horizon value $\mathcal E_{\rm CM}(r_h)$ versus $g$ for several $(s,\lambda)$ of the colliding pair; the dashed line marks $g_{\rm ext}$. (c) At $g=g_{\rm ext}$ with a critical particle ($\lambda_1=-\sqrt2$) and an uncharged partner, $\mathcal E_{\rm CM}\propto(r-r_h)^{-1}$ diverges.}
\label{fig:bsw}
\end{figure*}

Near a non-degenerate horizon $f\simeq f'(r_h)(r-r_h)$ and the radial momentum of a particle with $\mathcal E\neq-\lambda g/r_h$ behaves as $p_r\sim\mathcal E_{\rm eff}/f$, so that in $\mathcal E_{\rm CM}$ the terms $\propto1/f$ from $g^{tt}p_tp_t$ and $g^{rr}p_rp_r$ cancel. What remains is finite. For two uncharged, spinless, marginally bound particles with opposite angular momenta ($\mathcal L_{1,2}=\pm\mathcal L_{\rm cr}$), the limit is exactly
\begin{equation}
\mathcal E_{\rm CM}(r_h)=2\Bigl(1+\frac{\mathcal L_{\rm cr}^2}{r_h^2}\Bigr),
\label{eq:ecm_h}
\end{equation}
which equals $10$ for Schwarzschild ($\mathcal L_{\rm cr}=4M$, $r_h=2M$) and grows with $g$ only because $r_h$ shrinks: we find $\mathcal E_{\rm CM}(r_h)\simeq11.5$ at $g/M=0.5$ and $22.8$ at $g/M=0.768$, just below $g_{\rm ext}$ (Fig.~\ref{fig:bsw}(a,b)). Spin and charge shift these values by order unity (Fig.~\ref{fig:bsw}(b)) but do not change the picture: for every non-extremal $g<g_{\rm ext}$ examined, $\mathcal E_{\rm CM}(r_h)$ is finite, and it is the \emph{closeness to extremality} that makes it large, in the same way as for Kerr and Reissner--Nordstr\"om black holes.

The situation changes at the degenerate horizon $g=g_{\rm ext}$, where $f\simeq\tfrac12f''(r_0)(r-r_0)^2$. A particle with the critical charge $\lambda_1=-\mathcal E_1r_0/g_{\rm ext}$ ($=-\sqrt2$ for $\mathcal E_1=1$, $g=g_{\rm ext}$, $r_0=\sqrt{32/27}\,M$) has a vanishing leading term of $p_r$, and a collision with a generic second particle gives
\begin{equation}
\mathcal E_{\rm CM}\;\simeq\;\frac{C}{r-r_0}
\label{eq:bsw_div}
\end{equation}
here $C\simeq0.92\ (\text{uncharged, spinless partner with }\mathcal L_2=-3M),$ i.e.\ $E_{\rm CM}\propto(r-r_0)^{-1/2}$, the standard BSW scaling. This is verified numerically over six decades in Fig.~\ref{fig:bsw}(c): $\mathcal E_{\rm CM}$ reaches $9\times10^6$ at $r-r_0=10^{-7}M$ and the product $(r-r_0)\mathcal E_{\rm CM}$ stays constant, for $s=0$ as well as for $s=0.1M$. Hence \emph{the regular core does not cap the collision energy}: the energy is finite for non-extremal horizons, and divergent on the extremal horizon when a critical particle is present, exactly as for singular black holes. The finite maximum quoted in earlier discussions of regular black holes applies to the non-degenerate case, or to a sequence of collisions that stops at a finite distance from the horizon, and is a property of the horizon structure rather than of the regularity of the centre. The divergence is of course only formal: it requires fine-tuned critical particles, is limited by back-reaction and by the finite radiation/escape conditions~\cite{Berti:2009bk}, and the critical charge found above, $|\lambda_1|=\sqrt2$, is large in the sense discussed in Sec.~\ref{sec:discussion}.

\section{Constraints from twin-peak QPOs of microquasars}\label{sec:mcmc}

\subsection{Method}
We confront the characteristic frequencies of Sec.~\ref{subsec:eff} with the twin-peak high-frequency QPOs observed in four stellar-mass black-hole X-ray binaries~\cite{Remillard:2006fc,Strohmayer:2001,Remillard:2002,Morgan:1997}, following the Bayesian procedure used for Kerr black holes with the periastron-precession-type identification: the upper peak is the orbital frequency, $\nu_U=\nu_\phi$, and the lower one is $\nu_L=\nu_\phi-\nu_r$ ($\nu_r$ the radial epicyclic frequency), both evaluated at the same emission radius $r$ and both scaling as $1/M$. The parameters are $\boldsymbol\theta=(M,g,s,\lambda,r)$. The likelihood is Gaussian,
\begin{equation}
\ln\mathcal L=-\tfrac12\Bigl[\bigl(\tfrac{\nu_U-\nu_\phi}{\sigma_U}\bigr)^2+\bigl(\tfrac{\nu_L-\nu_\phi+\nu_r}{\sigma_L}\bigr)^2\Bigr]+\text{const.},
\end{equation}
with the priors: $M$ Gaussian around the dynamical mass~\cite{Greene:2001,Orosz:2011,Reid:2014} (truncated to the quoted range), $g/M$ uniform in $[0,0.75]$ (below $g_{\rm ext}=0.770$), $s/M$ and $\lambda$ uniform in $[-0.5,0.5]$, and $r$ uniform between $r_{\rm ISCO}(g,s,\lambda)$ and $r_{\rm ISCO}+11.5M$. The data and mass priors are listed in Table~\ref{tab:mcmc}. We use the affine-invariant sampler \texttt{emcee}~\cite{ForemanMackey:2013io} with 64 walkers and $3\times10^4$ steps (burn-in $8\times10^3$); the integrated autocorrelation time is $\lesssim260$ steps for all parameters, so every chain is longer than $110\,\tau$. Because each evaluation of $\nu_r$ from the full MPD flow takes tens of milliseconds, $\nu_\phi$, $\nu_r^2$ and $r_{\rm ISCO}$ were tabulated on a regular grid in $(g,s,\lambda,r-r_{\rm ISCO})$ from the exact solution and interpolated with cubic splines; against 25 random exact evaluations, the interpolation error is below $0.4\%$ in $\nu_\phi$ and $1.3\%$ in $\nu_r$, smaller than the observational errors.

\subsection{Results}
\begin{figure*}[t]
\centering
\includegraphics[width=0.97\linewidth]{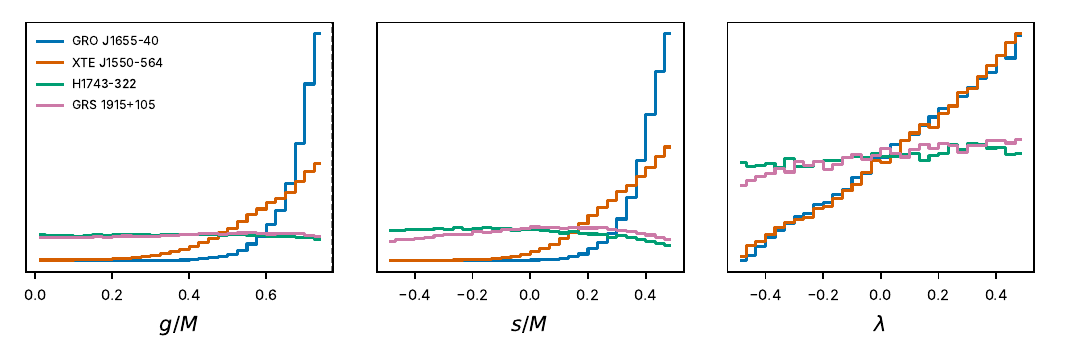}
\caption{Marginalised posteriors of $g/M$, $s/M$ and $\lambda$ for the four sources. The dashed line marks $g_{\rm ext}$. For GRO~J1655$-$40 and XTE~J1550$-$564, the posterior of $g$ is shifted to large values; for H1743$-$322 and GRS~1915+105, it is almost flat.}
\label{fig:mcmc_post}
\end{figure*}

\begin{figure*}[t]
\centering
\includegraphics[width=0.49\linewidth]{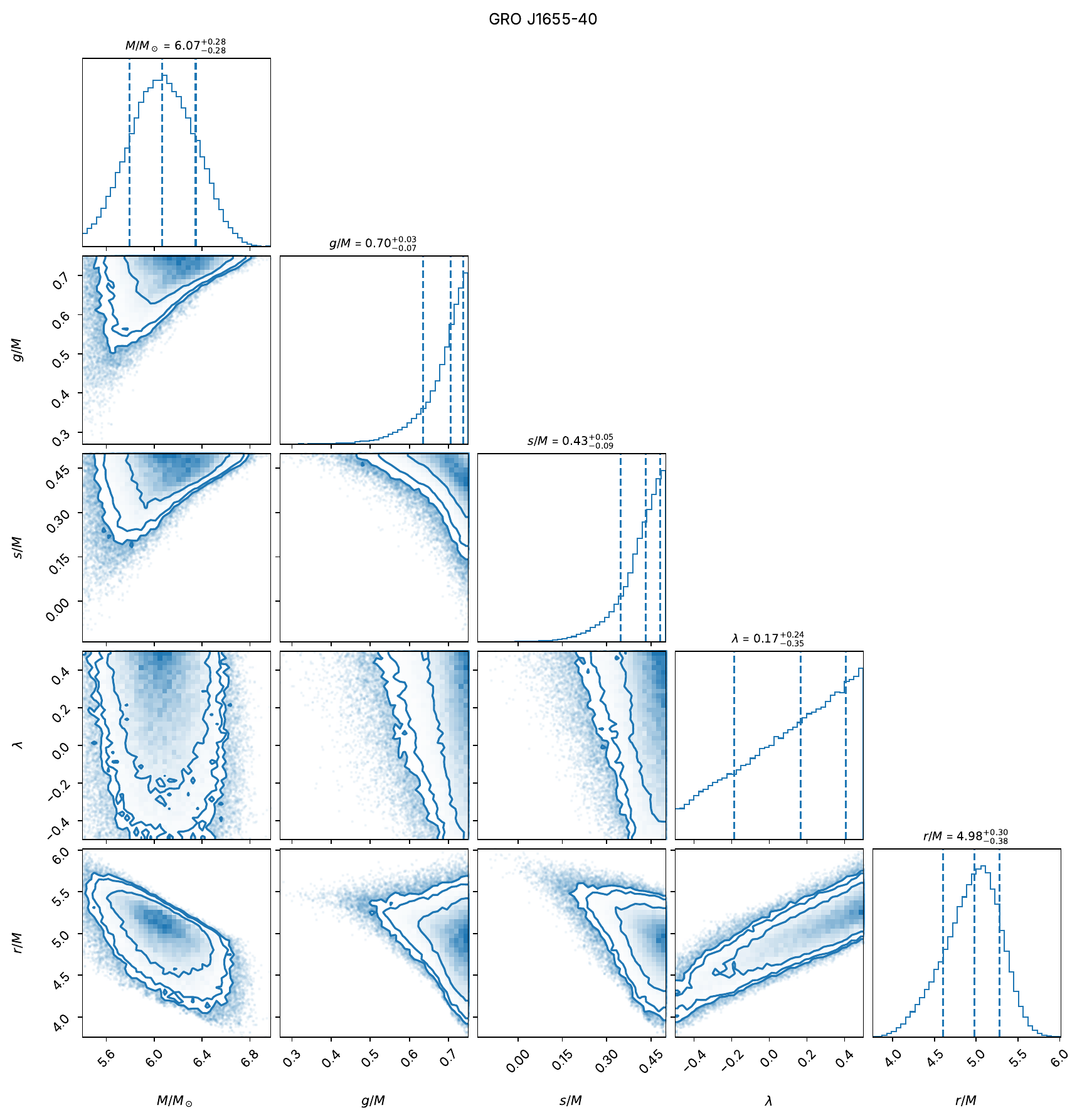}\hfill\includegraphics[width=0.49\linewidth]{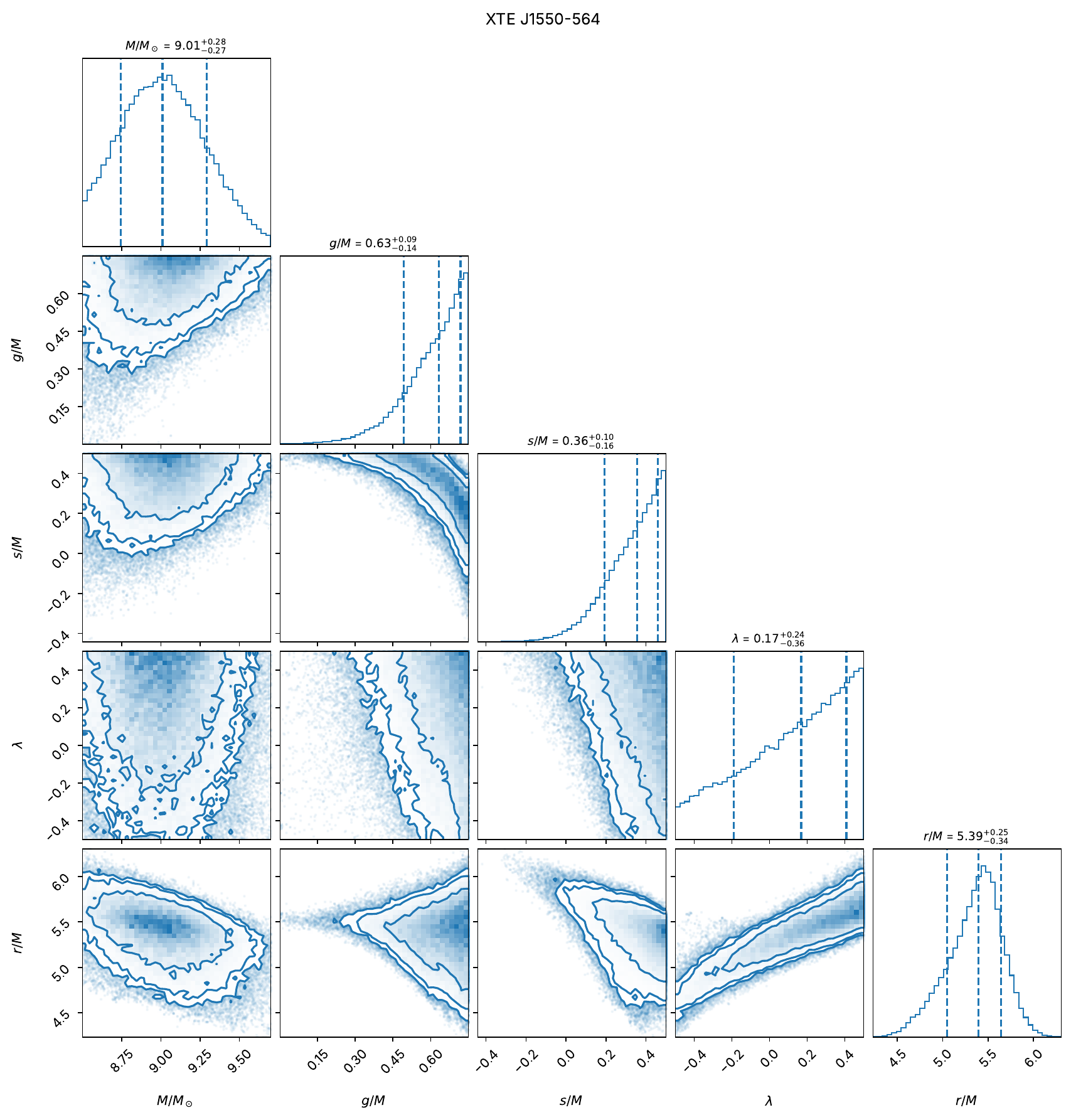}\\[2mm]
\includegraphics[width=0.49\linewidth]{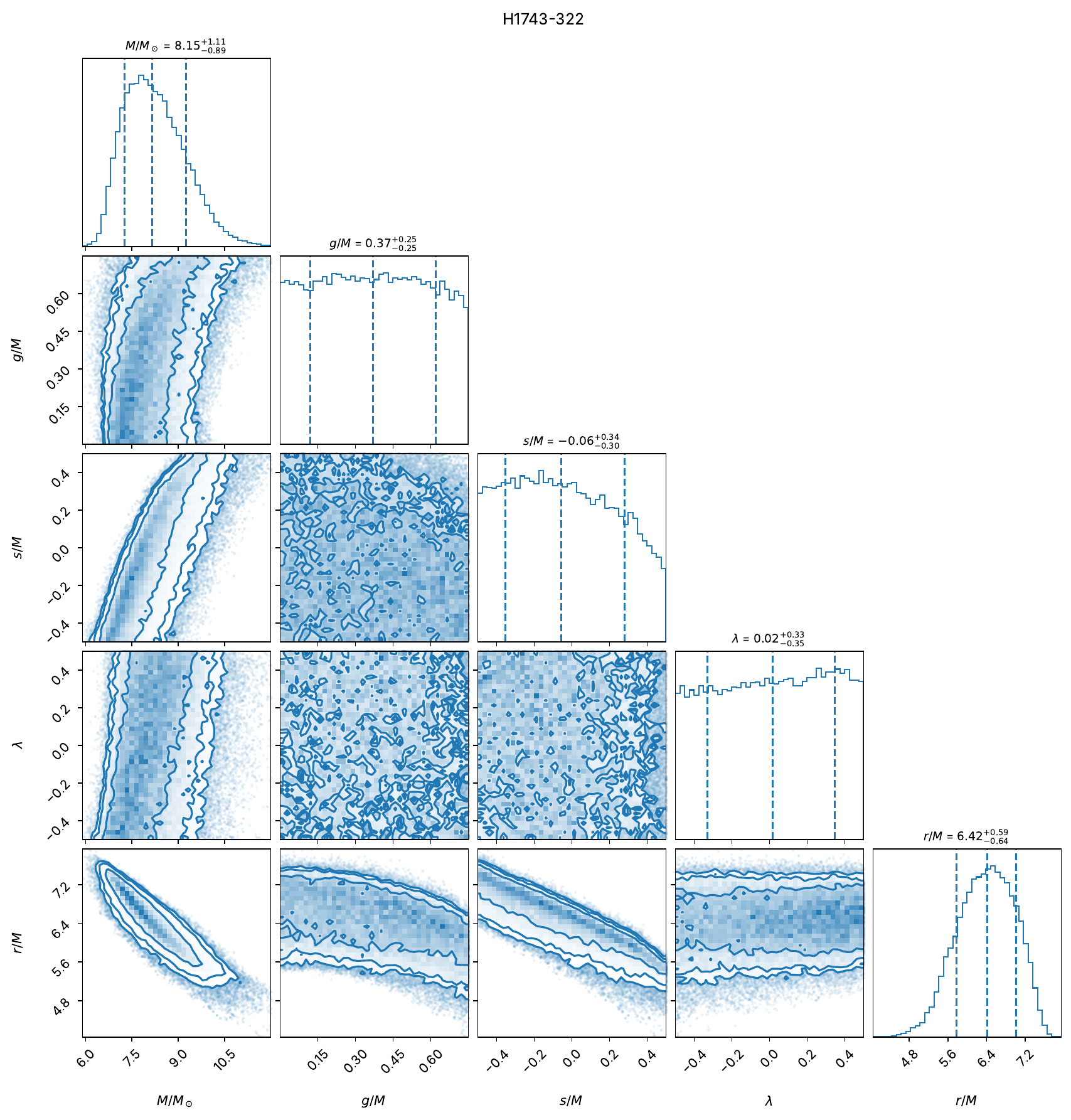}\hfill\includegraphics[width=0.49\linewidth]{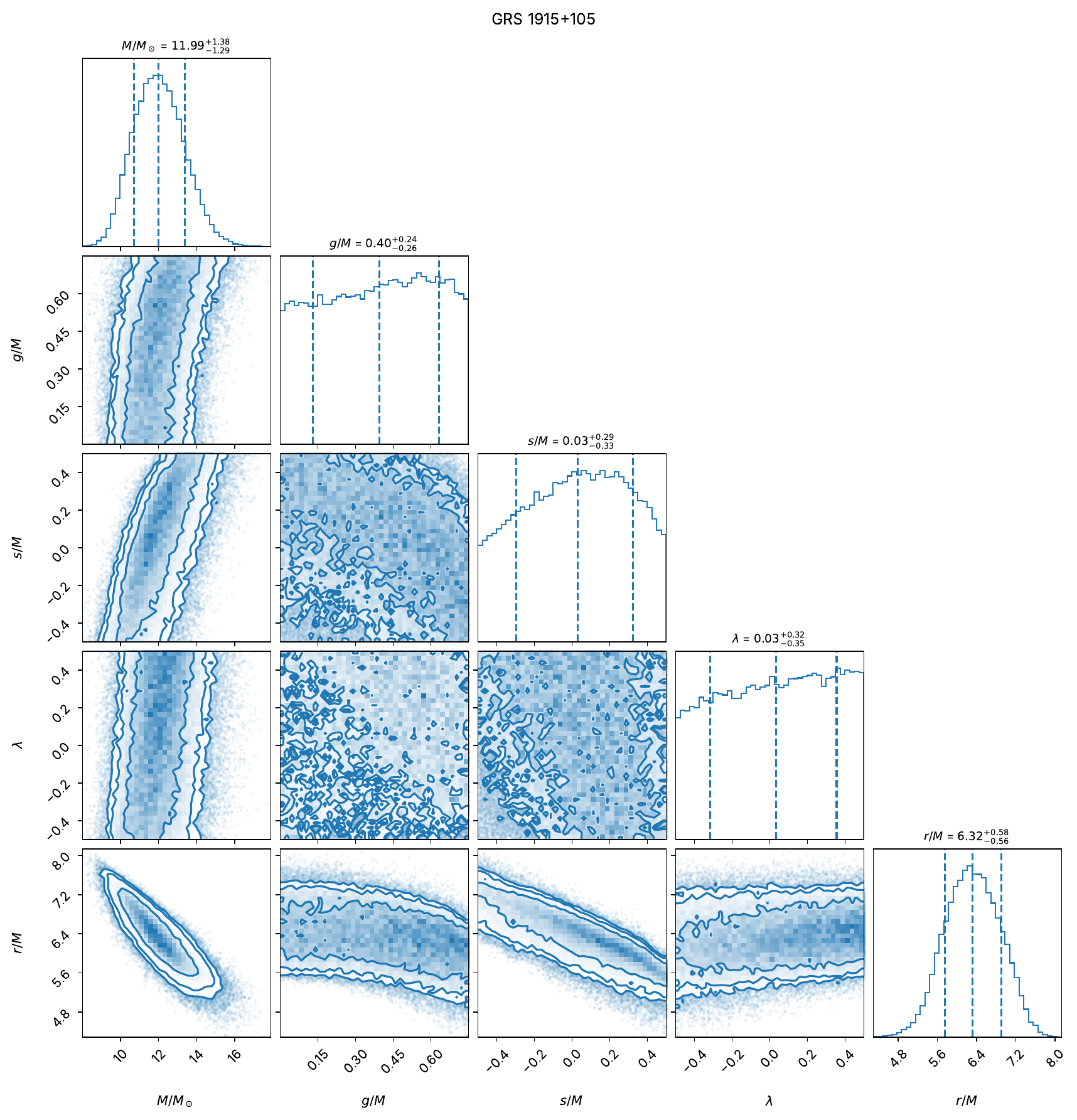}
\caption{Corner plots (68\%, 90\%, 95\% contours) of the parameters $(M,g,s,\lambda,r)$ for GRO~J1655$-$40, XTE~J1550$-$564, H1743$-$322 and GRS~1915+105.}
\label{fig:mcmc_corner}
\end{figure*}

Fig.~\ref{fig:mcmc_post} and Table~\ref{tab:mcmc} summarise the posteriors; the full corner plots are shown in Fig.~\ref{fig:mcmc_corner}. Three points deserve emphasis.

First, for GRO~J1655$-$40 and XTE~J1550$-$564, the Schwarzschild limit ($g=s=\lambda=0$) cannot reproduce both peaks with the dynamical mass: the best fit has $\chi^2=405$ and $110$, respectively. Both frequency pairs require a deeper potential than that of a non-rotating singular hole, and the posterior moves to $g/M=0.70^{+0.03}_{-0.07}$ and $0.63^{+0.09}_{-0.14}$ (with $g>0.53$ and $g>0.31$ at $95\%$). In the standard interpretation, this tension is absorbed by the Kerr spin; the Bardeen metric has none, so $g$ (together with the test-particle spin $s>0$) plays that role. The inferred values therefore measure the amount of additional compactness needed, not an independent detection of the regular core: Table~\ref{tab:isco} shows the same degeneracy between $g$ and $s$ for the ISCO.

Second, for H1743$-$322 and GRS~1915+105 the Schwarzschild limit fits the data (the mass priors are broad), and the posteriors of $g$ are almost flat: the data only exclude $g/M\gtrsim0.73$ at the $2\sigma$ level and give $g/M=0.37\pm0.25$ and $0.40\pm0.25$, i.e.\ they are prior-dominated and provide no upper bound beyond the extremal limit itself. The posteriors of $s$ and $\lambda$ coincide with their priors in all four sources except for the positive $s$ of the first two, so with two data points per source the data cannot constrain these parameters independently of $g$.

Third, the inferred spins $s\simeq0.4M$ for GRO~J1655$-$40 and XTE~J1550$-$564 lie at the edge of the validity of the pole--dipole approximation (Sec.~\ref{sec:discussion}), and the posteriors of $g$ for these two sources press against the upper end of the prior box. Both facts mean that these two results should be read as ``a rapidly compact configuration is required'', not as a measurement. A joint fit with the independent information on the black-hole spin of these sources, or with additional data (the hierarchy of QPO pairs and the line-profile spins), would be needed to make a quantitative statement.

\begin{table*}[t]
\caption{Twin-peak QPO data (upper $\nu_U$, lower $\nu_L$, in Hz), mass priors (Gaussian, $M_\odot$) and posterior medians with the 68\% credible intervals from the MCMC fit of the Bardeen--MPD model, $\nu_U=\nu_\phi$, $\nu_L=\nu_\phi-\nu_r$. The posteriors of $g$, $s$ and $\lambda$ are prior-limited ($0\le g\le0.75$, $|s|,|\lambda|\le0.5$) and the values for GRO~J1655$-$40 and XTE~J1550$-$564 are close to the edge of that box. The last column is the minimum $\chi^2$ of the Schwarzschild limit ($g=s=\lambda=0$, fitting $M$ within its prior range and $r$); the full model has $\chi^2\simeq0$ since it has more parameters (5) than data points (2).}
\label{tab:mcmc}
\begin{ruledtabular}
\begin{tabular}{l ccc c ccccc c}
source & $\nu_U$ & $\nu_L$ & $M_{\rm prior}$ & & $M/M_\odot$ & $g/M$ & $s/M$ & $\lambda$ & $r/M$ & $\chi^2_{\rm Schw}$\\
\hline
GRO J1655$-$40 & $450\pm3$ & $300\pm5$ & $6.3\pm0.3$ & & $6.07^{+0.28}_{-0.28}$ & $0.70^{+0.03}_{-0.07}$ & $0.43^{+0.05}_{-0.09}$ & $0.17^{+0.24}_{-0.35}$ & $4.98^{+0.30}_{-0.38}$ & 405 \\
XTE J1550$-$564 & $276\pm3$ & $184\pm5$ & $9.1\pm0.3$ & & $9.01^{+0.28}_{-0.27}$ & $0.63^{+0.09}_{-0.14}$ & $0.36^{+0.10}_{-0.16}$ & $0.17^{+0.24}_{-0.36}$ & $5.39^{+0.25}_{-0.34}$ & 110 \\
H1743$-$322 & $242\pm3$ & $166\pm5$ & $8\pm2$ & & $8.15^{+1.11}_{-0.89}$ & $0.37^{+0.25}_{-0.25}$ & $-0.06^{+0.34}_{-0.30}$ & $0.02^{+0.33}_{-0.35}$ & $6.42^{+0.59}_{-0.64}$ & 0 \\
GRS 1915+105 & $168\pm5$ & $113\pm5$ & $12.4\pm1.9$ & & $11.99^{+1.38}_{-1.29}$ & $0.40^{+0.24}_{-0.26}$ & $0.03^{+0.29}_{-0.33}$ & $0.03^{+0.32}_{-0.35}$ & $6.32^{+0.58}_{-0.56}$ & 0 \\
\end{tabular}
\end{ruledtabular}
\end{table*}

\section{Discussion: physical scales and limits of validity}\label{sec:discussion}

\emph{Pole--dipole validity.} The MPD equations with a Tulczyjew condition describe a spinning body only while its size, set by $s$ (spin per unit mass), is small against the curvature scale: $s\ll M$. Consequently the values $|s|\gtrsim M$ that appear in Figs.~\ref{fig:jcrit} and \ref{fig:sl} (shaded) are formal extrapolations. For a compact body of mass $m$ one has $s\lesssim m$, i.e.\ $s/M\lesssim m/M$, which makes the spin corrections of Sec.~IV of order the mass ratio, e.g.\ $10^{-5}$ in an extreme-mass-ratio inspiral. Only for comparable-size companions (or in a heuristic exploration of spin-curvature effects) are the quoted shifts of order unity.

\emph{Magnetic coupling and Dirac quantisation.} With the energy term $-\lambda g/r$ identified with the Coulomb interaction $-q\,Q_m/(mc^2r)$ between a particle of electric charge $q$ and a monopole of charge $Q_m$, the Dirac condition $qQ_m=n\hbar c/2$ gives $\lambda g=n\lambda_C/(2M)$, where $\lambda_C=\hbar/(mc)$ is the reduced Compton wavelength. For a proton and a $10\,M_\odot$ black hole ($M=GM_\odot/c^2\times10\simeq15\,$km) this is $\lambda g\simeq7\times10^{-21}\,n$, so couplings of order $0.1$ require either $n\sim10^{19}$ or a much lighter $m/q$, such as a dust grain with many elementary charges. This is an order-of-magnitude estimate under the stated identification, and the values $\lambda\sim0.1$--$1$ used here should be viewed as effective parameters that make the structure of the solutions visible rather than as astrophysically expected values. For the observable ISCO quantities they enter at first order only through $\mathcal L_{\rm ISCO}$ and $\mathcal E_{\rm ISCO}$ (Table~\ref{tab:coeff}), whereas $r_{\rm ISCO}$ is unchanged at first order.

\emph{Astrophysical observables.} The monopole parameter alone changes the radiative efficiency from $5.72\%$ to $6.09\%$ at $g/M=0.5$ and the ISCO frequency of a $10\,M_\odot$ black hole from $220$ to $241$ Hz (Table~\ref{tab:freq}); a prograde spin $s=0.3M$ gives a similar shift. Hence $g$, $s$ and $\lambda g$ are degenerate in the ISCO frequency and in $\eta$ and cannot be separated from these quantities alone; the 3:2 resonance location shifts differently ($\Delta\nu^{3:2}_\phi=+5.3\%$ against $\Delta\nu_{\rm ISCO}=+9.7\%$ at $g/M=0.5$), which in principle breaks part of the degeneracy. Since Kerr rotation shifts the same observables by far larger amounts, none of these corrections is separable from black-hole spin without independent information.

\emph{Collisions.} The collision energy is bounded only because the horizon is non-degenerate (Sec.~\ref{subsec:horizon}); it diverges on the extremal Bardeen horizon in the presence of a critical particle. In addition, the particles reaching the horizon are test particles in a fixed background; back-reaction and the actual escape of products are not included.

\section{Conclusion}\label{sec:conclusion}

We have studied magnetically charged spinning test particles around the Bardeen regular black hole, using the MPD equations with the Tulczyjew condition and a dual-field Coulomb coupling, and have verified the equatorial effective-potential description against direct integration of the full MPD system (Appendix~\ref{app:mpd}, Table~\ref{tab:val}).

(1) The ISCO is characterised analytically. Eq.~\eqref{eq:risco_pert} gives $r_{\rm ISCO}$ to second order in $(s,\lambda g,g^2)$ and recovers the known spinning-particle shift in the Schwarzschild limit. The magnetic coupling does not shift $r_{\rm ISCO}$ at first order but does shift $\mathcal L_{\rm ISCO}$ and $\mathcal E_{\rm ISCO}$; prograde spin and a larger $g$ both decrease the ISCO radius and increase the radiative efficiency, $\eta$ rising from $5.72\%$ to $6.09\%$ at $g/M=0.5$ and up to $8.19\%$ for $(g,s,\lambda)=(0.7,0.3,0.2)$.

(2) Orbital and radial epicyclic frequencies obtained from the linearised MPD flow give observable shifts of the ISCO frequency ($+9.7\%$ at $g/M=0.5$ for $10\,M_\odot$) and of the 3:2 resonance, which are however degenerate with spin and with the magnetic coupling.

(3) The superluminality condition is evaluated for the circular orbits themselves and gives, for each $(g,\lambda)$, the maximal spin $s_{\rm c}$ for which the ISCO is subluminal (Table~\ref{tab:sc}); attractive couplings relax and repulsive ones tighten the limit.

(4) The capture threshold $\mathcal L_{\rm cr}$ is lowered by prograde spin and by a repulsive coupling, raised by an attractive one, and depends only weakly on $g$.

(5) A Bayesian fit of the twin-peak QPOs of GRO~J1655$-$40, XTE~J1550$-$564, H1743$-$322 and GRS~1915+105 (Sec.~\ref{sec:mcmc}) shows that the first two require $g/M\gtrsim0.3$--$0.5$ (95\%) in the absence of black-hole rotation, whereas the latter two are consistent with the Schwarzschild limit; the parameters $s$ and $\lambda$ remain unconstrained.

(6) The collision energy at the horizon is finite for every non-degenerate horizon (Eq.~\eqref{eq:ecm_h}) and diverges, with the BSW scaling $E_{\rm CM}\propto(r-r_0)^{-1/2}$, on the extremal Bardeen horizon when a critical particle is present. The regular core therefore does not by itself cap the collision energy, contrary to a statement often made for regular black holes.

All values of the spin $|s|\gtrsim M$ and of the couplings $\lambda g\gtrsim10^{-2}$ should be regarded as formal (Sec.~\ref{sec:discussion}); the analytic coefficients and the validated numerical method are the robust results. Extensions include the dependence on the choice of spin supplementary condition, off-equatorial orbits and the inclusion of the electric field of the dual potential.

\section*{Acknowledgments}
 J.R. thanks Grant No. F-FA-2021-510 of the Uzbekistan Agency for Innovative Development. Y.C. and J.R. acknowledge the National Natural Science Foundation of China (NSFC) grant under Grant No. U2541210.

\section*{Data Availability Statement}  All data generated or analyzed during this study are included in this published article and its supplementary information files.

\section*{Code Availability} Code/Software used for the numerical verification presented in this work, including the numerical integration and calculations, is available from the corresponding author upon reasonable request.

\subsection*{Declaration of competing interest}
The authors declare no known competing financial interests or personal relationships that could have influenced the work reported in this paper.

\appendix
\section{Full MPD system and numerical verification}\label{app:mpd}

\emph{Equations.} With the Tulczyjew condition $S^{\mu\nu}p_\nu=0$ the four-velocity of the centroid is fixed by the nullspace of the linear system obtained by combining the spin equation with the supplementary condition,
\begin{eqnarray}
&&\Bigl[m^2\delta^\mu{}_\nu+p^\mu p_\nu+S^{\mu\sigma}\bigl(-\tfrac12R_{\sigma\nu\rho\lambda}S^{\rho\lambda}+\lambda\tilde F_{\sigma\nu}\bigr)\Bigr]u^\nu=0, \nonumber \\
&&u^\mu p_\mu=-m^2,
\label{eq:A1}
\end{eqnarray}
i.e.\ $u^\mu$ is the (one-dimensional) null vector of the matrix in brackets, which follows from differentiating $S^{\mu\nu}p_\nu=0$ and using Eqs.~\eqref{eq:A2}--\eqref{eq:A3}; it reduces to $u^\mu=p^\mu/m$ for $s=0$ and differs from it at $O(s\lambda g)$ and $O(s^2)$. The equations of motion are
\begin{eqnarray}
\frac{Dp_\mu}{d\tau}&=&-\tfrac12R_{\mu\nu\alpha\beta}u^\nu S^{\alpha\beta}+\lambda\,\tilde F_{\mu\nu}u^\nu,\label{eq:A2}\\
\frac{DS^{\mu\nu}}{d\tau}&=&p^\mu u^\nu-p^\nu u^\mu,\label{eq:A3}
\end{eqnarray}
with the dual-field coupling $\tilde A_t=g/r$, $\tilde F_{tr}=g/r^2$ and the constraints
\begin{equation}
S^{\mu\nu}p_\nu=0,\qquad p^\mu p_\mu=-m^2,\qquad S^2=\tfrac12S_{\mu\nu}S^{\mu\nu}=s^2m^2.
\label{eq:A4}
\end{equation}
The spin-tensor orientation is chosen so that, for the equatorial spin along $+z$, the Riemann-curvature convention of the text gives the effective potential of Eq.~\eqref{eq:Vpm}; the opposite orientation reproduces the same dynamics with $s\to-s$. The conserved quantities are $\mathcal E=-p_t+\tfrac12S^{\mu\nu}\nabla_\mu\xi^{(t)}_\nu+\lambda\tilde A_t$, and the analogous expression for $J$ with $\xi^{(\phi)}$.

\emph{Frequencies.} For a circular orbit the angular velocity is $\Omega=u^\phi/u^t$ with $u^\mu$ from Eq.~\eqref{eq:A1}. The radial epicyclic frequency is obtained from the Jacobian of the reduced nine-dimensional flow (the components $x^\mu$, $p_\mu$ and the antisymmetric part of $S^{\mu\nu}$ with the constraints imposed) linearised about the circular orbit; the smaller of the two imaginary eigenvalue pairs is the radial mode and is expressed in coordinate time via $dt/d\tau$. It was checked against the analytic $s=\lambda=0$ result $\Omega_r^2=\Omega^2(1-6M/r)$ and against time-domain integration (relative difference $<10^{-3}$ at the tested point).

\emph{Verification.} Table~\ref{tab:val} compares the effective-potential description with the direct integration of Eqs.~\eqref{eq:A1}--\eqref{eq:A4}: circular orbits remain circular to $10^{-13}$ and the conserved energy and angular momentum are preserved to better than $10^{-11}$, while the Tulczyjew normalisation drifts at the $10^{-6}$ level over $120\,M$.

\begin{table*}[t]
\caption{Verification of the effective-potential results against direct numerical integration of the full MPD system, Eqs.~(\ref{eq:A1})--(\ref{eq:A4}), at $g/M=0.5$. A circular orbit with the ISCO parameters is integrated over $\Delta\tau=60\,M$ (maximal radial drift and energy drift), and an eccentric orbit started at $1.03\,r_{\rm ISCO}$ with the same $(\mathcal E,\mathcal L)$ over $\Delta\tau=120\,M$ (drift of $\mathcal E$, $J$, of the normalisation $p^\mu p_\mu+m^2$ and of the spin magnitude).}
\label{tab:val}
\begin{ruledtabular}
\begin{tabular}{c c c c ccccc}
 & & \multicolumn{2}{c}{circular orbit} & \multicolumn{4}{c}{eccentric orbit}\\
$(s,\lambda)$ & $r_{\rm ISCO}$ & $\max|\delta r|$ & $|\delta\mathcal E|$ & $|\delta\mathcal E|$ & $|\delta J|$ & $|p^2+m^2|$ & $|\delta S|$\\
\hline
$(+0.3,+0.2)$ & $5.0828$ & $4.9\times10^{-13}$ & $5.5\times10^{-12}$ & $7.7\times10^{-13}$ & $5.3\times10^{-15}$ & $5.3\times10^{-6}$ & $2.3\times10^{-7}$ \\
$(-0.3,-0.2)$ & $6.0531$ & $1.2\times10^{-13}$ & $4.6\times10^{-12}$ & $7.0\times10^{-13}$ & $1.3\times10^{-15}$ & $3.4\times10^{-6}$ & $1.5\times10^{-6}$ \\
$(+0.5,+0)$ & $4.6251$ & $2.8\times10^{-13}$ & $1.8\times10^{-11}$ & $2.4\times10^{-12}$ & $1.3\times10^{-15}$ & $5.8\times10^{-6}$ & $7.8\times10^{-6}$ \\
\end{tabular}
\end{ruledtabular}
\end{table*}

\bibliographystyle{apsrev4-1}

\bibliography{References.bib}

@ARTICLE{2021A&AT...32...83N,
       author = {{Nuritdinov}, S.~N. and {Muminov}, A.~A. and {Botirov}, F.~U.},
        title = "{On the method for the analysis of compulsive phase mixing and its application in cosmogony}",
      journal = {Astronomical and Astrophysical Transactions},
         year = 2021,
        month = jan,
       volume = {32},
       number = {2},
        pages = {83-88},
          doi = {10.48550/arXiv.2002.09310},
archivePrefix = {arXiv},
       eprint = {2002.09310},
 primaryClass = {astro-ph.GA},
       adsurl = {https://ui.adsabs.harvard.edu/abs/2021A&AT...32...83N}
}

@article{Senovilla:2014gza,
    author = "Senovilla, Jos{\'e} M. M. and Garfinkle, David",
    title = "{The 1965 Penrose singularity theorem}",
    eprint = "1410.5226",
    archivePrefix = "arXiv",
    primaryClass = "gr-qc",
    doi = "10.1088/0264-9381/32/12/124008",
    journal = "Class. Quant. Grav.",
    volume = "32",
    number = "12",
    pages = "124008",
    year = "2015"
}

@article{Hawking:1970zqf,
    author = "Hawking, S. W. and Penrose, R.",
    title = "{The Singularities of gravitational collapse and cosmology}",
    doi = "10.1098/rspa.1970.0021",
    journal = "Proc. Roy. Soc. Lond. A",
    volume = "314",
    pages = "529--548",
    year = "1970"
}

@INPROCEEDINGS{1968qtr..conf...87B,
       author = {{Bardeen}, James},
        title = "{Non-singular general relativistic gravitational collapse}",
    booktitle = {Proceedings of the 5th International Conference on Gravitation and the Theory of Relativity},
         year = 1968,
        month = sep,
        pages = {87},
       adsurl = {https://ui.adsabs.harvard.edu/abs/1968qtr..conf...87B}
}

@article{Cai:2025ejd,
    author = "Cai, Lan-Lan and Lai, Meng-Yun and Zou, De-Cheng and Zhang, Lina and Huang, Hyat",
    title = "{Spontaneous scalarization of regular Hayward black holes in Einstein-nonlinear electromagnetic-scalar gravity}",
    journal = {arXiv e-prints},
    eprint = "2510.18354",
    archivePrefix = "arXiv",
    primaryClass = "gr-qc",
    month = "10",
    year = "2025"
}

@article{Ayon-Beato:1999kuh,
    author = "Ayon-Beato, Eloy and Garcia, Alberto",
    title = "{New regular black hole solution from nonlinear electrodynamics}",
    eprint = "hep-th/9911174",
    archivePrefix = "arXiv",
    doi = "10.1016/S0370-2693(99)01038-2",
    journal = "Phys. Lett. B",
    volume = "464",
    pages = "25",
    year = "1999"
}

@article{Dymnikova:2004zc,
    author = "Dymnikova, Irina",
    title = "{Regular electrically charged vacuum structures with de Sitter centre in nonlinear electrodynamics coupled to general relativity}",
    eprint = "gr-qc/0407072",
    archivePrefix = "arXiv",
    doi = "10.1088/0264-9381/21/18/009",
    journal = "Class. Quant. Grav.",
    volume = "21",
    pages = "4417--4428",
    year = "2004"
}

@article{Mathisson:1937zz,
    author = "Mathisson, Myron",
    title = "{Neue mechanik materieller systemes}",
    journal = "Acta Phys. Polon.",
    volume = "6",
    pages = "163--200",
    year = "1937"
}

@article{Papapetrou:1951pa,
    author = "Papapetrou, Achille",
    title = "{Spinning test particles in general relativity. 1.}",
    doi = "10.1098/rspa.1951.0200",
    journal = "Proc. Roy. Soc. Lond. A",
    volume = "209",
    pages = "248--258",
    year = "1951"
}

@article{Dixon-1970,
    author  = "Dixon, W. G.",
    title   = "{Dynamics of extended bodies in general relativity. I. Momentum and angular momentum}",
    doi     = "10.1098/rspa.1970.0020",
    journal = "Proc. Roy. Soc. Lond. A",
    volume  = "314",
    number  = "1519",
    pages   = "499--527",
    year    = "1970"
}

@article{PhysRevD.6.1476,
  title = {Electromagnetic Fields and Massive Bodies},
  author = {Wald, Robert M.},
  journal = {Phys. Rev. D},
  volume = {6},
  issue = {6},
  pages = {1476--1479},
  numpages = {0},
  year = {1972},
  month = {Sep},
  publisher = {American Physical Society},
  doi = {10.1103/PhysRevD.6.1476},
  url = {https://link.aps.org/doi/10.1103/PhysRevD.6.1476}
}

@article{Semerak_1999,
 doi = {10.1088/0264-9381/16/11/402},
 url = {https://doi.org/10.1088/0264-9381/16/11/402},
year = {1999},
month = {nov},
publisher = {},
volume = {16},
number = {11},
pages = {3769},
author = {O Semerak},
title = {Gravitomagnetic 
clock effect and extremely accelerated 
observers},
journal = {Classical and Quantum Gravity}
}

@article{Blaschke:2022pgf,
    author = "Blaschke, Martin and Stuchl{\'\i}k, Zden{\v{e}}k and Hensh, Sudipta",
    title = "{Evolution of braneworld Kerr-Newman naked singularities}",
    eprint = "2205.07558",
    archivePrefix = "arXiv",
    primaryClass = "gr-qc",
    doi = "10.1103/PhysRevD.105.084069",
    journal = "Phys. Rev. D",
    volume = "105",
    number = "8",
    pages = "084069",
    year = "2022"
}

@article{Wang:2025oek,
    author = "Wang, Rui and Shi, Qi-Long and Xiong, Wei and Li, Peng-Cheng",
    title = "{Tidal Love numbers for regular black holes}",
    journal = {arXiv e-prints},
    eprint = "2512.05767",
    archivePrefix = "arXiv",
    primaryClass = "gr-qc",
    month = "12",
    year = "2025"
}

@article{Rayimbaev:2022hrn,
    author = "Rayimbaev, Javlon and Bardiev, Dilshodbek and Abdulxamidov, Farrux and Abdujabbarov, Ahmadjon and Ahmedov, Bobomurat",
    title = "{Magnetized and Magnetically Charged Particles Motion around Regular Bardeen Black Hole in 4D Einstein Gauss{\textendash}Bonnet Gravity}",
    doi = "10.3390/universe8100549",
    journal = "Universe",
    volume = "8",
    number = "10",
    pages = "549",
    year = "2022"
}

@article{Rayimbaev:2022mrk,
    author = "Rayimbaev, Javlon and Majeed, Bushra and Jamil, Mubasher and Jusufi, Kimet and Wang, Anzhong",
    title = "{Quasiperiodic oscillations, quasinormal modes and shadows of Bardeen{\textendash}Kiselev Black Holes}",
    eprint = "2202.11509",
    archivePrefix = "arXiv",
    primaryClass = "gr-qc",
    doi = "10.1016/j.dark.2021.100930",
    journal = "Phys. Dark Univ.",
    volume = "35",
    pages = "100930",
    year = "2022"
}

@article{doi:10.1142/S0218271822500559,
author = {Rayimbaev, Javlon and Bardiev, Dilshodbek and Mirzaev, Temurbek and Abdujabbarov, Ahmadjon and Khalmirzaev, Akram},
title = {Shadow and massless particles around regular Bardeen black holes in 4D Einstein Gauss–Bonnet gravity},
journal = {International Journal of Modern Physics D},
volume = {31},
number = {07},
pages = {2250055},
year = {2022},
doi = {10.1142/S0218271822500559},
URL = {https://doi.org/10.1142/S0218271822500559},
eprint = {https://doi.org/10.1142/S0218271822500559}
}

@article{Zeng:2025wlb,
    author = "Zeng, Hui and Meng, Yuan",
    title = "{Images from disk and spherical accretions of Bardeen black hole surrounded by perfect fluid dark matter}",
    journal = {arXiv e-prints},
    eprint = "2512.05147",
    archivePrefix = "arXiv",
    primaryClass = "gr-qc",
    month = "12",
    year = "2025"
}

@article{Nascimento:2025get,
    author = "Nascimento, F. F. and Bezerra, V. B. and Toledo, J. M. and Morais, P. H. and Rocha, J. C.",
    title = "{Charged Bardeen black hole with a cosmological constant and surrounded by quintessence and a cloud of strings}",
    journal = {arXiv e-prints},
    eprint = "2511.02201",
    archivePrefix = "arXiv",
    primaryClass = "gr-qc",
    month = "11",
    year = "2025"
}

@article{Mahanta:2025uph,
    author = "Mahanta, Gunindra Krishna",
    title = "{Thermodynamic geodesics in Bardeen regular black hole: conventional vs. modified geometrothermodynamics metrics}",
    eprint = "2510.22398",
    archivePrefix = "arXiv",
    primaryClass = "gr-qc",
    doi = "10.1140/epjc/s10052-025-14947-8",
    journal = "Eur. Phys. J. C",
    volume = "85",
    number = "10",
    pages = "1203",
    year = "2025"
}

@article{Benkrane:2025kdq,
    author = "Benkrane, Abdelhakim",
    title = "{Non-rotating hairy Bardeen black hole: thermodynamic properties, entropic force, and circular photon motion}",
    doi = "10.1007/s10509-025-04499-4",
    journal = "Astrophys. Space Sci.",
    volume = "370",
    number = "10",
    pages = "110",
    year = "2025"
}

@article{Ma:2025tzm,
    author = "Ma, Meng-Sen and He, Yun and Wang, Xiao-Ming and Li, Huai-Fan",
    title = "{From singular to regular: Revisiting thermodynamics of Bardeen-Ads black holes}",
    eprint = "2510.06576",
    archivePrefix = "arXiv",
    primaryClass = "hep-th",
    doi = "10.1016/j.physletb.2025.139961",
    journal = "Phys. Lett. B",
    volume = "870",
    pages = "139961",
    year = "2025"
}

@article{Cordeiro:2025ydg,
    author = "Cordeiro, Daniela S. J. and Junior, Ednaldo L. B. and Junior, Jos{\'e} Tarciso S. S. and Lobo, Francisco S. N. and Ramos, Jorde A. A. and Rodrigues, Manuel E. and Rubiera-Garcia, Diego and da Silva, Lu{\'\i}s F. Dias and Vieira, Henrique A.",
    title = "{Regular Bardeen black holes within non-minimal scalar-linear electrodynamic couplings}",
    journal = {arXiv e-prints},
    eprint = "2509.24052",
    archivePrefix = "arXiv",
    primaryClass = "gr-qc",
    month = "9",
    year = "2025"
}

@article{Jumaniyozov:2025nnj,
    author = "Jumaniyozov, Shokhzod and Rayimbaev, Javlon and Akhmedov, Munisbek and Turaev, Yunus and Usanov, Sulton and Palvanov, Satimbay",
    title = "{Dynamics of charged-spinning particles around charged Kalb{\textendash}Ramond black holes}",
    doi = "10.1016/j.dark.2025.102161",
    journal = "Phys. Dark Univ.",
    volume = "50",
    pages = "102161",
    year = "2025"
}

@article{Jumaniyozov:2025irx,
    author = "Jumaniyozov, Shokhzod and Rayimbaev, Javlon and Turaev, Yunus",
    title = "{Dynamics of spinning particles around a charged black-bounce spacetime}",
    doi = "10.1140/epjc/s10052-025-14834-2",
    journal = "Eur. Phys. J. C",
    volume = "85",
    number = "11",
    pages = "1247",
    year = "2025"
}

@article{Oteev:2025yjh,
    author = "Oteev, Tursinbay and Stuchl{\'\i}k, Zden{\v{e}}k and Sharibaev, Murat and Rayimbaev, Javlon and Ibragimov, Inomjon and Turaev, Yunus and Vapayev, Murodbek",
    title = {{Spin effects on charged particle motion in magnetized Reissner{\textendash}Nordstr{\"o}m spacetime}},
    doi = "10.1140/epjc/s10052-025-14974-5",
    journal = "Eur. Phys. J. C",
    volume = "85",
    number = "10",
    pages = "1204",
    year = "2025"
}

@article{Oteev:2025jvf,
    author = "Oteev, Tursinbay and Rayimbaev, Javlon and Ahmedov, Bobomurat and Ibragimov, Inomjon and Vapayev, Murodbek and Muminov, Sokhibjan",
    title = "{Collisions and circular motion of spinning-charged particle around magnetized black holes in modified gravity}",
    doi = "10.1016/j.dark.2025.102118",
    journal = "Phys. Dark Univ.",
    volume = "50",
    pages = "102118",
    year = "2025"
}

@article{Chen:2025ncm,
    author = "Chen, Yi-Ping and Hsieh, Tien and Lee, Da-Shin",
    title = "{Motions of spinning particles in the Kerr-Newman black hole exterior and gravitational wave emission. I. Periodic orbits}",
    journal = {arXiv e-prints},
    eprint = "2510.05603",
    archivePrefix = "arXiv",
    primaryClass = "gr-qc",
    month = "10",
    year = "2025"
}

@article{Iosifidis:2025ano,
    author = "Iosifidis, Damianos",
    title = "{Lagrangian Dynamics of Spinning Pole-Dipole-Quadrupole Particles in Metric-Affine Geometries}",
    journal = {arXiv e-prints},
    eprint = "2509.14757",
    archivePrefix = "arXiv",
    primaryClass = "gr-qc",
    month = "9",
    year = "2025"
}

@article{Abdukayumova:2025ztr,
    author = "Abdukayumova, Gulnisa and Atamurotov, Farruh and Abdujabbarov, Ahmadjon and Channuie, Phongpichit and Mustafa, G.",
    title = "{Dynamics of spinning particles around the Reissner-Nordstrom-like black hole}",
    doi = "10.1016/j.nuclphysb.2025.117121",
    journal = "Nucl. Phys. B",
    volume = "1019",
    pages = "117121",
    year = "2025"
}

@article{Oteev:2025csb,
    author = "Oteev, Tursinbay and Stuchl{\'\i}k, Zden{\v{e}}k and Rayimbaev, Javlon and Ibragimov, Inomjon and Sharibaev, Murat and Abdujabbarov, Ahmadjon",
    title = "{Circular motion and collisions of charged spinning test particles around magnetized Schwarzschild black hole}",
    doi = "10.1140/epjc/s10052-025-14660-6",
    journal = "Eur. Phys. J. C",
    volume = "85",
    number = "9",
    pages = "953",
    year = "2025"
}

@article{Jumaniyozov:2025uwo,
    author = "Jumaniyozov, Shokhzod and Khan, Saeed Ullah and Rayimbaev, Javlon and Ibragimov, Inomjon and Abdujabbarov, Ahmadjon and Stuchl{\'\i}k, Zden{\v{e}}k",
    title = "{Collisions and circular motion of spinning magnetized particles orbiting magnetized Kerr black holes}",
    doi = "10.1103/bl45-hbjm",
    journal = "Phys. Rev. D",
    volume = "112",
    number = "4",
    pages = "044068",
    year = "2025"
}

@article{Umarov:2025wzm,
    author = "Umarov, Dilmurod and Atamurotov, Farruh and Ghosh, Sushant G. and Abdujabbarov, Ahmadjon and Mustafa, G.",
    title = "{Dynamics of spinning particles around static black holes in effective quantum gravity}",
    doi = "10.1140/epjc/s10052-025-14541-y",
    journal = "Eur. Phys. J. C",
    volume = "85",
    number = "7",
    pages = "800",
    year = "2025"
}

@article{Alimova:2025izs,
    author = "Alimova, Asalkhon and Atamurotov, Farruh and Abdujabbarov, Ahmadjon and Mustafa, G. and Channuie, Phongpichit",
    title = "{Impact of quantum-corrected parameter on spinning particle motion around a black hole}",
    doi = "10.1140/epjc/s10052-025-14385-6",
    journal = "Eur. Phys. J. C",
    volume = "85",
    number = "6",
    pages = "646",
    year = "2025"
}

@article{Mannobova:2025uqf,
    author = "Mannobova, Sojida and Atamurotov, Farruh and Abdujabbarov, Ahmadjon and Alkahtani, Badr S. and Mustafa, G.",
    title = "{Spinning particle motion around asymptotically safe gravity exhibiting regular black holes}",
    doi = "10.1140/epjc/s10052-025-14305-8",
    journal = "Eur. Phys. J. C",
    volume = "85",
    number = "5",
    pages = "586",
    year = "2025"
}

@article{Jumaniyozov:2025xxh,
    author = "Jumaniyozov, Shokhzod",
    title = "{Thermodynamic fluctuations and radiation properties around Schwarzschild black holes immersed in Hernquist dark matter halo}",
    doi = "10.1140/epjc/s10052-025-15025-9",
    journal = "Eur. Phys. J. C",
    volume = "85",
    number = "11",
    pages = "1267",
    year = "2025"
}

@article{Jumaniyozov:2025lox,
    author = "Jumaniyozov, Shokhzod and Rayimbaev, Javlon and Turaev, Yunus and Ibragimov, Inomjon and Abdulazizov, Bakhrom and Akhmedov, Munisbek and Djumanov, Sherzod",
    title = "{Rotating charged black holes in Kalb{\textendash}Ramond gravity: electromagnetic fields, circular motion and collisions of charged particles}",
    doi = "10.1140/epjc/s10052-025-14952-x",
    journal = "Eur. Phys. J. C",
    volume = "85",
    number = "10",
    pages = "1201",
    year = "2025"
}

@article{Shermatov:2025qad,
    author = {Shermatov, Abubakir and Rayimbaev, Javlon and Murodov, Sardor and L{\"u}tf{\"u}o{\u{g}}lu, Bekir Can and Ahmedov, Bobomurat and Zahid, Muhammad and Ibragimov, Inomjon and Shermatov, Bahran},
    title = "{Phantom black holes in f(R,T) gravity: From circular orbits to QPO tests}",
    doi = "10.1016/j.dark.2025.102110",
    journal = "Phys. Dark Univ.",
    volume = "50",
    pages = "102110",
    year = "2025"
}

@article{Saydullayev:2025oop,
    author = "Saydullayev, Sirojiddin and Nishonov, Isomiddin and Dusaliyev, Muysin and Xoldorov, Obid and Murodov, Sardor and Karshiboev, Shavkat and Urinov, Sunnatillo and Rahmatov, Bekzod",
    title = "{Black hole surrounded by perfect fluid dark matter in STV gravity: particle dynamics, thermodynamics, gravitational weak lensing and EHT tests}",
    doi = "10.1140/epjc/s10052-025-14780-z",
    journal = "Eur. Phys. J. C",
    volume = "85",
    number = "9",
    pages = "1081",
    year = "2025"
}

@article{NOORIGASHTI2026117244,
title = {Thermodynamic signatures in black hole geometry and harmonic oscillations with nonlinear electromagnetic fields and phantom global monopole},
journal = {Nuclear Physics B},
volume = {1022},
pages = {117244},
year = {2026},
issn = {0550-3213},
doi = {https://doi.org/10.1016/j.nuclphysb.2025.117244},
url = {https://www.sciencedirect.com/science/article/pii/S055032132500450X},
author = {Saeed {Noori Gashti} and Yassine Sekhmani and Mohammad Ali S Afshar and Mohammad Reza Alipour and Mohammadreza {Khodajou Masouleh} and Behnam Pourhassan and Jafar Sadeghi and Javlon Rayimbaev}
}

@article{RAHMATOV2025102102,
title = {QPO tests and charged particles around regular Ayón-Beato-Garcia black holes},
journal = {Physics of the Dark Universe},
volume = {50},
pages = {102102},
year = {2025},
issn = {2212-6864},
doi = {https://doi.org/10.1016/j.dark.2025.102102},
url = {https://www.sciencedirect.com/science/article/pii/S221268642500295X},
author = {Bekzod Rahmatov and Sardor Murodov and Javlon Rayimbaev and Sokhibjan Muminov and Inomjon Ibragimov and Rashid Eshburiev}
}

@article{khan2025circular,
  title={Circular motion and acceleration of charged particles around magnetized rotating black holes in scalar-tensor-vector gravity},
  author={Khan, Saeed Ullah and Rayimbaev, Javlon and Chen, Zhi-Min and Stuchl{\'\i}k, Zden{\v{e}}k},
  journal={Chinese Physics C},
  volume={49},
  number={9},
  pages={095102},
  doi={10.1088/1674-1137/add8fc},
  year={2025}
}

@article{Banados:2009pr,
  author="Ba\~nados, M. and Silk, J. and West, S. M.",
  title="{Kerr black holes as particle accelerators to arbitrarily high energy}",
  journal="Phys. Rev. Lett.", volume="103", pages="111102", year="2009", eprint="0909.0169", archivePrefix="arXiv"}

@article{Zaslavskii:2010jd,
  author="Zaslavskii, O. B.",
  title="{Acceleration of particles as universal property of rotating black holes}",
  journal="JETP Lett.", volume="92", pages="571", year="2010", eprint="1007.3678", archivePrefix="arXiv"}

@article{Berti:2009bk,
  author="Berti, E. and Cardoso, V. and Gualtieri, L. and Pretorius, F. and Sperhake, U.",
  title="{Comment on ``Kerr black holes as particle accelerators to arbitrarily high energy''}",
  journal="Phys. Rev. Lett.", volume="103", pages="239001", year="2009"}

@article{Jefremov:2015gza,
  author="Jefremov, P. I. and Tsupko, O. Yu. and Bisnovatyi-Kogan, G. S.",
  title="{Innermost stable circular orbits of spinning test particles in Schwarzschild and Kerr space-times}",
  journal="Phys. Rev. D", volume="91", pages="124030", year="2015", eprint="1503.07060", archivePrefix="arXiv"}

@article{Poincare:1896,
  author="Poincar\'e, H.",
  title="{Remarques sur une exp\'erience de M. Birkeland}",
  journal="C. R. Acad. Sci. Paris", volume="123", pages="530", year="1896"}

@incollection{Novikov:1973,
  author="Novikov, I. D. and Thorne, K. S.",
  title="{Astrophysics of black holes}",
  booktitle="Black Holes (Les Astres Occlus)", publisher="Gordon and Breach", address="New York", year="1973", pages="343"}

@article{Bardeen:1972fi,
  author="Bardeen, J. M. and Press, W. H. and Teukolsky, S. A.",
  title="{Rotating black holes: locally nonrotating frames, energy extraction, and scalar synchrotron radiation}",
  journal="Astrophys. J.", volume="178", pages="347", year="1972"}

@article{ForemanMackey:2013io,
 author="Foreman-Mackey, D. and Hogg, D. W. and Lang, D. and Goodman, J.",
 title="{emcee: The MCMC Hammer}", journal="Publ. Astron. Soc. Pac.", volume="125", pages="306", year="2013", eprint="1202.3665", archivePrefix="arXiv"}

@article{Remillard:2006fc,
 author="Remillard, R. A. and McClintock, J. E.",
 title="{X-ray properties of black-hole binaries}", journal="Ann. Rev. Astron. Astrophys.", volume="44", pages="49", year="2006"}

@article{Strohmayer:2001,
 author="Strohmayer, T. E.", title="{Discovery of a 450 Hz quasi-periodic oscillation from the microquasar GRO J1655-40 with the Rossi X-Ray Timing Explorer}", journal="Astrophys. J. Lett.", volume="552", pages="L49", year="2001"}

@article{Remillard:2002,
 author="Remillard, R. A. and others", title="{X-ray oscillations in the black hole binary XTE J1550-564}", journal="Astrophys. J.", volume="580", pages="1030", year="2002"}

@article{Orosz:2011,
 author="Orosz, J. A. and others", title="{The mass of the black hole in XTE J1550-564}", journal="Astrophys. J.", volume="730", pages="75", year="2011"}

@article{Reid:2014,
 author="Reid, M. J. and others", title="{The parallax and proper motion of GRS 1915+105}", journal="Astrophys. J.", volume="796", pages="2", year="2014"}

@article{Greene:2001,
 author="Greene, J. and Bailyn, C. D. and Orosz, J. A.", title="{Optical and infrared photometry of GRO J1655-40}", journal="Astrophys. J.", volume="554", pages="1290", year="2001"}

@article{Morgan:1997,
 author="Morgan, E. H. and Remillard, R. A. and Greiner, J.", title="{RXTE observations of QPOs in the black hole candidate GRS 1915+105}", journal="Astrophys. J.", volume="482", pages="993", year="1997"}
\end{document}